\documentclass[a4paper,oneside,12pt,notitlepage]{report}

\usepackage[a4paper, margin=0.8in]{geometry}
\usepackage{amsmath,amssymb}
\usepackage{enumitem,placeins}
\usepackage{graphicx}

\usepackage{color}
\definecolor{grey}{rgb}{0.6,0.6,0.6}
\definecolor{darkblue}{rgb}{0,0,.3}
\definecolor{darkgreen}{rgb}{0,.5,0}
\definecolor{deepred}{rgb}{0.4,0,0}
\usepackage[colorlinks,linkcolor=darkblue,urlcolor=deepred,citecolor=darkgreen]{hyperref}
\usepackage{bookmark}

\begin{document}
\title{A Thousand Problems in Cosmology:\\ Horizons}
\title{A Thousand Problems in Cosmology:\\ Horizons}
\author{Yu.L.~Bolotin$^{1,\sharp}$ and I.V. Tanatarov$^{1,2,\flat}$ \\[1em]
\small $^1$Kharkov Institute of Physics and Technology,\\
\small 1 Akademicheskaya, Kharkov 61108, Ukraine\\
\small $^2$Kharkov V.N. Karazin National
University,\\
\small 4 Svoboda Square, Kharkov 61077, Ukraine\\[1em]
$^\sharp$ ybolotin@gmail.com \\
$^\flat$ igor.tanatarov@gmail.com }

\maketitle

\begin{abstract}
This is one chapter of the collection of problems in cosmology, in which we assemble the problems, with solutions, that concern one of the most distinctive features of general relativity and cosmology---the horizons.

The first part gives an elementary introduction into the concept in the cosmological context, then we move to more formal exposition of the subject and consider first simple, and then composite models, such as $\Lambda$CDM. The fourth section elevates the rigor one more step and explores the causal structure of different simple cosmological models in terms of conformal diagrams. The section on black holes relates the general scheme of constructing conformal diagrams for stationary black hole spacetimes. The consequent parts focus on more specific topics, such as the various problems regarding the Hubble sphere, inflation and holography.

The full collection is available in the form of a wiki-based resource at \href{www.universeinproblems.com}{universeinproblems.com}. The cosmological community is welcome to contribute to its development. 
\end{abstract}

\tableofcontents


\newpage
\chapter*{Chapter: horizons}
\setcounter{chapter}{1}

In this chapter we assemble the problems that concern one of the most distinctive features of General Relativity and Cosmology --- the horizons. The first part gives an elementary introduction into the concept in the cosmological context, following and borrowing heavily from the most comprehensible text by E. Harrison \cite{Harrison}; the figures in this section are also taken from \cite{Harrison}. Then we move to more formal exposition of the subject, making use of the seminal works of W. Rindler \cite{Rindler} and G.F.R. Ellis, T. Rothman \cite{Ellis}, and consider first simple, and then composite models, such as $\Lambda$CDM. The fourth section elevates the rigor one more step and explores the causal structure of different simple cosmological models in terms of conformal diagrams, following mostly the efficient approach of V. Mukhanov \cite{Muchanov}. The section on black holes relates the general scheme of constructing conformal diagrams for stationary black hole spacetimes, following mostly the excellent textbook of K. Bronnikov and S. Rubin \cite{BrRubin}. The consequent sections focus on more specific topics, such as the various problems regarding the Hubble sphere, inflation and holography. 

\section{Simple English}
A vague definition of a horizon can be accepted as the following: it is a frontier between things observable and things unobservable.

\textit{Particle horizon.} If the Universe has a finite age, then light travels only a finite distance in that time and the volume of space from which we can receive information at a given moment of time is limited. The boundary of this volume is called the particle horizon.

\textit{Event horizon.} The event horizon is the complement of the particle horizon. The event horizon encloses the set of points from which signals sent at a given moment of time will never be received by an observer in the future.

Space-time diagram is a representation of space-time on a two-dimensional plane, with one timelike and one spacelike coordinate. It is typically used for spherically symmetric spacetimes (such as all homogeneous cosmological models), in which angular coordinates are suppressed.

\begin{enumerate}
\item Draw a space-time diagram that shows behaviour of worldlines of comoving observers in~a
\begin{enumerate}
\item stationary universe with beginning
\item expanding universe in comoving coordinates
\item expanding universe in proper coordinates
\end{enumerate}
\paragraph{No solution}

\item Suppose there is a static universe with homogeneously distributed galaxies, which came into being at some finite moment of time. Draw graphically the particle horizon for some static observer.
\paragraph{Solution.} The worldline $O$ represents our Galaxy from which we observe the universe. At the present moment we look out in the space and back in time and see other galaxies on our backward lightcone. Worldline $X$ determines the particle horizon. Objects (galaxies) beyond $X$ have not yet been observed by the present moment.
\begin{figure}[htb]
\center
\includegraphics*[height=0.535\textwidth]{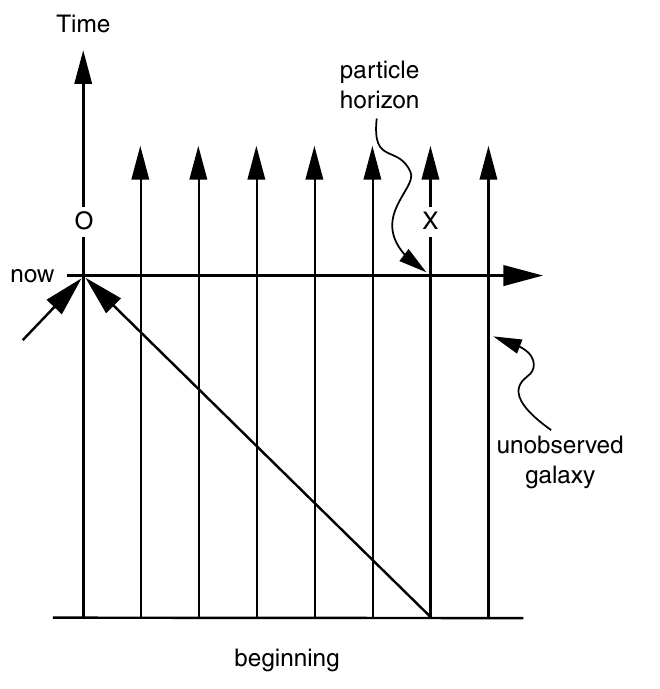}
\hfill
\includegraphics*[height=0.5\textwidth]{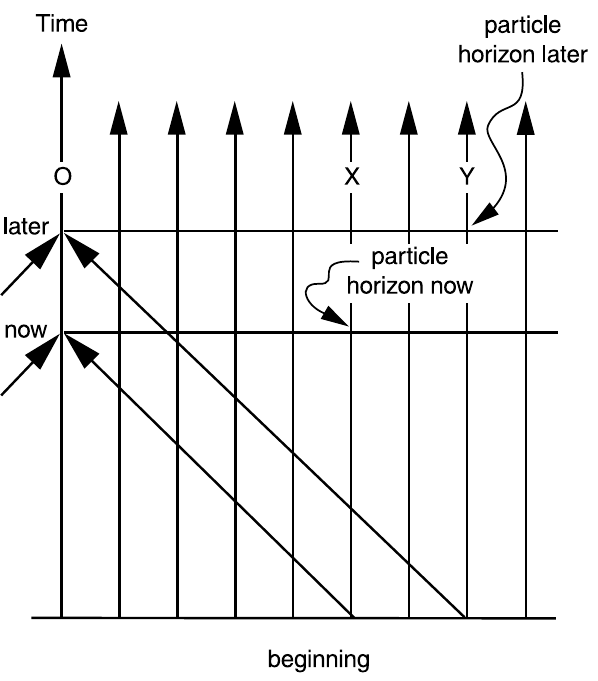}
\parbox{0.8\textwidth}{\caption{\label{Conf-Harr12} Left: Stationary Universe with a beginning. Right: evolution of horizon with time \cite{Harrison}.}}
\end{figure}

\item How does the horizon for the given observer change with time?
\paragraph{Solution.} 
At present the observer $O$ sees no farther than $X$ (see Fig. \ref{Conf-Harr12}). At some later moment he sees beyond $X$ up to some worldline $Y$. The particle horizon thus recedes in the static universe and as time passes the part of the Universe we observe grows ever larger. 

\item Is there an event horizon in the static Universe? What if the Universe ends at some finite time?
\paragraph{Solution.} If the Universe is eternal and galaxies shine forever, no event horizons exist. However, it does exist in a Universe that lives for some finite time (has an ``end''). For an observer in such a Universe the event horizon is the lightcone built on its worldline at the last possible moment.
\begin{figure}[hb]
\center
\includegraphics*[width=0.6\textwidth]{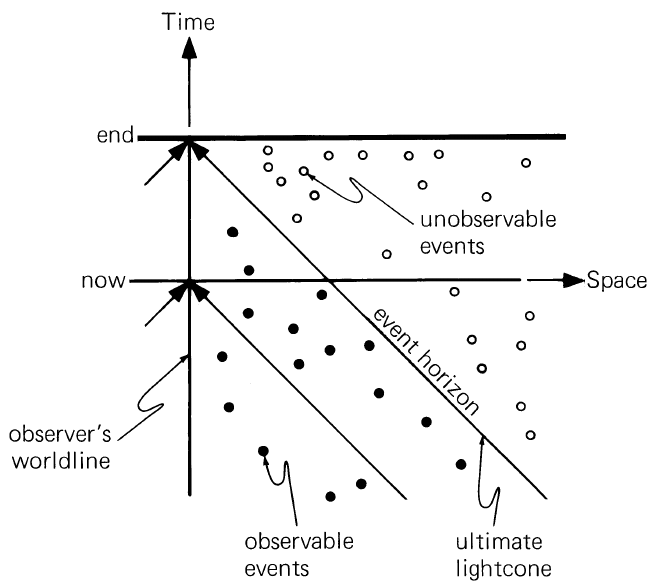}
\caption{Stationary Universe with an ending \cite{Harrison}.}
\end{figure}
 
Inside this ultimate lightcone are the events that have been seen by the end of the Universe, and outside are the events that can never been seen. The reason is that the lightcone cannot advance farther into time and all events outside of it remain unseen.

\item \emph{The horizon riddle.} Consider two widely separated observers, A and B (see Fig. \ref{Conf-Harr-Hor}). Suppose  they have overlapping horizons, but each can apparently see things that the other cannot.

\begin{figure}[!hbt]
\center
\includegraphics*[width=0.45\textwidth]{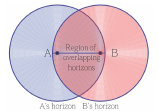}
\parbox{0.86\textwidth}{\caption{\label{Conf-Harr-Hor} The horizon riddle: can two observers with overlapping horizons pass information to each other regarding things outside of the other's horizon \cite{Harrison}?}}
\end{figure}

We ask: Can B communicate to A information that extends A's knowledge of things beyond his horizon? If so, then a third observer C may communicate to B information that extends her horizon, which can then be communicated to A. Hence, an unlimited sequence of observers B, C, D, E,\ldots may extend A's knowledge of the Universe to indefinite  limits. According to this argument A has no true horizon. This is the horizon riddle. Try to resolve it for the static Universe.

\paragraph{Solution.} Suppose, for example,  that luminous galaxies originated 10 billion years ago and the particle horizon is therefore at distance 10 billion light years. Observers A and B see each other and have overlapping horizons. Suppose A and B are separated by a distance of 6 billion light years (see Fig. \ref{Conf-Harr6}).
\begin{figure}[!htb]
\center
\includegraphics*[width=0.6\textwidth]{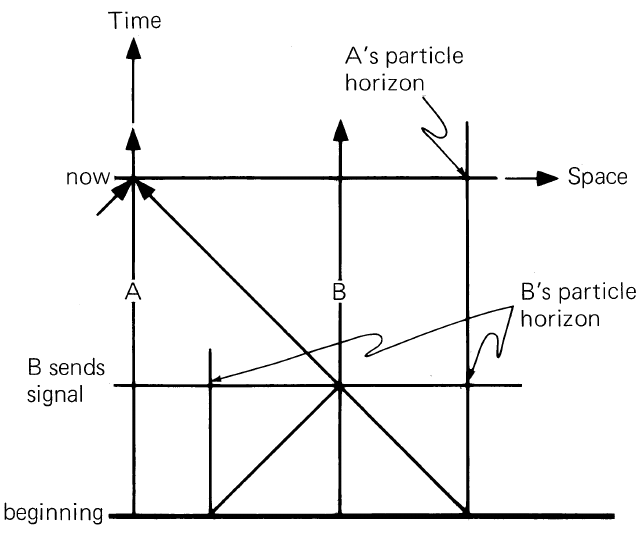}
\caption{\label{Conf-Harr6} The horizon riddle \cite{Harrison}.}
\end{figure}
B sends out information that travels at the speed of light and takes 6 billion years to reach A. Hence A receives from B information that was sent 6 billion years ago, when the Universe was 4 billion years old. But B's particle horizon in the past at the time when the information was sent was only 4 billion light years distant. Thus B's horizon at that time did not extend beyond A's present horizon. In other words, B communicates information to A by sending it at the speed of light on A's backward lightcone. But when B sends the information, her horizon extends no farther than A's horizon, and he cannot see farther than A.

\item Suppose observer O in a stationary universe with beginning sees A in some direction at distance $L$ and B in the opposite direction, also at distance $L$. How large must $L$ be in order for A and B to be unaware of each other's existence at the time when they are seen by O?

\paragraph{Solution.} Consider two visible bodies at equal distances in opposite directions from us, as shown by world lines A and B on Figure \ref{Conf-Harr8}.

\begin{figure}[!hb]
\center
\includegraphics*[width=0.6\textwidth]{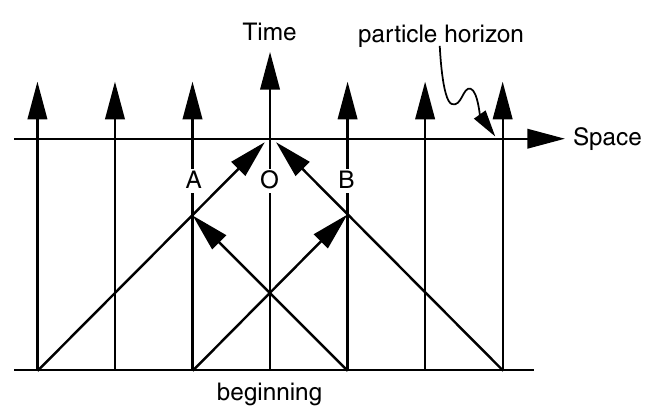}
\parbox{0.65\textwidth}{\caption{\label{Conf-Harr8} Two observers unaware of each other's existence by the time seen by some third observer \cite{Harrison}.}}
\end{figure}

We see these bodies, but can they see each other? Let $t$ be the time that it takes for light to travel to us from A and B. The time it takes the light to travel from A to B, or from B to A, is obviously $2t$. Hence when the Universe is older than $3t$, we not only see A and B, but they also see each other. If the Universe is younger than $3t$, and older than $t$, we see A and B, but they cannot yet see each other. There is thus a maximum distance beyond which the observed bodies A and B do not know of each other's existence. By examining Figure \ref{Conf-Harr8}, we see that this maximum distance is one third of the distance to the particle horizon. The answer to our question is that bodies at opposite directions and equal distances from us, which are larger than one third of the distance to the particle horizon, cannot at present see each other.

\item Draw spacetime diagrams in terms of comoving coordinate and conformal time and determine whether event or particle horizons exist for:
\begin{enumerate}
\item the universe which has a beginning and an end in conformal time. The closed Friedman universe that begins with Big Bang and ends with Big Crunch belongs to this class.

\item the universe which has a beginning but no end in conformal time.  The Einstein--de Sitter universe and the Friedman universe of negative curvature, which begin with a Big Bang and expand forever, belong to this class.

\item the universe which has an end but no beginning in conformal time. The de Sitter and steady-state universes belong to this class. 


\item the universe which has no beginning and no ending in conformal time, as in the last figure of Fig. \ref{Conf-Harr15-18}. The Einstein static and the Milne universes are members of this class.
\end{enumerate}
Conformal time is the altered time coordinate $\eta=\eta (t)$, defined in such a way that lightcones on the spacetime diagram in terms of $\eta$ and comoving spatial coordinate are always straight diagonal lines, even when the universe is not stationary.
\paragraph{Solution.} See Figure \ref{Conf-Harr15-18}
\begin{figure}[htb]
\center
\includegraphics*[width=0.48\textwidth]{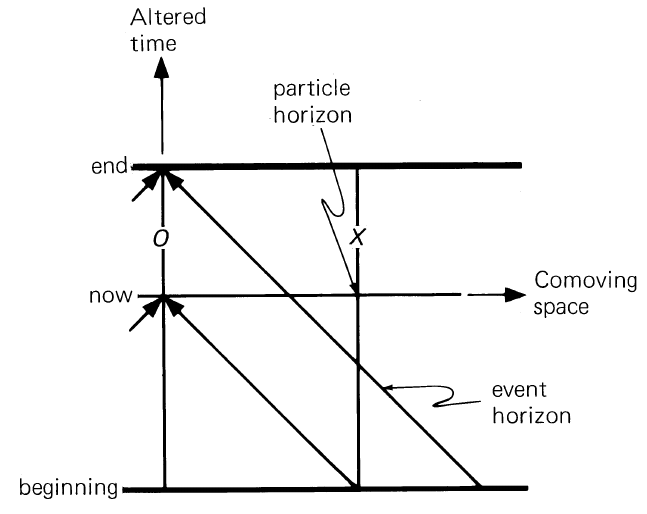}\hfill
\includegraphics*[width=0.48\textwidth]{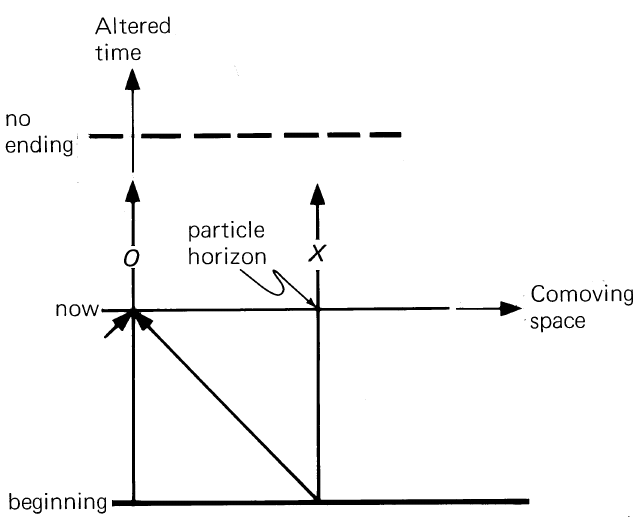}\hfill
\includegraphics*[width=0.48\textwidth]{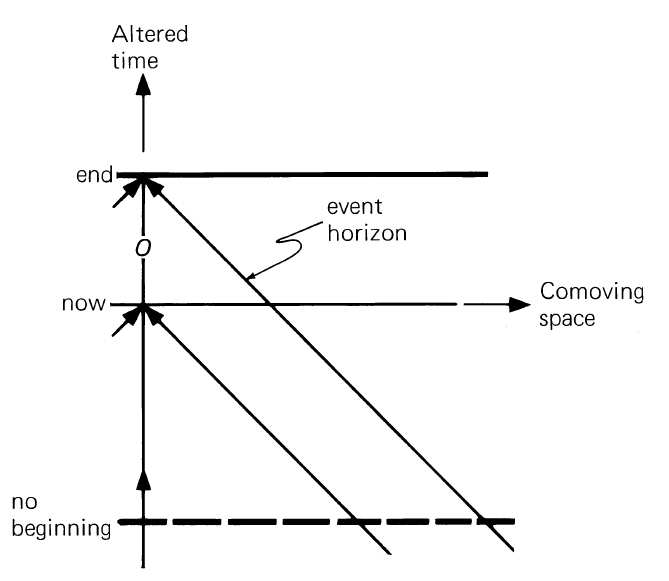}\hfill
\includegraphics*[width=0.48\textwidth]{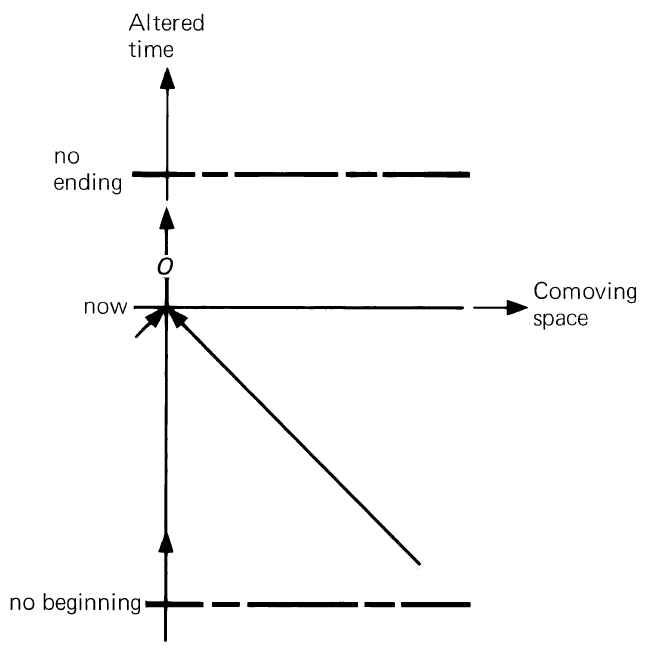}
\caption{\label{Conf-Harr15-18} Spacetime diagrams for universes with and without beginnings and endings \cite{Harrison}.}
\end{figure}

\begin{enumerate}
\item The world line X is at the particle horizon. Notice the existence of an event horizon.
\item The particle horizon is at world line X, and no event horizon exists.
\item Only the event horizon exists in this case.
\item There are no particle or event horizons.
\end{enumerate}

\FloatBarrier

\item Draw the spacetime diagram in terms of comoving space and ordinary time or the universe with an end but no beginning in conformal time.
\paragraph{Solution.} There are universes that expand forever and yet have endings in conformal time. The de Sitter is of this kind and therefore has event horizons. The figure (\ref{Conf-Harr22}) shows such a universe in a spacetime diagram of comoving space and cosmic time
\begin{figure}[hbt]
\center
\includegraphics*[width=0.6\textwidth]{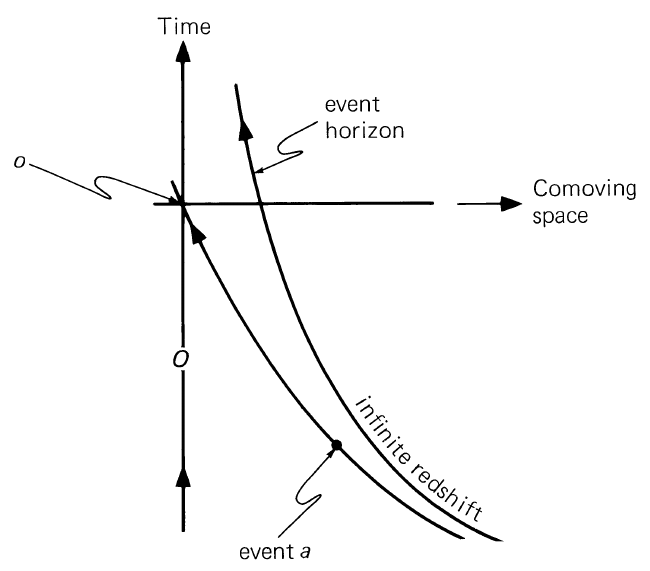}
\parbox{0.6\textwidth}{\caption{\label{Conf-Harr22} A universe with an end but no beginning in conformal time, drawn in terms of cosmic time \cite{Harrison}.}}
\end{figure}
The event horizon is as shown. As the moment now advances into the infinite future, the observer’s backward lightcone approaches the event horizon more and more closely. For example, the observer O at moment $O$ sees event $a$. As the moment of observation advances into the unlimited future, the lightcone moves upward  and approaches more and more slowly but never reaches the event horizon. The event horizon is the observer’s lightcone in the infinite future. Events outside this horizon can never be observed.

\FloatBarrier

\item Formulate the necessary conditions in terms of conformal time for a universe to provide a comoving observer with
\begin{enumerate}
\item a particle horizon
\item an event horizon
\end{enumerate}
\paragraph{Solution.}
\begin{enumerate}
\item The necessary condition for the existence of a particle horizon is that conformal time has a beginning (see items a) and b) of problem 7). The observer's lightcone stretches back and terminates at the lower boundary where the universe begins. When conformal time has no beginning, there is no lower boundary (see items c) and e) of problem 7). In this case the lightcone stretches back without limit and intersects all world lines in the universe. In these universes there are no particle horizons. Note that beginning in conformal time does not necessarily mean beginning in ordinary time.
\item The necessary condition for the existence of an event horizon is that conformal time has an ending. The event horizon in cosmology is thus nothing more than the observer's ultimate lightcone at the end of conformal time. All the events inside the event horizon (the ultimate lightcone) are at some time observed, and all events outside are never observed. Note that an end in conformal time does not necessarily mean an end in ordinary time. 
\end{enumerate}

\item Consider two galaxies, observable at present time, $A$ and $B$. Suppose at the moment of detection of light signals from them (now) the distances to them are such that $L_{det}^{A}<L^{B}_{det}$. In other words, if those galaxies had equal absolute luminosities, the galaxy $B$ would seem to be dimmer. Is it possible for galaxy $B$ (the dimmer one) to be closer to us at the moment of its signal's emission than galaxy $A$ (the brighter one) at the moment of $A$'s signal's emission?

\paragraph{Solution.} Yes, it is possible. The corresponding spacetime diagram is shown on \ref{Conf-Harr10}. Worldlines branch out radially in all directions from the ``big bang''. Spatial slices of constant cosmic time are represented by spherical surfaces perpendicular to the world lines, while time is measured along the radial worldlines. An arbitrary worldline  is chosen as the observer and labeled $O$. At some instant in time -- let it be ``now'' -- the observer's lightcone stretches out and back and intersects other worldlines such as $X$ and $Y$. Because of the expansion of space, the lightcone does not stretch out straight as in a static universe, but contracts back into the big bang. All worldlines and all backward lightcones converge into the big bang. 
\begin{figure}[!hb]
\center
\includegraphics*[width=0.6\textwidth]{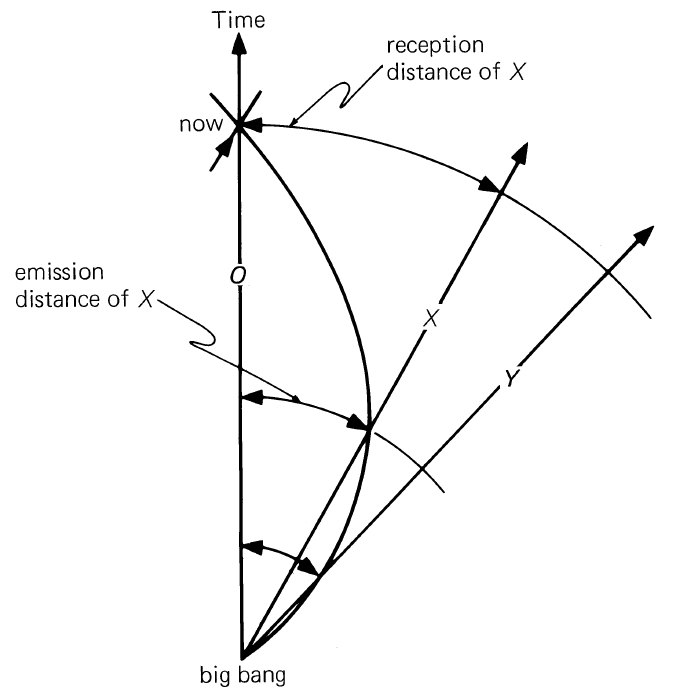}
\parbox{0.87\textwidth}{\caption{\label{Conf-Harr10}  The reception and emission distances of galaxies $X$ and $Y$. Although galaxy $Y$ has a greater reception distance, its emission distance is smaller than that of $X$. Thus  $Y$, which is now farther away than $X$, was closer to us than $X$ at the time of emission (which is different for $X$ and $Y$) of the light we now see \cite{Harrison}.}}
\end{figure}

\item Show on a spacetime diagram the difference in geometry of light cones in universes with and without particle horizons.

\paragraph{Solution.} A spacetime diagram of comoving space (in which all worldlines are parallel) and cosmic time is shown on Fig. \ref{Conf-Harr13}. Some universes have particle horizons and in their case the lightcone stretches out and back to the beginning at a finite comoving distance indicated by the world line $X$ . In universes without particle horizons, the lightcone stretches out to an unlimited distance and intersects all world lines.
\begin{figure}[hb]
\center
\includegraphics*[width=0.6\textwidth]{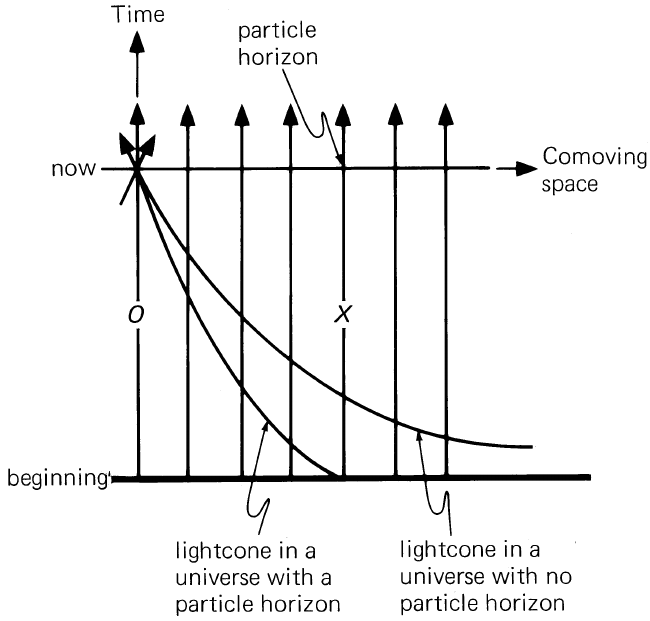}
\caption{\label{Conf-Harr13} Lightcones in universes with and without particle horizons \cite{Harrison}.}
\end{figure}
\end{enumerate}

\FloatBarrier

\section{Simple Math}
The problems of this section need basic understanding of Friedman equations, definitions of proper, comoving and conformal coordinates, the cosmological redshift formula and simple cosmological models (see Chapters 2 and 3). 

Let us make our definitions a little more strict.

A \textbf{particle horizon}, for a given observer $A$ and cosmic instant $t_0$ is a surface in the instantaneous three-dimensional section $t=t_0$ of space-time, which divides all comoving particles\footnote{Rindler uses the term ``fundamental observers''} into two classes: those that have already been observable by $A$ up to time $t_0$ and those that have not. 

An \textbf{event horizon}, for a given observer $A$, is a hyper-surface in space-time, which divides all events into two non-empty classes: those that have been, are, or will be observable by $A$, and those that are forever outside of $A$'s possible powers of observation. It follows from definition that event horizon, and its existence, depend crucially on the observer's (and the whole Universe's) future as well as the past: thus it is said to be an essentially \emph{global concept}. It is formed by null geodesics. 

The following notation is used hereafter: $L_p (t_0)$ is the proper distance from observer $A$ to its particle horizon, measured along the slice $t=t_0$. For brevity, we will also call this distance simply ``the particle horizon in proper coordinates'', or just ``particle horizon''. The corresponding comoving distance $l_p$ is the particle horizon in comoving coordinates.

Likewise, $L_e$ is the proper distance from an observer to its event horizon (or, rather, its section with the hypersurface $t=t_0$), measured also along the slice $t=t_0$. It is called ``space event horizon at time $t_0$'', or just the event horizon, for brevity. The respective comoving distance is denoted $l_e$.

\begin{enumerate}[resume]
\item The proper distance $D_p (t_0)$ between two comoving observers is the distance measured between them at some given moment of cosmological time $t=t_0$. It is the quantity that would be obtained if all the comoving observers between the given two measure the distances between each other at one moment $t=t_0$ and then sum all of them up. Suppose one observer detects at time $t_{0}$ the light signal that was emitted by the other observer at time $t_e$. Find the proper distance between the two observers at $t_0$ in terms of $a(t)$. 

\paragraph{Solution.} By definition, measuring distance along the slice $t=const$, we have for some given time $t_0$
\begin{equation}
D_p (r, t_0)=\int |ds| = a_0 \int_0^r \frac{dr}{\sqrt {1 - k r^2} },
	\quad {a_0} \equiv a (t_0)
\end{equation}

On the other hand, along the worldline of the light ray connecting the two observers $ds=0$, so we have
\[\frac{dt}{a} = \frac{dr}{\sqrt {1 - k r^2 }}.\]
Integrating along the worldline from the time of emission $t_e$ to the time of detection $t_0$, we get
\[\int_{t_e}^{t_0} \frac{dt'}{a(t')}  
	= \int_0^r \frac{dr}{\sqrt {1 - k r^2}}  = \frac{D_p}{a_0},\]
Therefore the proper distance between two observers in an expanding Universe is
\begin{equation}
	D_p (t_e, t_0)= a_0\int_{t_e}^{t_0} \frac{dt'}{a(t')}. 
\end{equation}

\item Show, that the proper distance $L_p$ to the particle horizon at time $t_0$ is
\begin{equation}
L_p (t_0)=\lim\limits_{t_e \to  0}D_p (t_e ,t_0). \label{LpDp}
\end{equation}
\paragraph{Solution.} As define above, particle horizon is the proper distance between the observer that receives the light signal at present and the comoving particle that emitted this light at the very beginning of the Universe (which may correspond to $t\to -\infty$). Thus (\ref{LpDp}).

\item The past light cone of an observer at some time $t_0$ consists of events, such that light emitted in each of them reaches the selected observer at $t_0$. Find the past light cone's equation in terms of proper distance vs. emission time $D_{plc} (t_e)$. What is its relation to the particle horizon?
\paragraph{Solution.} 
\begin{equation}
D_{plc}(t_e ,t_0) =a(t_e) \int\limits_{t_e}^{t_0} \frac{c\,dt}{a(t)}.
	\label{Dplc}
\end{equation}
The corresponding comoving distance $d(t_e ,t_0)=D(t_e, t_0)/a(t_e)$ at $t_e \to t_{in}$ ($t_{in}$ is the time of creation singularity or infinite past, whichever is realized) gives us the observer that only now (at $t_0$) just becomes observable, thus determining the particle horizon at $t_0$:
\[L_p (t_0)=a(t_0) d_{plc}(t_{in}, t_0).\]

\item The simplest cosmological model is the one of \emph{Einstein-de Sitter}, in which the Universe is spatially flat and filled with only dust, with $a(t)\sim t^{2/3}$. Find the past light cone distance $D_{plc}$ (\ref{Dplc}) for Einstein-de Sitter.
\paragraph{Solution.}
\begin{equation}
D_{plc}(t_e, t_0)=3c (t_e^{2/3}t_0^{1/3}-t_e ).
\end{equation}

\item Demonstrate that in general $D_{plc}$ can be non-monotonic. For the case of Einstein-de Sitter show that its maximum -- the maximum emission distance -- is equal to $8/27 L_H$, while the corresponding redshift is $z=1.25$.
\paragraph{Solution.} From $dD_{plc}/dt_e =0$ we see that maximum exists and lies at
\begin{equation}
\frac{t_e}{t_0}=\frac{8}{27},\qquad D_{plc}^{max}=\frac{4}{9}ct_0 .
\end{equation}
The redshift of light signal emitted at maximum emission distance is then
\begin{equation}
z=\frac{a(t_0)}{a(t_e)}-1=\Big(\frac{t_0}{t_e}\Big)^{2/3}-1 =\frac{9}{4}-1=1.25 .
\end{equation}

\item In a matter dominated Universe we see now, at time $t_0$, some galaxy, which is now on the Hubble sphere. At what time in the past was the photon we are registering emitted?
\paragraph{Solution.} The emission event is at the intersection of the particle's worldline with the past light cone, so
\begin{equation}
	 \tfrac{3}{2}ct_0 \;\Big(\frac{t_e}{t_0}\Big)^{2/3}
	 		=3c (t_e^{2/3}t_0^{1/3}-t_e ),
\end{equation} 
from which
\[t_e =\frac{t_0}{8}.\]

\item Show that the particle horizon in the Einstein-de Sitter model recedes at three times the speed of light.
\paragraph{Solution.} In Einstein-de Sitter $a\sim t^{2/3}$, thus $L_p =3ct$ and $H=2/3t$, so
\begin{equation}
\dot{L}_p =c+ \frac{2}{3t}\cdot 3ct =3c.
\end{equation}

\item Does the number of observed galaxies in an open Universe filled with dust increase or decrease with time?

\item Draw the past light cones $D_{plc}(t_e)$ for Einstein-de Sitter and a Universe with dominating radiation on one figure; explain their relative position.

\item Find the maximum emission distance and the corresponding redshift for power law expansion $a(t)\sim (t/t_0 )^n$.
\paragraph{Solution.} The comoving distance along a null geodesic is $\int d\eta =\int dt/a(t)$. Then at the time of emission $t_e$ the proper distance between the comoving emitter and detector is
\begin{equation}
L (t_e ,t_0)=a(t_e )\int\limits_{t_e}^{t_0}\frac{dt}{a(t)}=\frac{t_0}{1-n}\big[x^n -1\big],
\end{equation}
where $t_0$ is the time of detection (present), and
\begin{equation}
x=\frac{t_e}{t_0}.
\end{equation}
This is the past light cone of an event at $t_0$ given in terms of proper distance and cosmic time.

The maximum of $L(t_e ,t_0)$ is at
\begin{equation}
x_m=n^{1/(1-n)}, \qquad L_{\max}=L(x_m t_0, t_0)
	=\frac{t_0}{n}x_m =\frac{t_0}{n}n^{1/(1-n)}.
\end{equation}
The corresponding redshift is given by the general relation
\begin{equation}
z_{\max}+1 =\frac{a(t_0)}{a(xt_0)}=n^{n/(n-1)}.
\end{equation}

For radiation domination ($n=1/2$) we have
\begin{equation}
L_{\max}=\tfrac{1}{2}ct_0=\tfrac{1}{2}R_H,\qquad z_{\max}=1,
\end{equation}
and for matter domination ($n=2/3$)
\begin{equation}
L_{\max}=\tfrac{4}{9}ct_0 =\tfrac{8}{27}R_H ,\qquad z=1.25.
\end{equation}

\item Show that the most distant point on the past light cone was exactly at the Hubble sphere at the moment of emission of the light signal that is presently registered.
\paragraph{Solution.} The recession velocity of the emitter at the time of emission $t_e =x_m t_0$ is
\begin{equation}
v(t_e ,L_{\max})=H(x_m t_0)\cdot L_{\max}=\frac{n}{x_m t_0}\cdot \frac{t_0}{n}x_m =1 ,
\end{equation}
so by definition the emitter (i.e. galaxy) was at that moment exactly on the Hubble sphere.

\item Show that the comoving particle horizon is the age of the Universe in conformal time
\paragraph{Solution} Starting from the definition
\begin{equation}
\frac{L_p (t)}{a(t)} = \int_0^t \frac{dt'}{a(t')}  = \int_0^\eta  d\eta '  = \eta .
\end{equation}

\item Show that
\begin{align}
\frac{dL_p}{dt}&=L_p (z)H(z)+1;\\
\frac{dL_e}{dt}&=L_e (z)H(z)-1.
\end{align}
\paragraph{Solution.} From the definitions (here $c=1$)
\begin{align}
\dot{L_p}&=\frac{d}{dt}\Big(a(t)\int\limits^{t}\frac{dt}{a(t)}\Big)= +1+HL_p , \label{dotLp}\\
\dot{L_e}&=\frac{d}{dt}\Big(a(t)\int\limits_{t}\frac{dt}{a(t)}\Big)= -1+HL_e .\label{dotLe}
\end{align}

\item Find $\ddot{L_p}$ and $\ddot{L_e}$.
\paragraph{Solution.} Differentiating $L_p$ and $L_e$ from (\ref{dotLp}-\ref{dotLe}) once again, we get
\begin{align}
\frac{d^2 L_p}{dt^2} &= H(+1-q H L_p),\\
\frac{d^2 L_e}{dt^2} &= H(-1-qH L_e),
\end{align}
where $q=-\frac{\ddot{a}/a}{H^2}$ is the deceleration parameter.

\item Show that observable part of the Universe expands faster than the Universe itself. In other words, the observed fraction of the Universe always increases.
\paragraph{Solution.} The particle horizon at distance $L_p$ recedes with velocity $\dot{L}_p$ found in the previous problem, while the galaxies at the particle horizon recede at velocity $HL_p$, hence the horizon overtakes the galaxies with the speed of light $c$.

\item Show that the Milne Universe has no particle horizon.
\paragraph{Solution.} In the Milne Universe $a\sim t$, $H=1/t$, $q=0$,  the lightcone reaches the beginning of time $t=0$ at an infinite comoving distance and there is no particle horizon. The observable universe fills the entire actual Universe and all galaxies are in principle visible. In other words, all galaxies are visible at some stage in their evolution.

\item Consider a universe which started with the Big Bang, filled with one matter component. How fast must $\rho(a)$ decrease with $a$ for the particle horizon to exist in this universe?
\paragraph{Solution.} The comoving particle horizon is $\int_{0}dt a^{-1}(t)$, where $t=0$ is assumed to correspond to the Big Bang singularity. Its existence depends on whether this integral converges at small times or not. Then using the Friedman equation,
\begin{equation}
dt=\frac{da}{\dot{a}}=\frac{da}{\sqrt{\rho a^2}},
\end{equation}
so the particle horizon is
\begin{equation}
l_p = \int\limits_0 \frac{da /a}{\sqrt{\rho a^2}}.
\end{equation}
The integral converges as long as $\rho a^2$ diverges at small $a$. That is, if equation of state is such that $\rho$ changes faster than $\sim 1/a^2$, then light can only propagate a finite distance between Big Bang and now. In particular, the particle horizon exists in models, which are dominated at early times either by radiation $\rho\sim 1/a^4$ or matter $\rho\sim 1/a^3$. 

\item Calculate the particle horizon for a universe with dominating
\begin{enumerate}
\item radiation;
\item matter.
\end{enumerate}
\paragraph{Solution.} In radiation-dominated universe $a\sim t^{1/2}$, in matter-dominated $a\sim t^{2/3}$. Then integration of $L =a(t)\int dt/a(t)$ gives
\begin{equation}
L_{p}^{(rad)}=2t,\qquad L_p^{(mat)}=3t.
\end{equation}

\item Consider a flat universe with one component with state equation  $p =w \rho$. Find the particle horizon at present time $t_0$. 
\paragraph{Solution.} The particle horizon $L_p$ at the current moment $t_0$ is (see \ref{LpDp})
\[L_p =\lim\limits_{t_e \to 0} D_p (t_e ,t_0)\]	  
The proper distance can be written as
\begin{equation}
	D_p = a_0\int_t^{t_0} \frac{dt'}{a(t')}
	=- \int_{z(t)}^{z(t_0)} \frac{dz'}{H(z')} 
		=\int_0^z \frac{dz'}{H(z')}. 
\end{equation}
The density in terms of redshift $z$ is
\[\rho =\rho_{0}(1+z)^{3(1+w)},\]
and
\[H(z)=H_0 \sqrt{(1+z)^{3(1+w)}},\]
so
\begin{equation}
L_p (t_0)=H_0^{-1}\int \limits_{0}^{\infty}
	\frac{dz}{\sqrt{(1+z)^{3(1+w_i)}}}
\end{equation}

\item Show that in a flat universe in case of domination of one matter component with equation of state $p=w\rho$, $w>-1/3$
\begin{equation}
L_{p} (z)=\frac{2}{H(z) (1+3w)},\qquad \dot{L_p}(z)=\frac{3(1+w)}{(1+3w)}.
\end{equation}
\paragraph{Solution.} In the considered case
\begin{equation}
L_p (z) = \frac{1}{H(z)}\int_0^\infty  \frac{dz'}{\sqrt{(1 + z')^{3(1 + w)}}}
	  = \frac{2}{H(z)(1 + 3w)}.
\end{equation}
Note that $L_p >0$, i.e. the horizon exists, only if $w>-1/3$. 

On differentiating by time and using
\begin{equation}
H^2 = \frac{1}{3}\rho ,\qquad \dot H =- \frac{1}{2}\rho (1 + w),
\end{equation}
we get
\begin{equation}
\frac{dL_p}{dt} = \frac{3(1 + w)}{1 + 3w}.
\end{equation}

\item Show that in a flat universe in case of domination of one matter component with equation of state $p=w\rho$, $w<-1/3$
\begin{equation}
L_{e} (z)=-\frac{2}{H(z) (1+3w)},\qquad \dot{L_p}(z)=-\frac{3(1+w)}{(1+3w)}.
\end{equation}
\paragraph{Solution.} In the considered case
\begin{equation}
L_e (z) = \frac{1}{H(z)}\int_{-1}^{0}  \frac{dz'}{\sqrt {(1 + z')^{3(1 + w)}}}
	  = -\frac{2}{H(z)(1 + 3w)},
\end{equation}
and 
\begin{equation}
\frac{dL_e}{dt} = -\frac{3(1 + w)}{1 + 3w}.
\end{equation}

Note that $L_e >0$, i.e. the event horizon exists, only if $w<-1/3$. In particular, when we have the cosmological constant, $w=-1$, there is only the event horizon
\[L_e =H .\] 


\end{enumerate}

\begin{enumerate}[resume]
\item Estimate the particle horizon size at matter-radiation equality. 
\end{enumerate}

\section{Composite models}

\begin{enumerate}[resume]

\item Consider a flat universe with several components $\rho=\sum_i \rho_i$, each with density $\rho_i$ and partial pressure $p_i$ being related by the linear state equation  $p_i =w_i \rho_i$. Find the particle horizon at present time $t_0$. 
\paragraph{Solution.} The particle horizon $L_p$ at the current moment $t_0$ is (see \ref{LpDp})
\[L_p =\lim\limits_{t_e \to 0} D_p (t_e ,t_0)\]	  
The proper distance can be written as
\[D_p = a_0\int_t^{t_0} \frac{dt'}{a(t')}
	=- \int_{z(t)}^{z(t_0)} \frac{dz'}{H(z')} 
		=\int_0^z \frac{dz'}{H(z')}. \]
The $i$-th energy density in terms of redshift $z$ is
\[\rho_i =\rho_{0i}(1+z)^{3(1+w_i)},\]
and
\[H(z)=H_0 \sqrt{\sum \Omega_{0i}(1+z)^{3(1+w_i)}},\]
so
\begin{equation}
L_p (t_0)=H_0^{-1}\int \limits_{0}^{\infty}
	\frac{dz}{\sqrt{\sum \Omega_{0i}(1+z)^{3(1+w_i)}}}
\end{equation}

\item Suppose we know the current material composition of the Universe $\Omega_{i0}$, $w_i$ and its expansion rate as function of redshift $H(z)$. Find the particle horizon $L_p (z)$ and the event horizon $L_p (z)$ (i.e the distances to the respective surfaces along the surface $t=const$) at the time that corresponds to current observations with redshift $z$.
\paragraph{Solution.} For the particle horizon we rewrite the previous result in terms of redshifts
\begin{equation}
L_p (z)=H^{-1} (z)\int \limits_{0}^{\infty}
	\frac{dz'}{\sqrt{\sum \Omega_{i}(z)(1+z')^{3(1+w_i)}}},
\end{equation}
and take into account how the partial densities $\Omega_i$ depend on time: they are defined to satisfy
\begin{equation}
H^2 (z)=H_0^2 \sum \Omega_{i0} (1+z)^{3(1+w_i)}.
\end{equation}
at any $z$ (or, equivalently, $t$), thus $\Omega_i (z)$ by definition is the ratio of the $i$th term of the sum to the whole sum at any moment of time:
\begin{equation}
\Omega_i (z) = \Omega _{i0} \frac{H_0^2 }{H^2 (z)}(1 + z)^{3(1+w_i)}.
\end{equation}
Then for the particle horizon (and for event horizon in the same way) we obtain
\begin{align}
L_p (z) &= \frac{1}{H(z)}\int_0^\infty  
	\frac{dz'}{\sqrt {\sum\limits_i \Omega_i (z)(1 + z')^{3(1 +w_i )}}}; \label{LpZ}\\
L_e (z) &= \frac{1}{H(z)}\int_{-1}^{0}
	\frac{dz'}{\sqrt {\sum\limits_i \Omega_i (z)(1 + z')^{3(1 +w_i )}}} \label{LeZ}.
\end{align}

\item When are $L_p$ and $L_e$ equal? It is interesting to know whether both horizons might have or not the same values, and if so, how often this could happen.
\paragraph{No solution.}
\item Find the particle and event horizons for any redshift $z$ in the standard cosmological model -- $\Lambda$CDM.
\paragraph{Solution.} Using the general formulae (\ref{LpZ}--\ref{LeZ}), one obtains
\begin{align}
L_p(z) &= \frac{1}{H(z)}\int_0^\infty  
	\frac{dz'}{\sqrt {\Omega_m (z) (1 + z')^3 + \Omega_\Lambda (z)}},\\
L_e(z) &= \frac{1}{H(z)}\int_{-1}^{0}
	\frac{dz'}{\sqrt {\Omega_m (z) (1 + z')^3 + \Omega_\Lambda (z)}},\\
&\Omega_m (z)=\Omega _{m0}\frac{H_0^2}{H(z)^2} (1 + z)^3 ,\\
&\Omega_\Lambda (z) = \Omega_{\Lambda 0} \frac{H_0^2}{H(z)^2}.
\end{align}

\item Express the particle $L_p (z)$ and event $L_e (z)$ horizons in $\Lambda$CDM through the hyper-geometric function.
\paragraph{Solution.} 
\begin{align}
L_p (z) &= \frac{2\sqrt {A(z)}}{H_0 \sqrt {\Omega _{\Lambda 0}}} F\left(\frac{1}{2},\frac{1}{6},\frac{7}{6}; - A(z) \right),\\
L_e (z) &= \frac{1}{H_0 \sqrt {\Omega _{\Lambda 0}}} F\left(\frac{1}{2},\frac{1}{3},\frac{4}{3}; - \frac{1}{A(z)} \right),\\
&A(z) = \frac{\Omega_{\Lambda 0}}{\Omega _{m0}} (1 + z)^{-3}.
\end{align}
\end{enumerate}

\section{Causal structure}
The causal structure is determined by propagation of light and is best understood in terms of \emph{conformal diagrams}. In this section we construct and analyze those for a number of important model cosmological solutions (which are assumed to be already known), following mostly the exposition of \cite{Muchanov}.

In terms of comoving distance $\tilde{\chi}$ and conformal time $\tilde{\eta}$ (in this section they are denoted by tildes) the two-dimensional radial part of the FLRW metric takes form
\begin{equation}
	ds_{2}^{2}=a^{2}(\tilde{\eta})\big[d\tilde{\eta}^2 -d\tilde{\chi}^2\big].
	\label{ds2Dconf}
\end{equation}
In the brackets here stands the line element of two-dimensional Minkowski flat spacetime. Coordinate transformations that preserve the \emph{conformal} form of the metric
\begin{equation*}
ds_2^2 =\Omega^{2}(\eta,\chi)\big[d\eta^2 -d\chi^2 \big],
\end{equation*}
are called conformal transformations, and the corresponding coordinates $(\eta,\chi)$ -- conformal coordinates. 

\begin{enumerate}[resume]
\item Show that it is always possible to construct $\eta(\tilde{\eta},\tilde{\chi})$, $\chi(\tilde{\eta},\tilde{\chi})$, such that the conformal form of metric (\ref{ds2Dconf}) is preserved, but $\eta$ and $\chi$ are bounded and take values in some finite intervals. Is the choice of $(\eta,\chi)$ unique?
\paragraph{Solution.} Suppose $\tilde{\eta},\tilde{\chi}$ span infinite or semi-infinite values. Then we can always make the following sequence of coordinate transformations:
\begin{enumerate}
\item Pass to null coordinates
\begin{equation}
u=\tilde{\eta}-\tilde{\chi},\qquad v=\tilde{\eta}+\tilde{\chi};
\end{equation}
\item Bring their range of values to a finite interval by some appropriate function, i.e.
\begin{equation}
U=\arctan u,\qquad V=\arctan v.
\end{equation}
\item Go back to timelike and spacelike coordinates (this is not really necessary at this point and is done mostly for aesthetic reasons):
\begin{equation}
T=V+U,\qquad R=V-U.
\end{equation}
Now the range of $(T,R)$ obviously covers some bounded region on the plane, while the radial part of the line element preserves its conformal form:
\begin{equation}
ds_2^2 \sim d\tilde{\eta}^2-d\tilde{\chi}^2 \sim du\, dv \sim dU\, dV \sim dT^2 -dR^2 .
\end{equation}
\end{enumerate}
As the choice of function $\arctan$ was rather arbitrary (though convenient), the choice of conformal coordinates is not unique.
\end{enumerate}
In this section we will reserve notation $\eta$ and $\chi$ and names ``conformal coordinates'' and ``conformal variables'' to such variables that can only take values in a bounded region on $\mathbb{R}^2$; $\tilde{\eta}$ and $\tilde{\chi}$ can span infinite or semi-infinite intervals. Spacetime diagram in terms of conformal variables $(\eta,\chi)$ is called conformal diagram. Null geodesics $\eta=\pm \chi + const$ are diagonal straight lines on conformal diagrams.

\begin{enumerate}[resume]
\item Construct the conformal diagram for the closed Universe filled with
\begin{enumerate}
\item radiation;
\item dust;
\item mixture of dust and radiation.
\end{enumerate}
Show the particle and event horizons for the observer at the origin $\chi=0$ (it will be assumed hereafter that the horizons are always constructed with respect to this chosen observer).

\paragraph{Solution.} 
\begin{enumerate}
\item Solution for $a(\eta)$ in a radiation dominated closed Universe is
\begin{equation}
a=a_m \sin\eta ,\qquad \eta\in(0,\pi),\quad \chi \in [0,\pi].
\end{equation}
As the ranges spanned by $\eta$ and $\chi$ are finite, they are already conformal coordinates. The conformal diagram is a square $\eta,\chi\in [0,\pi]$. Edges $\eta=0$ and $\eta=\pi$ correspond to the Big Bang and Big Crunch singularities respectively; worldline $\chi=\pi$ is the point on the three-sphere that is situated at the opposite pole with respect to observer at $\chi=0$.

\begin{figure}[!ht]
\center
\includegraphics*[width=0.48\textwidth]{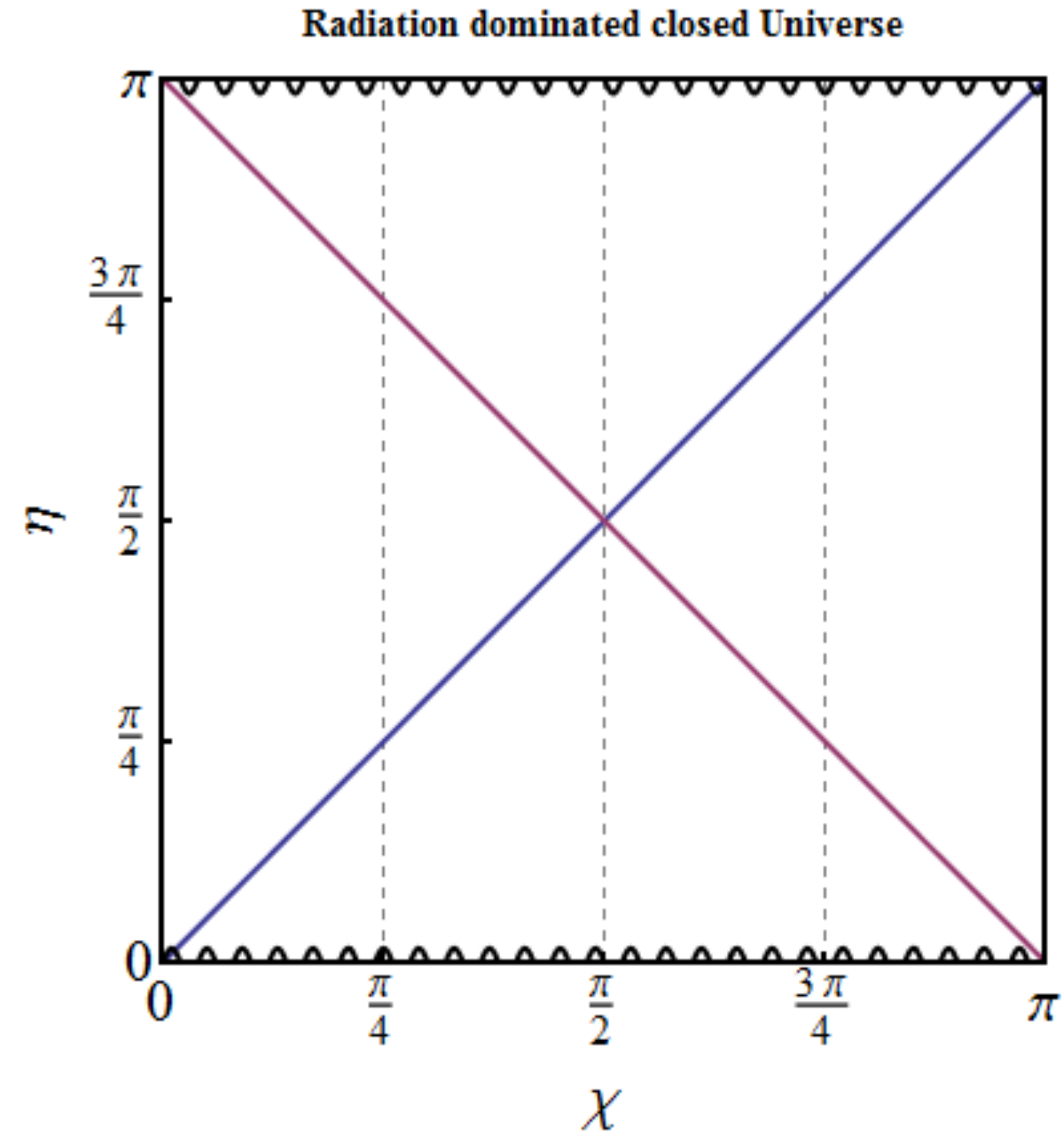}\hfill
\includegraphics*[width=0.48\textwidth]{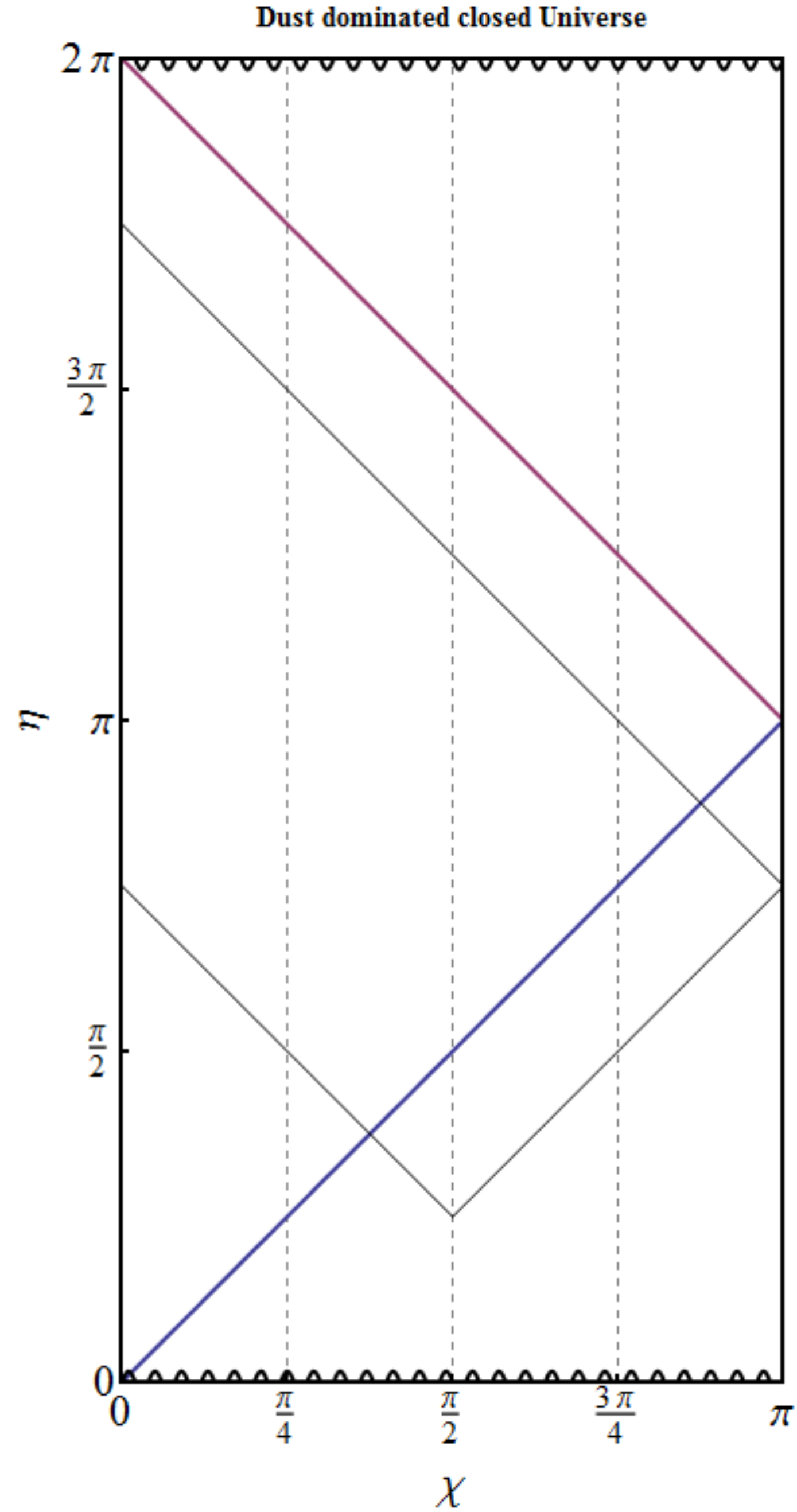}
\parbox{0.9\textwidth}{\caption{Conformal diagrams of radiation (left) and dust (right) dominated closed Universes. Particle horizon is shown in blue, event horizon in red. Thin lines in the dust dominated universe show light rays that realize the first and second images of a galaxy at different cosmic times, from opposite directions.}}
\end{figure}

The particle horizon is 
\[\eta=\chi,\]
and event horizon is
\[\eta=\eta_{max}-\chi =\pi -\chi.\]
Both exist for all cosmological times: by the finite moment of the Big Crunch $\eta=\pi$ the event horizon is collapsed into a point (which is natural, as there is no more time left), while the particle horizon extends to the whole Universe. Thus only at the finite moment the whole of the Universe becomes observable. The farther the point, though, the younger will it look, of course, and the opposite pole will only be ``observed'' by our observer at the last moment of the Universe as it was at its creation.

\item Solution for $a(\eta)$ in a dust dominated closed Universe is
\begin{equation}
a=a_m (1-\cos\eta) ,\qquad \eta\in(0,2\pi),\quad \chi \in [0,\pi].
\end{equation}
The difference from the previous case is that $\eta$ spans twice the range, and $\eta_{max}=2\pi =2\chi_{max}$. 

Therefore the event horizon is given by 
\[\eta=\eta_{max}-\chi=2\pi -\chi,\]
so it exists only in the second, contracting, phase $\eta>\pi$. The particle horizon is given by $\eta=\chi$ again, but now it exists only during the expanding phase $\eta<\pi$. It encloses the full Universe at the moment of maximal expansion $\eta=\pi$ and for later times does not exist.

\item Though the full analytic solution is more complicated, it is clear that the main features remain the same as in the previous considered cases. At early and late times, close to the singularities, the dynamics is determined by the radiation component. If there is enough dust, then at large enough scale factors (which may or may not be achieved depending on the initial conditions), which would correspond to the epoch around the maximal expansion, it will be dominating. The influence of dust is that dynamics is slowed down, so that depending on the ratio of densities
\[\eta_{max}\in [\pi,2\pi].\]
Thus qualitatively the picture will be the same as in a dust dominated Universe: the conformal diagram is a rectangle, the event horizon exists only starting from some time $\eta_{e}=\eta_{max}-\pi$, while particle horizon, on the contrary, vanishes at $\eta_{p}=\pi$.
\begin{figure}[!hb]
\center
\includegraphics*[width=0.48\textwidth]{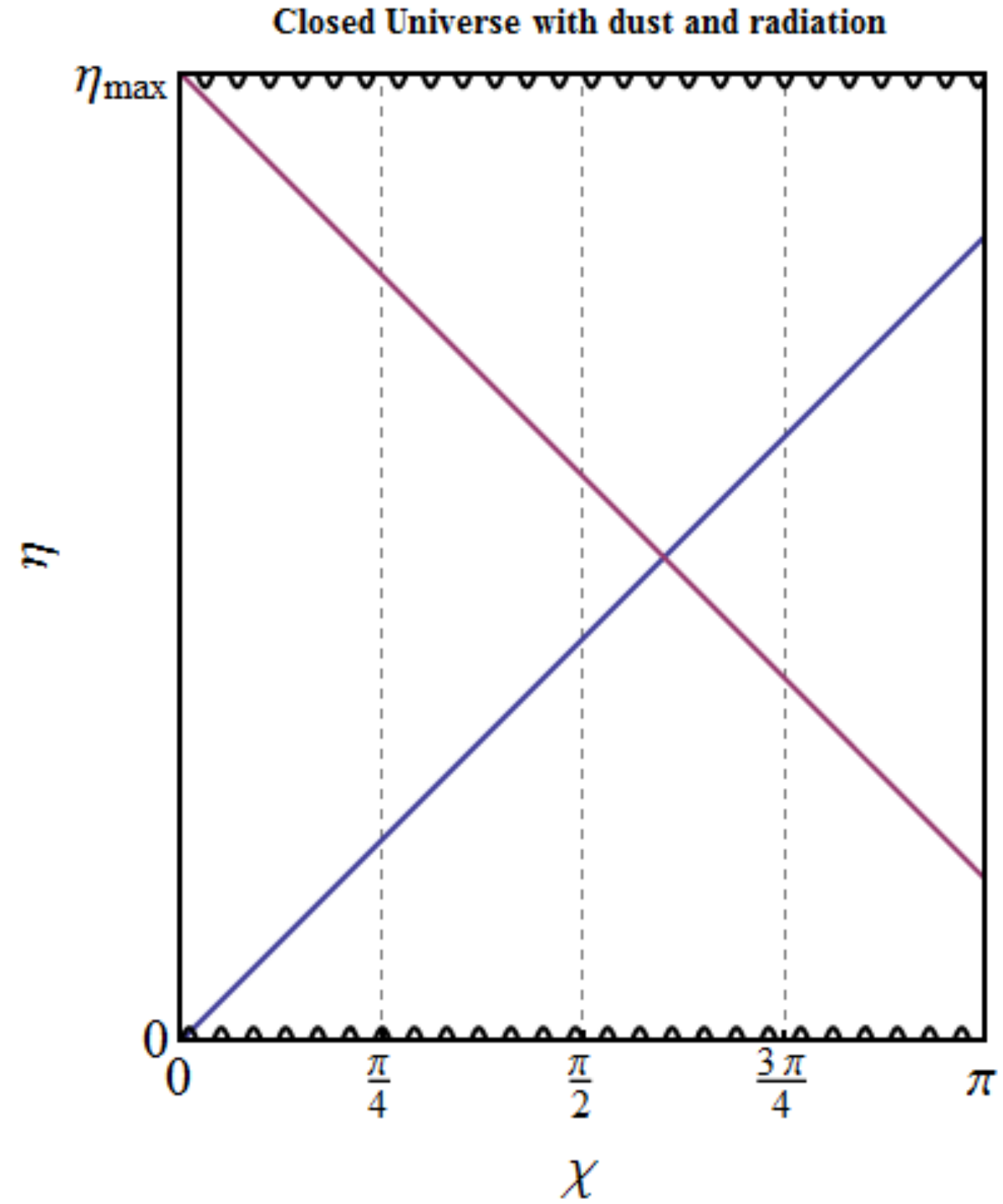}
\parbox{0.53\textwidth}{\caption{A closed Universe filled with a mix of dust and radiation.}}
\end{figure}
\end{enumerate}

\item \emph{Closed dS.} Construct the conformal diagram for the de Sitter space in the closed sections coordinates. Provide reasoning that this space is (null) geodesically complete, i.e. every (null) geodesic extends to infinite values of affine parameters at both ends.

\paragraph{Solution.} The scale factor in the closed dS Universe is
\begin{equation}
a(t)=H_\Lambda^{-1}\cosh (H_\Lambda t),\qquad t\in (-\infty, +\infty),
\end{equation}
so on integration, for conformal time we obtain
\begin{equation}
\eta (t)=\int\limits_{-\infty}^{t}\frac{dt}{a(t)}=\arcsin\big[\tanh (H_\Lambda t)\big]+\frac{\pi}{2} \in (0,\pi).
\end{equation}
We choose the integration constant here so that $\eta=0$ corresponds to $t=-\infty$ and $\eta=\pi$ to $t=+\infty$. The full metric then can be written as
\begin{equation}
ds^{2}_{dS}=\frac{H_\Lambda^2}{\sin^2 \eta}\big[d\eta^2 -d\chi^2 -\sin^2 \chi d\Omega^2 \big]. \label{dSclosed}
\end{equation}

\begin{figure}[!hb]
\center
\includegraphics*[width=0.48\textwidth]{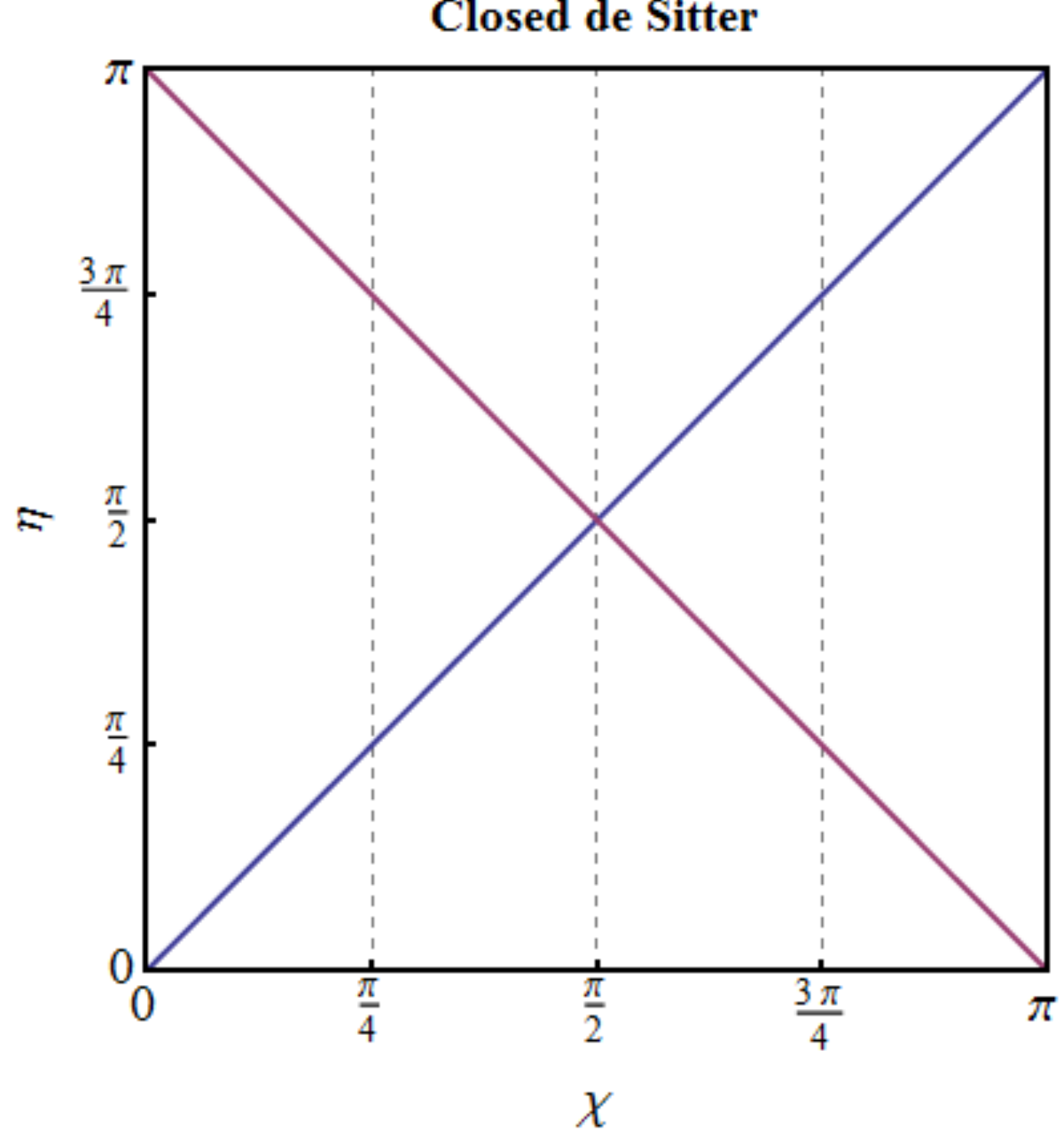}
\parbox{0.82\textwidth}{\caption{The de Sitter Universe. The closed sections coordinates cover the whole space, which is geodesically complete. There are no singularities: the horizontal boundaries of the diagram correspond to infinite past and future.}}
\end{figure}

As the values of $\eta$ span a finite interval, $(\eta,\chi)$ are already conformal coordinates. The conformal diagram is again a square
\begin{equation}
\eta\in [0,\pi],\qquad \chi \in[0,\pi],
\end{equation}
with the difference from the radiation dominated Universe that the edges $\eta=0,\pi$ do not represent a singularity anymore, but instead correspond to infinite (and regular) past and future respectively. Both horizons are given again by
\begin{equation}
\eta_{e}=\pi-\chi,\qquad \eta_{p}=\chi
\end{equation}
and exist at all times.

The spacelike boundaries of the conformal diagram correspond to $t\to \pm\infty$, and therefore to infinite values of affine parameter. This can be shown if one remembers the general formula for the cosmological redshift:
\begin{equation}
\text{const}=\omega a =\frac{dt}{d\lambda}a,\quad \Rightarrow\quad 
	\lambda=\text{const}\cdot\int\limits^{t}dt \cosh(H_\Lambda t)
		\underset{t\to\pm\infty}{\longrightarrow}\infty .
\end{equation}
The timelike boundary of the diagram corresponds to the opposite pole, there is no real boundary there, the same as on a sphere: as particles propagate across the pole, their radial coordinate begins to decrease again, while the worldline on the conformal diagram is reflected from $\chi=\pi$. Thus by definition the spacetime is (null) geodesically complete.

\item \emph{Static dS.} Rewrite the metric of de Sitter space (\ref{dSclosed}) in terms of ``static coordinates'' $T,R$:
\begin{equation}
\tanh (H_\Lambda T)=-\frac{\cos\eta}{\cos\chi},\qquad 
H_\Lambda R =\frac{\sin\chi}{\sin\eta}.
\end{equation}
\begin{enumerate}
\item What part of the conformal diagram in terms of $(\eta,\chi)$ is covered by the static coordinate chart $(T,R)$?
\item Express the horizon's equations in terms of $T$ and $R$
\item Draw the surfaces of constant $T$ and $R$ on the conformal diagram.
\item Write out the coordinate transformation between $(\eta,\chi)$ and $(T,R)$ in the regions where $|\cos\eta|>|\cos\chi|$. Explain the meaning of $T$ and $R$.
\end{enumerate}
\paragraph{Solution.} Let us introduce dimensionless coordinates $t=H_\Lambda T$ and $r=H_\Lambda R$. The inverse relations then are
\begin{equation}
\sin^2 \eta =\frac{1}{\cosh^2 t-r^2 \sinh^2 t},
	\quad \sin^2 \chi =\frac{r^2}{\cosh^2 t-r^2 \sinh^2 t}.
\end{equation}
Using them, after some algebra from (\ref{dSclosed}) we get 
\begin{equation}
ds_{dS}^2 =\big[1-H_\Lambda^2 R^2\big]dT^2 -\frac{dR^2}{1-H_\Lambda^2 R^2}-R^2 d\Omega^2 , \label{dSstatic}
\end{equation}
which resembles the Schwarzschild line element (and this is not a coincidence).

\begin{figure}[!hb]
\center
\includegraphics*[width=0.6\textwidth]{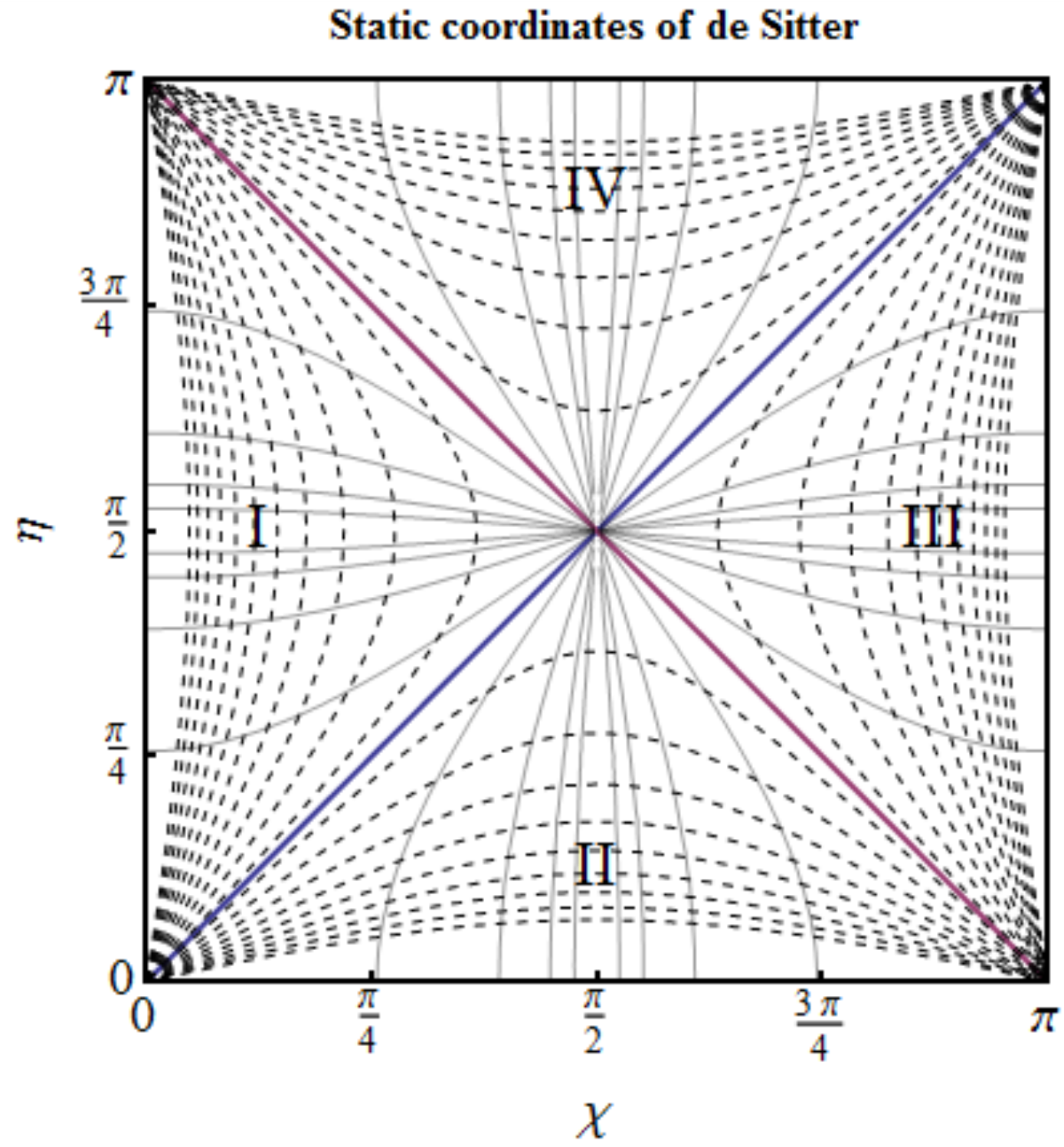}
\parbox{0.9\textwidth}{\caption{The de Sitter Universe with contour lines of the ``static'' coordinates $(T,R)$. The solid lines are $T=const$, and the dashed ones $R=const$. The coordinates become singular on both horizons, so the whole spacetime is divided by them into four sectors, each separately covered by the regular coordinate chart $(R,T)$. Spacetime is actually static only in the left and right sectors (I and III), where $T$ is a timelike coordinate and $R$ spacelike: the static patches are bounded by the horizons.}}
\end{figure}

\begin{enumerate}
\item  As $|\tanh (H_\Lambda T)|<1$, we have $|\cos\eta| \leq |\cos\chi|$. This condition cuts out two out of four sectors from the square conformal diagram, I and III: one is $\eta\in [\chi, \pi-\chi]$ and the other is $\eta\in [\pi-\chi, \chi]$. In both $|\sin\eta|\geq |\sin\chi|$, so $R \leq H_{\Lambda}^{-1}$;
\item The particle horizon $\eta=\chi$ corresponds to $R=H_\Lambda^{-1}$ and $T=-\infty$. It is one part of the boundary of the region (in two parts) covered by coordinates $(T,R)$. The event horizon $\eta=\pi-\chi$ corresponds to $R=H_\Lambda^{-1}$ and $T=+\infty$ and is the other part of the boundary of this region.
\item $T=const$ is $\cos\eta =t\cos\chi $ and $R=const$ is $\sin\eta =r^{-1}\sin\chi$.
\item In regions II and IV the needed relation is obtained if we simply replace $\tanh$ with $\coth$ in the first relation:
\begin{equation}
\coth (H_\Lambda T)=-\frac{\cos\eta}{\cos\chi},\qquad H_\Lambda R=\frac{\sin\chi}{\sin\eta},
\end{equation}
as can be checked explicitly by substitution into (\ref{dSstatic}), which again gives (\ref{dSclosed}). In these regions $R$ is a timelike coordinate, and $T$ is spacelike. The geodesics of comoving massive particles are $\chi=const$, one of them $\chi=\pi/2$ corresponds to $T=0$. Thus in the lower part of the diagram, where $R\in (+\infty,H_\Lambda^{-1})$, the spacetime is contracting; in the upper part, where $R\in (H_\Lambda^{-1},\infty)$, it is expanding. The coordinate frame is not static. The relations between $(T,R)$ and $(\eta,\chi)$ in static and non-static regions mirros those in Schwarzshild black hole solution between the static and the global (Kruskal-Szekeres) coordinates. Here $\eta$ and $\chi$ are the global coordinates.
\end{enumerate}

\item \emph{Flat dS.} The scale factor in flat de Sitter is $a(t)=H_\Lambda^{-1} e^{H_\Lambda t}$.

\begin{enumerate}
\item Find the range of values spanned by conformal time $\tilde{\eta}$ and comoving distance $\tilde{\chi}$ in the flat de Sitter space
\item Verify that coordinate transformation
\begin{equation}
\tilde{\eta}=\frac{-\sin\eta}{\cos\chi-\cos\eta},\qquad
\tilde{\chi}=\frac{\sin\chi}{\cos\chi-\cos\eta}
\end{equation}
bring the metric to the form of that of de Sitter in closed slicing (it is assumed that $\tilde{\eta}=0$ is chosen to correspond to infinite future).
\item Which part of the conformal diagram is covered by the coordinate chart $(\tilde{\eta},\tilde{\chi})$? Is the flat de Sitter space geodesically complete?
\item Where are the particle and event horizons in these coordinates?
\end{enumerate}

\paragraph{Solution.} As $t\in(-\infty,+\infty)$,
\begin{equation}
\tilde{\eta} =\int\frac{dt}{a(t)}=H_\Lambda \int\limits_{+\infty}^{t}dt\; e^{-H_\Lambda t} =-e^{-H_\Lambda t} \in (-\infty,0).
\end{equation}
Here we choose $+\infty$ as the lower limit, because at $-\infty$ the integral diverges.
\begin{enumerate}
\item Direct calculation yields (\ref{dSclosed}), with $\eta \in (0,\pi)$, $\chi\in (0,\pi)$;
\item The upper triangle $\eta>\chi$, above the particle horizon, on which $\tilde{\eta}\to -\infty$. It is not geodesically complete, as geodesics are cut at the particle horizon.
\item The particle horizon is the boundary of the patch of full dS space covered by flat slicing coordinates, and event horizon in these coordinates exists only in the latter part of evolution, for $\eta >\pi/2$.
\end{enumerate} 
\begin{figure}[!hb]
\center
\includegraphics*[width=0.48\textwidth]{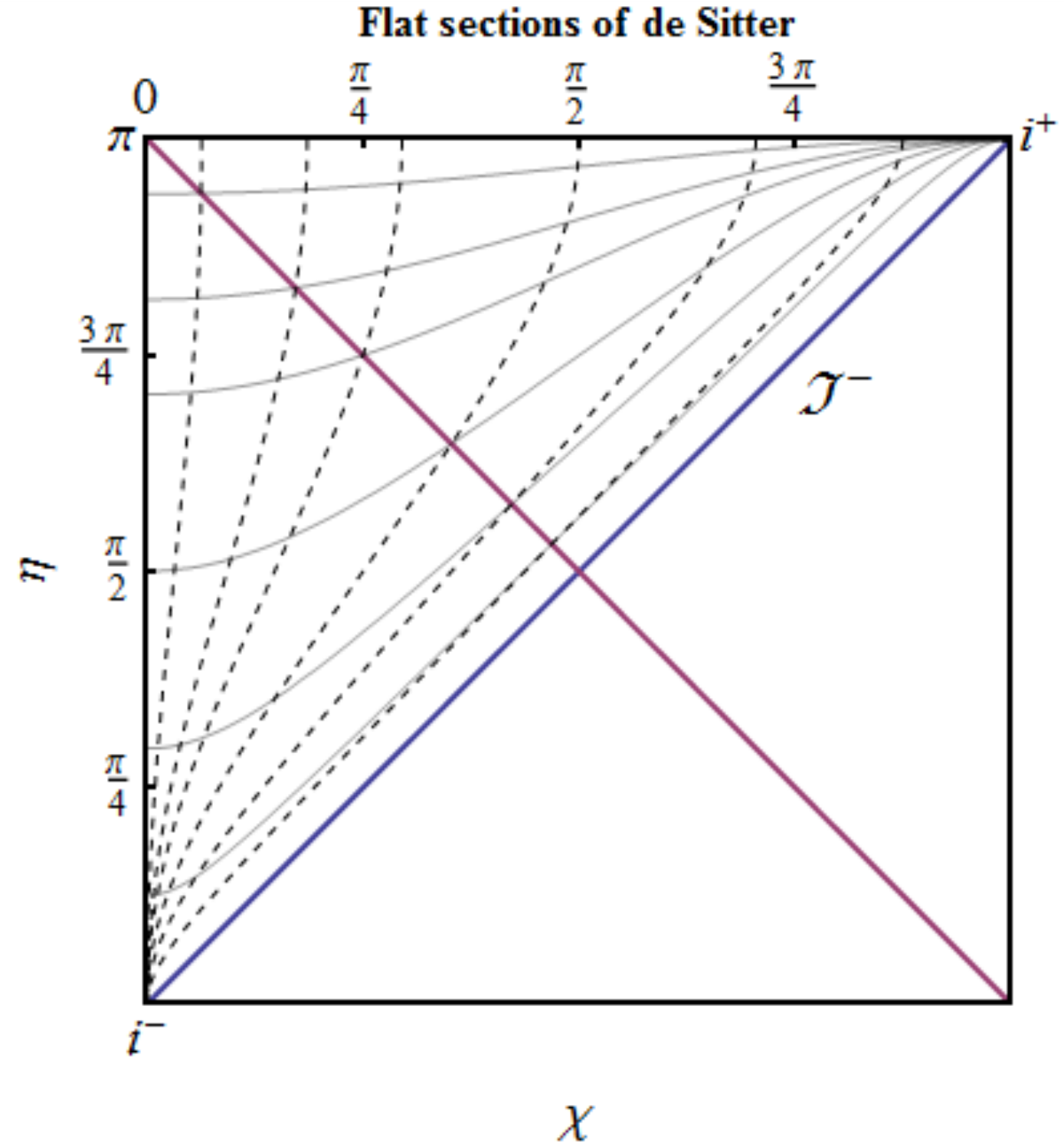}
\parbox{0.9\textwidth}{\caption{The de Sitter Universe in flat sections' coordinates, which cover only half of it. The boundary -- the particle horizon -- consists of three different infinities.}}
\end{figure}

\item \emph{Infinities.} What parts of the spacetime's boundary on the conformal diagram of flat de Sitter space corresponds to
\begin{enumerate}
\item spacelike infinity $i^0$, where $\tilde{\chi}\to +\infty$;
\item past timelike infinity $i^-$, where $\tilde{\eta}\to -\infty$ and from which all timelike worldlines emanate
\item past lightlike infinity $J^-$, from which all null geodesics emanate?
\end{enumerate}
\paragraph{Solution.}
\begin{enumerate}
\item The one point at the left of the diagram;
\item the one point at the bottom;
\item the particle horizon.
\end{enumerate}

\item \emph{Open dS.} Consider the de Sitter space in open slicing, in which $a(t)=H_\Lambda \sinh (H_\Lambda t)$, so conformal time is
\begin{equation}
\tilde{\eta} =\int\limits_{+\infty}^{t} \frac{dt}{a(t)},
\end{equation}
where again the lower limit is chosen so that the integral is bounded.
\begin{enumerate}
\item Find $\tilde{\eta}(t)$ and verify that coordinate transformation from $(\tilde{\eta},\tilde{\chi})$ to $\eta,\chi$, such that
\begin{equation}
\tanh\tilde{\eta}=\frac{-\sin\eta}{\cos\cos\chi},\qquad
\tanh\tilde{\chi}=\frac{\sin\chi}{\cos\eta}
\end{equation}
brings the metric to the form of de Sitter in closed slicing.
\item What are the ranges spanned by $(\tilde{\eta},\tilde{\chi})$ and $(\eta,\chi)$? Which part of the conformal diagram do they cover?
\end{enumerate}
\paragraph{Solution.} After getting
\begin{equation}
\sinh \tilde{\eta}=-\frac{1}{\sinh(H_\Lambda t)},
\end{equation}
the first part is checked straightforwardly; in the open de Sitter $\tilde{\chi}\in [0,+\infty)$, and $\tilde{\eta}\in(-\infty,0)$. The region covered by coordinates $(\tilde{\eta},\tilde{\chi})$ is $\{\eta>\chi+\pi/2\}$, only one eighth part of the full diagram.
\begin{figure}[!hb]
\center
\includegraphics*[width=0.48\textwidth]{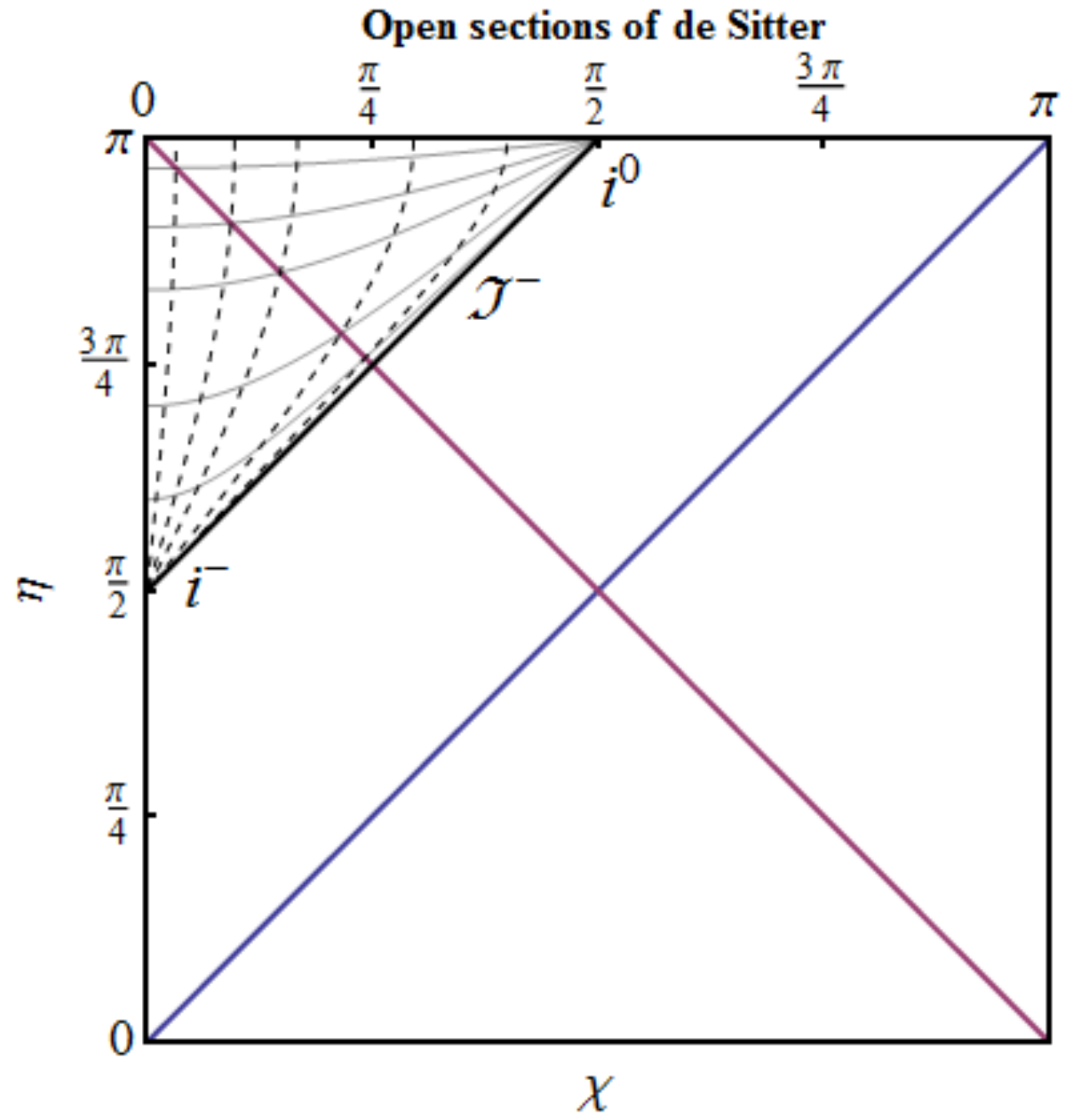}
\parbox{0.6\textwidth}{\caption{The de Sitter Universe in open sections' coordinates, which cover only $1/8^{\text{th}}$ of the full diagram.}}
\end{figure}

\item \emph{Minkowskii 1.} Rewrite the Minkowski metric in terms of coordinates $(\eta,\chi)$, which are related to $(t,r)$ by the relation
\begin{equation}
\tanh \tilde{\eta}=\frac{\sin\eta}{\cos\chi},\qquad
\tanh \tilde{\chi}=\frac{\sin\chi}{\cos\eta}
\end{equation}
that mirrors the one between the open and closed coordinates of de Sitter. Construct the conformal diagram and determine different types of infinities. Are there new ones compared to the flat de Sitter space?
\paragraph{Solution.} Coordinate transformation gives
\begin{equation}
ds^2 =\frac{1}{\cos^2 \chi -\sin^2 \eta}\big[d\eta^2 -d\chi^2 -\Psi^{2}(\eta,\chi)d\Omega^2 \big].
\end{equation}
Here $r\in [0,+\infty)$ and $t\in (-\infty,+\infty)$. Comparing with the relation between $(\tilde{\eta},\tilde{\chi})$ with conformal coordinates $(\eta,\chi)$ in the open de Sitter universe, where $\tilde{\eta}\in (-\infty,0)$, we see that the difference is that $\tilde{\eta}$ spans $(-\infty,0)$, while now $t$ spans twice the range, $(-\infty,\infty)$. Therefore the conformal diagram is composed of two triangles, one the same as for open de Sitter and one for its time-reversed copy. Accordingly there now appear future timelike infinity $i^+$ and future lightlike infinity $J^+$.
\begin{figure}[htb]
\center
\includegraphics*[width=0.48\textwidth]{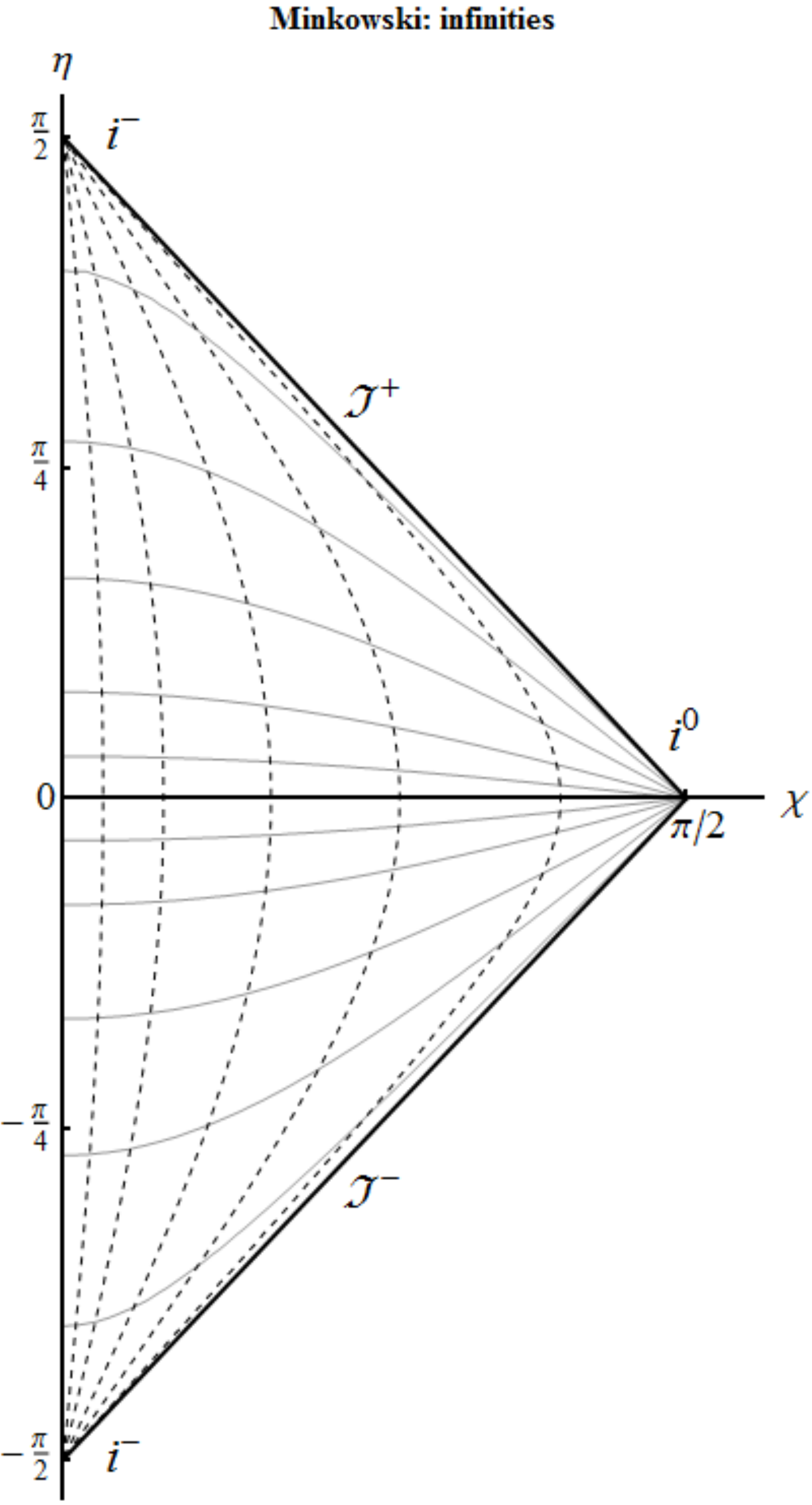}\hfill
\includegraphics*[width=0.48\textwidth]{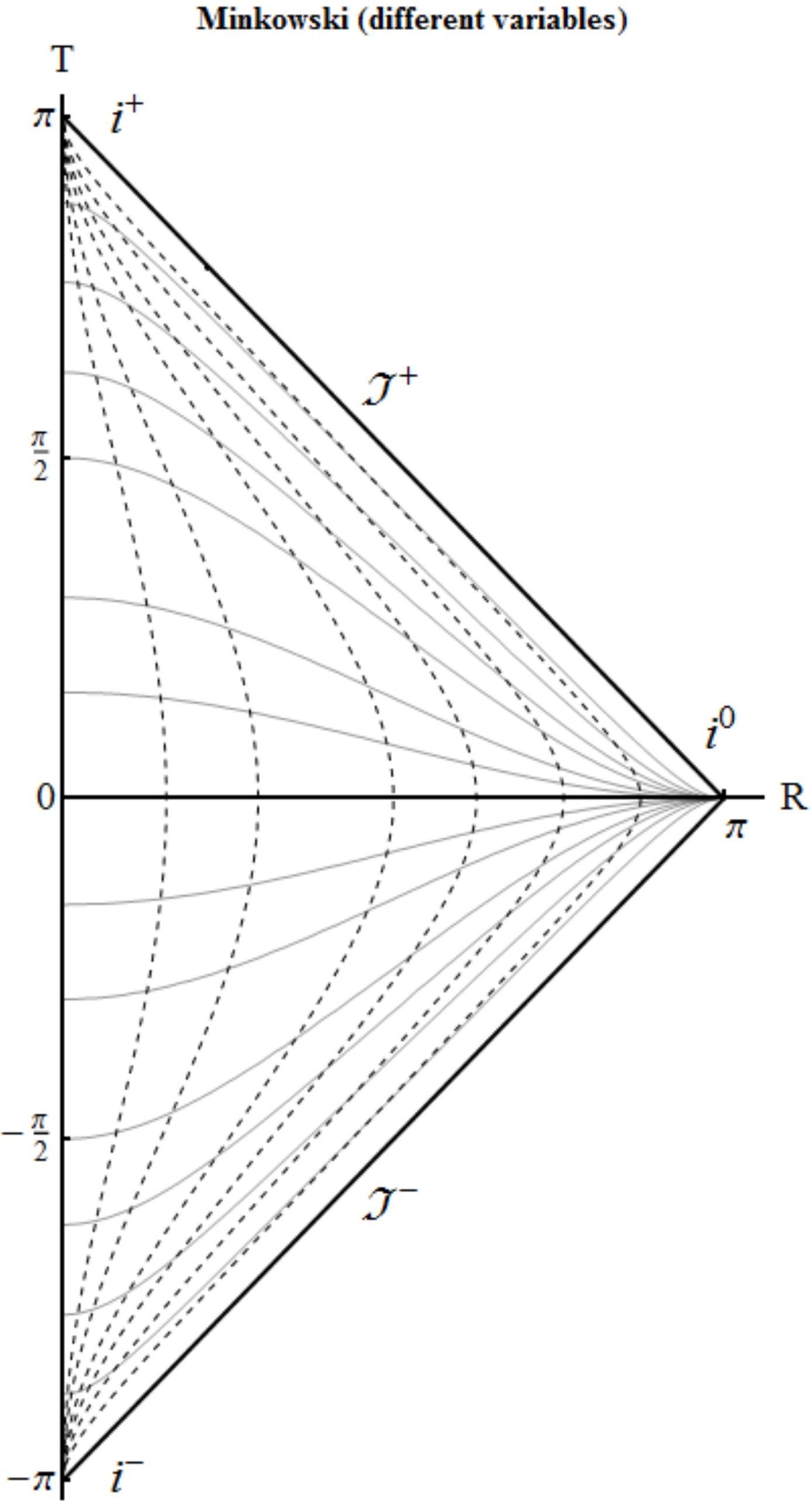}
\parbox{0.98\textwidth}{\caption{The Minkowski spacetime in two different pairs of conformal coordinates and the full set of infinities. Thin dashed and solid lines show the images of coordinate grid $(t,r)$.}}
\end{figure}
 
\item \emph{Minkowski 2.} The choice of conformal coordinates is not unique. Construct the conformal diagram for Minkowski using the universal scheme: first pass to null coordinates, then bring their span to finite intervals with $\arctan$ (one of the possible choices), then pass again to timelike and spacelike coordinates. 
\paragraph{Solution.} We start from spherical coordinates
\begin{equation}
ds^2 =dt^2 -dr^2 -r^2 d\Omega^2 .
\end{equation}
\begin{enumerate}
\item The first step is introducing null coordinates
\begin{equation}
u=t-r,\quad v=t+r ,
\end{equation}
so that
\begin{equation}
ds^2 =4du\,dv -\frac{(v-u)^2}{4}d\Omega^2 .
\end{equation}
\item Then bring the range of values to finite intervals
\begin{equation}
U=\arctan u, \qquad V=\arctan v,
\end{equation}
so that
\begin{equation}
ds^2 =\frac{1}{4\cos^2 U \cos^2 V}\big[4dU\,dV -\sin^2 (V-U)d\Omega^2\big].
\end{equation}
The whole spacetime is simply the half of the square $U,V\in (-\pi/2,\pi/2)$, in which $r>0$, i.e. $v>u\quad \Leftrightarrow\quad V>U$: on the plane $(U,V)$ it is the triangle \begin{equation}
	-\pi/2<U<V<\pi/2.
\end{equation}
\item Finally, go back to spacelike and timelike coordinates
\begin{equation}
T=V+U,\qquad R=V-U
\end{equation}
so that metric becomes
\begin{equation}
ds^2 =\frac{1}{[\cos T +\cos R]^2}\big[dT^2 -dR^2 -\sin^2 R\; d\Omega^2\big].
\end{equation}
The triangle is shrunken by $\sqrt{2}$ and rotated by $3\pi/4$ clockwise, thus turning into 
\begin{equation}
	\{R>0,\quad |T|<\pi/2 -R\}.
\end{equation}
This is the same form as obtained by the other construction (up to scaling, which is purely decorative).
\end{enumerate}

\item Draw the conformal diagram for the Milne Universe and show which part of Min\-kow\-ski space's diagram it covers.
\paragraph{Solution.} Minkowski metric $ds^2 =dT^2 -dR^2$, rewritten in terms of $(\tau, r)$ such that
\begin{equation}
T=\tau \cosh r ,\quad R=\tau \sinh r ,
\end{equation}
is the metric of the Milne Universe. As Minkowski space is complete, the Milne Universe then is a \emph{part} of Minkowski in different variables. This part is where $T>R$. The boundary $T=R$ is the future light cone, so the whole Milne Universe is represented by the triangle homothetic to the whole space but 4 times smaller in area, with the common node of future infinity.
\begin{figure}[!hb]
\center
\includegraphics*[width=0.55\textwidth, viewport=0 300 400 740]{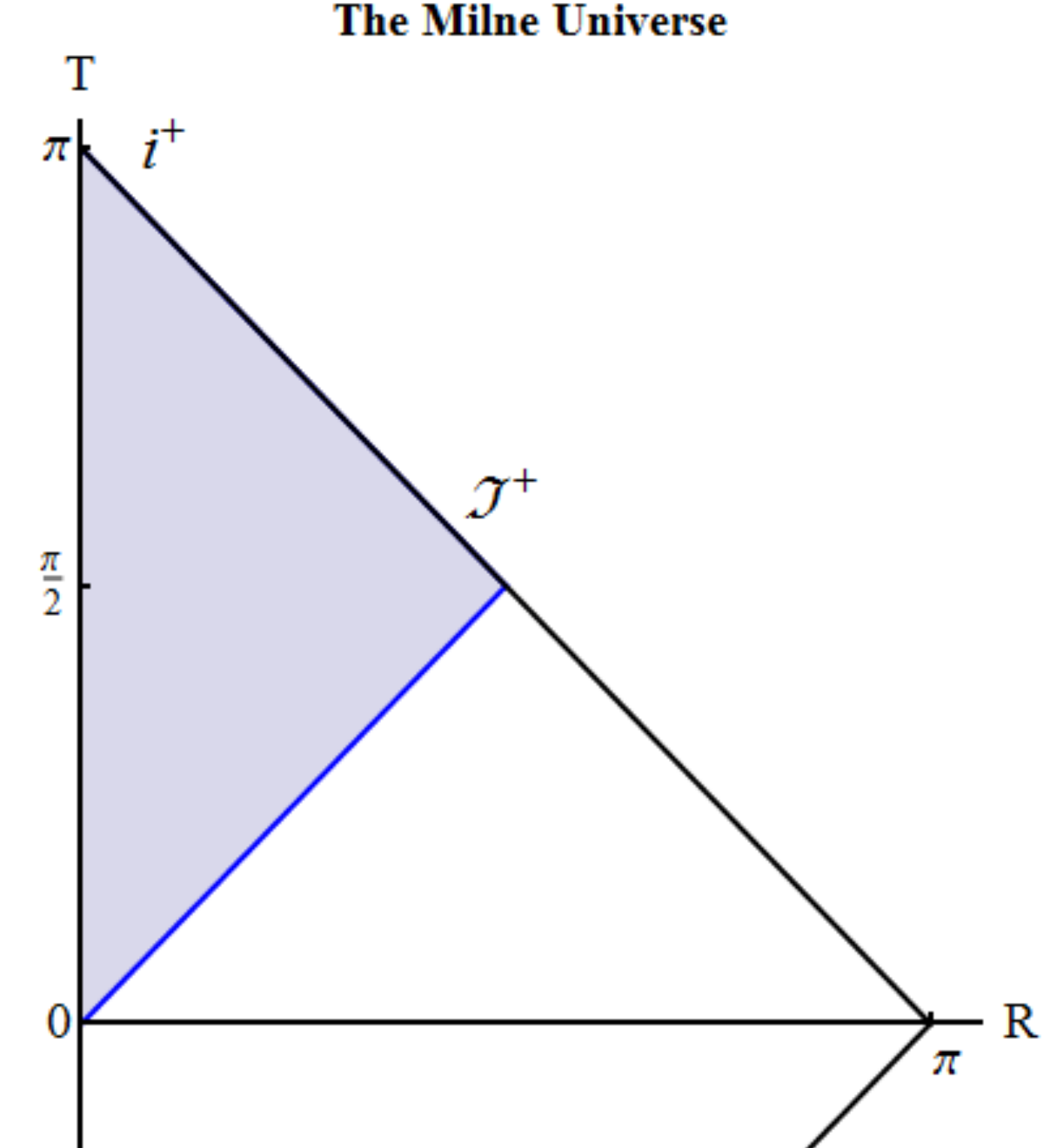}
\parbox{0.6\textwidth}{\caption{The Milne Universe on (a part of) Minkowski's conformal diagram. It has the same shape and shares the same future infinity, but covers one eighth of the area. The boundary is the past horizon.}}
\end{figure}

\item Consider open or flat Universe filled with matter that satisfies strong energy condition $\varepsilon +3p>0$. What are the coordinate ranges spanned by the comoving coordinate $\tilde{\chi}$ and conformal time $\tilde{\eta}$? Compare with the Minkowski metric and construct the diagram. Identify the types of infinities and the initial Big Bang singularity.
\paragraph{Solution.} For flat Universe $\tilde{\chi}=r$, for the open $\tilde{\chi}=\sinh r$, so in both cases $\tilde{\chi}\in (0,+\infty)$. 

Strong energy condition implies that $w>-1/3$, so 
\begin{equation}
\rho \sim a^{-3(1+w)}=a^{-n},
\end{equation}
where $n>2$. From the first Friedman equation then after simple manipulations we obtain that
\begin{equation}
\frac{da}{dt}\sim a^{-\theta},
\end{equation}
where $\theta$ is some positive number. Therefore both
\begin{equation}
t\sim \int da\; a^{\theta},
	\quad\text{and}\quad 
		\eta =\int \frac{da}{a}a^{\theta}
\end{equation}
converge at $a\to 0$ and diverge at $a\to \infty$. Consequently, the integration constant can be chosen so that $\eta\in (0,+\infty)$.

The conformal structure is the same as that of the \emph{upper} half of Minkowski spacetime. The Big Bang singularity at $\eta=0$ is at the cut, and there are spacelike infinity, future infinity and future null infinity.

\begin{figure}[htb]
\center
\includegraphics*[width=0.48\textwidth]{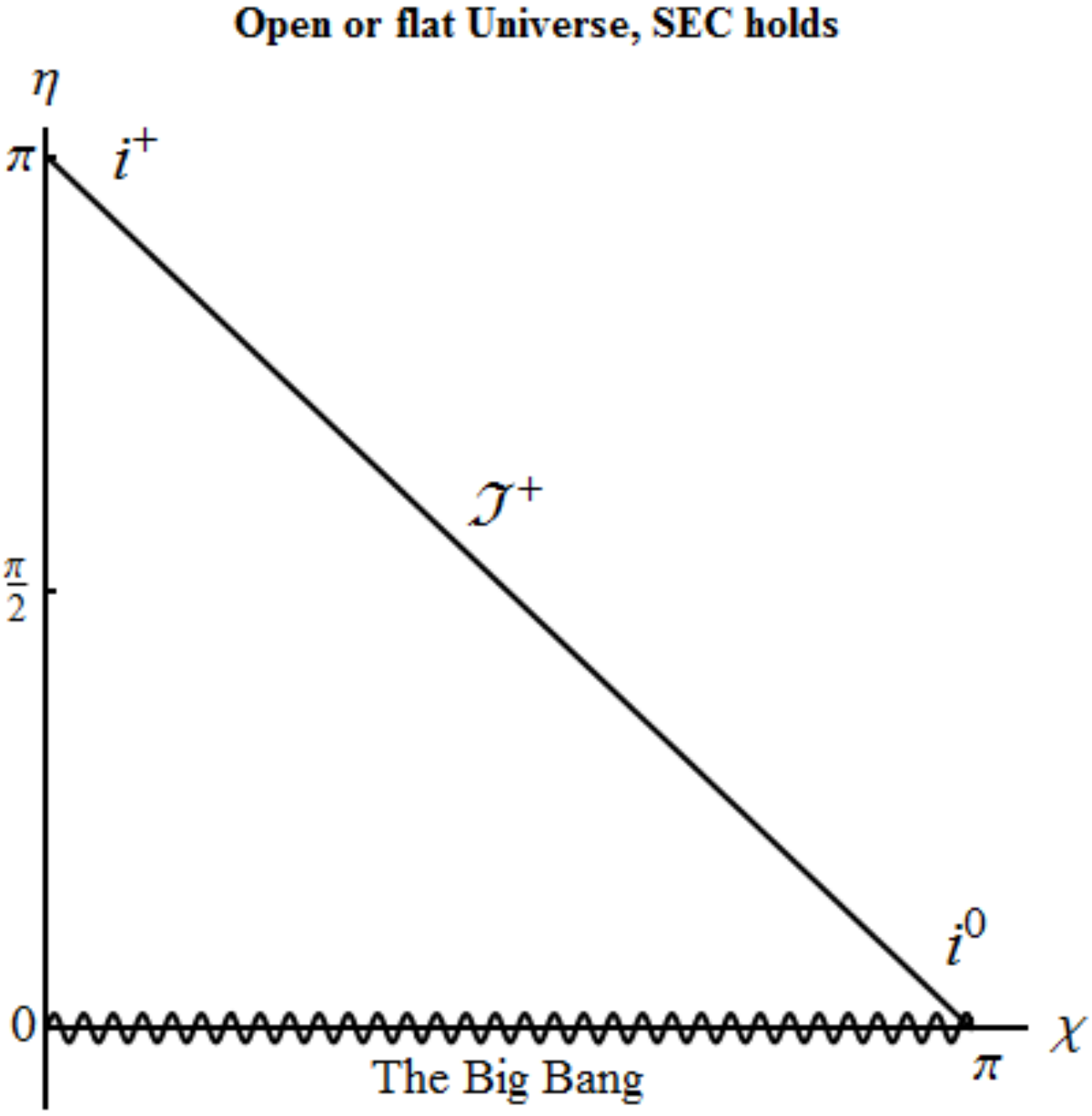}\hfill
\includegraphics*[width=0.48\textwidth]{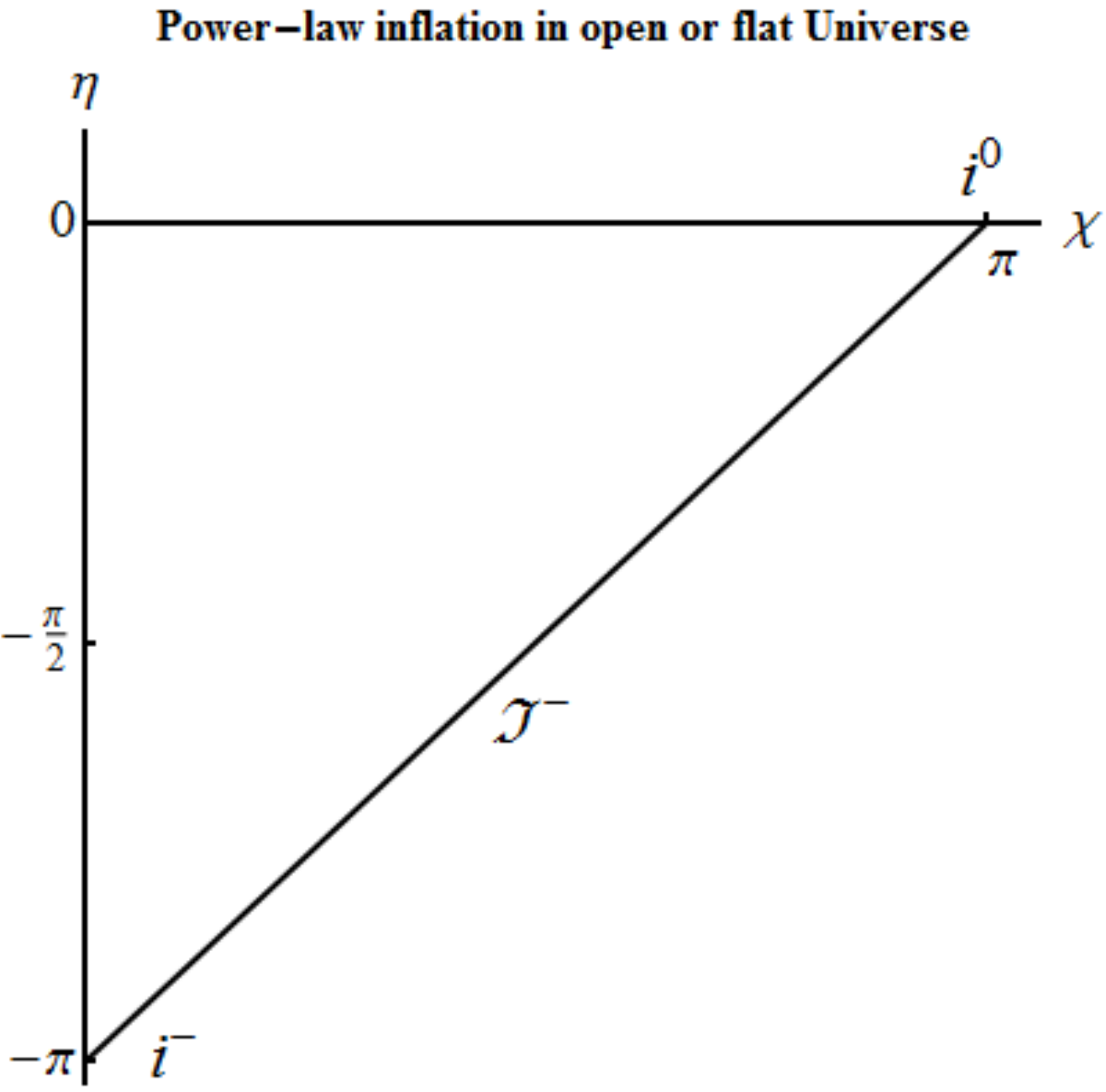}
\parbox{0.7\textwidth}{\caption{Conformal diagrams for open or flat Universes. The one on the left is for ones filled with matter which satisfies the strong energy condition (SEC), and the one on the right for ones in which SEC does not hold, such as in case of power-law inflation.}}
\end{figure}

\item Draw the conformal diagram for open and flat Universes with power-law scale factor $a(t)\sim t^{n}$, with $n>1$. This is the model for the power-law inflation. Check whether the strong energy condition is satisfied.
\paragraph{Solution.} As seen in the previous problem, if strong energy condition were satisfied, we would have $\dot{a}\sim a^{-\theta}$ with some positive $\theta$; this is not the case, so the condition is violated. As $t\in (0,+\infty)$, 
\begin{equation}
\eta\sim \int \frac{dt}{t^n}
\end{equation}
diverges at small $a$ (thus also small $t$) and converges at $a\to \infty$ (thus as large $t$). So integration constant can be chosen so that $\eta\in (-\infty,0)$. 

The conformal structure is the same as \emph{lower half} of Minkowski spacetime. The cut is regular future infinity, and from Minkowski there are spacelike infinity, past null infinity, and past infinity. The point of past infinity corresponds to Big Bang and is singular.

\end{enumerate}

\FloatBarrier

\section{Conformal diagrams: stationary black holes}
In the context of black hole spacetimes there are many subtly distinct notions of horizons,
 with the most useful being different from those used in cosmology. In particular, particle horizons do not play any role. The event horizon is defined not with respect to some selected observer, but with respect to \emph{all} external observers: in an asymptotically flat spacetime\footnote{Meaning the spacetime possesses the infinity with the same structure as that of Minkowski, which is important.} a future event horizon is the hypersurface which separates the events causally connected to future infinity and those that are not. Likewise the past event horizon delimits the events that are causally connected with past infinity or not. Another simple but powerful concept is the  \emph{Killing horizon}: in a spacetime with a Killing vector field $\xi^\mu$ it is a (hyper-)surface, on which $\xi^\mu$ becomes lightlike. We will see explicitly for the considered examples that the Killing horizons are in fact event horizons by constructing the corresponding conformal diagrams. In general, in the frame of GR a Killing horizon in a stationary spacetime is (almost) always an event horizon and also coincides with most other notions of horizons there are.

This section elaborates on the techniques of constructing conformal diagrams for stationary black hole solutions. The construction for the Schwarzschild black hole, or its variation, can be found in most textbooks on GR; for the general receipt see \cite{BrRubin}.

\subsection{Schwarzschild-Kruskal black hole solution}
\begin{enumerate}[resume]
\item \emph{Schwarzschild exterior.} The simplest black hole solution is that of Schwarzschild, given by
\begin{equation}
ds^{2}=f(r)dt^2 -\frac{dr^2}{f(r)}-r^2 d\Omega^2,\qquad f(r)=1-\frac{r_g}{r},
	\label{SchwBH}
\end{equation}
where $r_g$ is the gravitational radius, and $d\Omega^2$ is the angular part of the metric, which we will not be concerned with. The surface $r=r_g$ is the horizon. Focus for now only on the external part of the solution, 
\[\big\{-\infty<t<+\infty,\;\;
	 r_g<r<+\infty)\big\}.\]
The general procedure of building a conformal diagram for the $(t,r)$ slice, as discussed in the cosmological context earlier, works here perfectly well, but needs one additional step in the beginning:
\begin{enumerate}
\item use a new radial coordinate to bring the metric to conformally flat form;
\item pass to null coordinates;
\item shrink the ranges of coordinate values to finite intervals with the help of $\arctan$;
\item return to timelike and spacelike coordinates.
\end{enumerate}
Identify the boundaries of Schwarzschild's exterior region on the conformal diagram and compare it with Minkowski spacetime's.
\paragraph{Solution.} 
\begin{enumerate}
\item In terms of the new  (``tortoise'') coordinate 
\begin{equation}
x=\int \frac{dr}{f(r)}=r+r_g \ln |r-r_g |
\end{equation}
we get
\begin{equation}
ds^2 =f(r(x))\big[dt^2 -dx^2\big].
\end{equation}
The horizon is pushed to infinity: $x\to -\infty$.
\item Introduce
\begin{equation}
 v=t+x,\quad u=t-x
\end{equation}
The horizon is at $v\to-\infty$ when $u=const$ (past horizon) and at $u\to -\infty$ when $v=const$ (future horizon). The asymptotically flat infinity is in the opposite direction: at $v\to+\infty$ when $u=const$ and at $u\to +\infty$ when $v=const$ 
\begin{figure}[ht]
\center
\includegraphics*[width=0.7\textwidth]{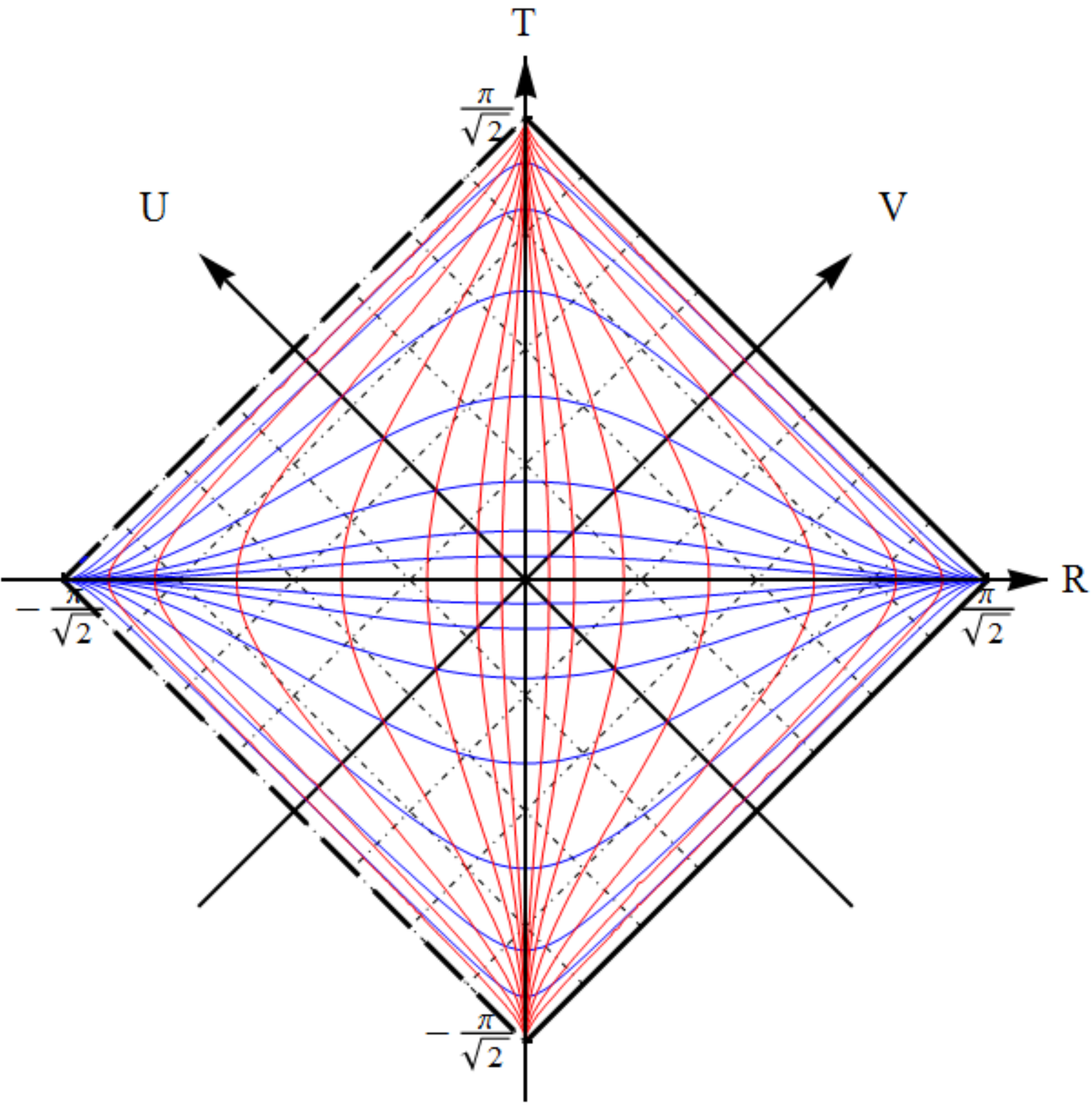}
\parbox{0.7\textwidth}{\caption{Conformal diagram for the exterior region of Schwarzschild. The dashed part of the boundary is the horizon; dash-dotted thin lines represent null geodesics; thin red lines are lines of constant $r$, thin blue lines are lines of constant $t$.}}
\end{figure}
\item Pass to $V=\arctan v$ and $U=\arctan u$; the horizon now becomes two lines at $V=-\pi/2$ and $U=-\pi/2$. The infinity is $V=+\pi/2$ and $U=+\pi/2$. The full exterior region is the square enclosed by those four lines.
\item Transformation to $T=V+U$, $R=V-U$ rotates this square by $\pi/4$ and this is the conformal diagram block that we need. There is a past and future horizon. Note that while past should be below and future above (we expect the spacetime to be oriented, so that the direction of future is always uniquely determined), there is no rule that says that the horizons must be on the left and infinities on the right. Thus there are two mirror-reflected variants, with the horizons on the left and on the right.
\end{enumerate}

\item \emph{Schwarzschild interior.} The region $r\in (0,r_g)$ represents the black hole's interior, between the horizon $r=r_g$ and the singularity $r=0$. Construct the conformal diagram for this region following the same scheme as before. 
\begin{enumerate}
\item Which of the coordinates $(t,r)$ are timelike and which are spacelike?
\item Is the singularity spacelike or timelike?
\item Is the interior solution static?
\end{enumerate}
The Schwarzschild black hole's interior is an example of the \emph{T-region}, where $f(r)<0$, as opposed to the \emph{R-region}, where  $f(r)>0$.
\paragraph{Solution.} The procedure is exactly the same as for the exterior region, the only thing that is different is the ranges of coordinates used and reversal of the timelike and spacelike coordinates:
\begin{enumerate}
\item Now $t$ is spacelike and $r$ a timelike coordinate:
\begin{equation}
ds^2 =|f(r)|^{-1}dr^2 -|f(r)|dt^2 .
\end{equation}
\item The singularity $r=0$ is therefore a spacelike one-dimensional object: one does not ``reach'' it or not, instead one lives until the specified time.
\item The interior solution in \emph{not} static, as the metric essentially depends on the new time coordinate $r$ (or $-r$).
\end{enumerate} 
The null coordinates then are (note the second sign)
\begin{equation}
v=t+x,\qquad u=x-t.
\end{equation}
The horizon is at $v\to -\infty$ when $u=const$ and at $u\to -\infty$ when $v=const$. After shrinking with $\arctan$ it is pulled to $v=-\pi/2$ and $u=-\pi/2$. The third part of the boundary, corresponding to the singularity $r=0$, is $v+u=const$. 

Finally, we return to timelike and spacelike coordinates by introducing (there are two ways of choosing the sign)
\begin{equation}
T=\pm (v+u),\qquad R=V-U.
\end{equation}
The resulting conformal diagram  is a 45 degrees right triangle with the right angle pointing either up (then the singularity is in the past) or down (then the singularity is in future). Note that we will know which of the horizons is past and which is future only when we match the blocks with the ones that have the corresponding infinities (although one can guess already where those will be).
\begin{figure}[!hb]
\center
\includegraphics*[width=0.7\textwidth]{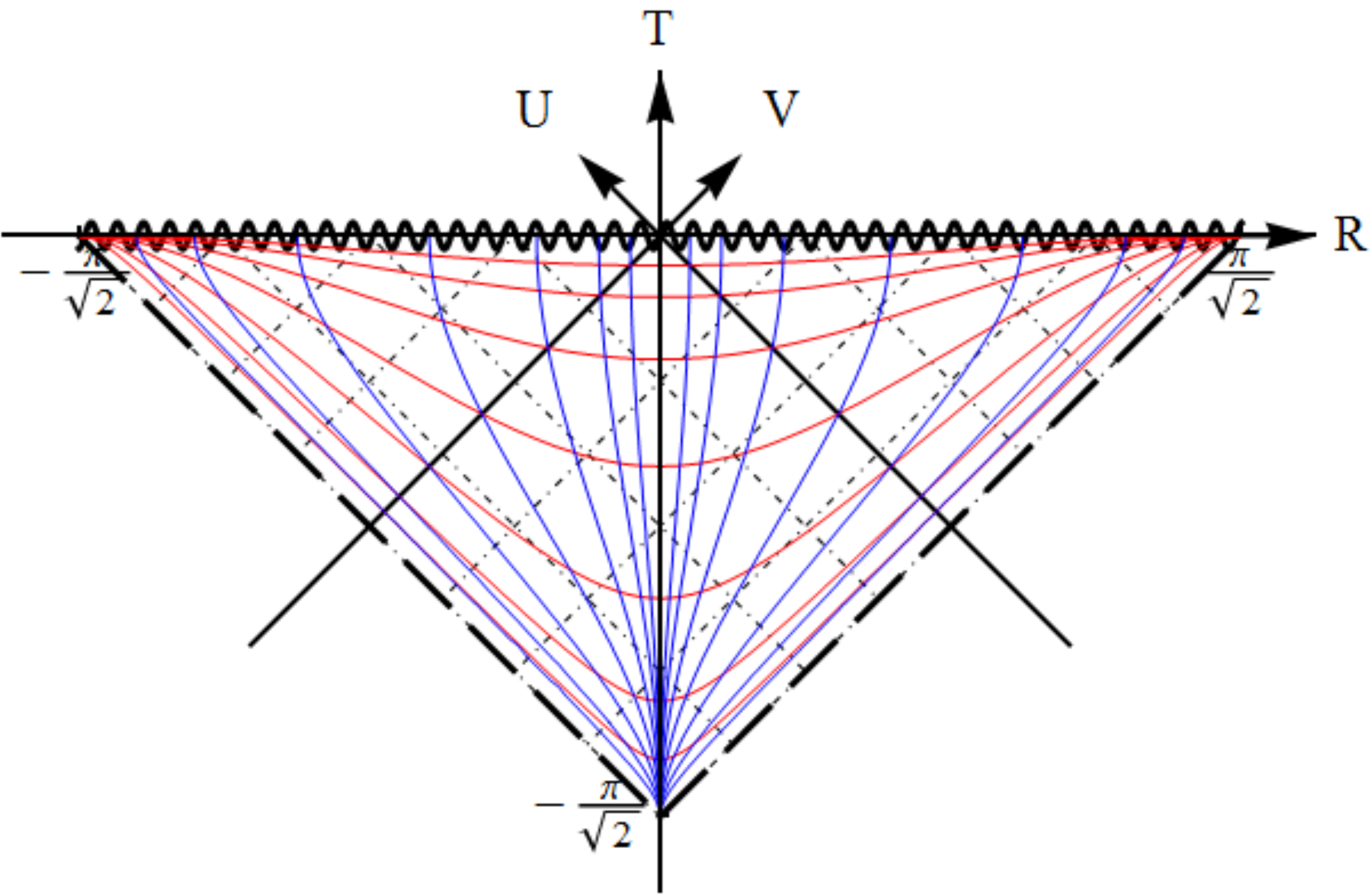}
\parbox{0.75\textwidth}{\caption{Conformal diagram for the interior region of Schwarzschild with future singularity. The dashed part of the boundary is the horizon(s); dash-dotted thin lines represent null geodesics; thin red lines are lines of constant $r$, thin blue lines are lines of constant $t$.}}
\end{figure}

\item \emph{Geodesic incompleteness and horizon regularity.} Consider radial motion of a massive particle and show that the exterior and interior parts of the Schwarzschild are not by themselves geodesically complete, i.e. particle's worldlines are terminated at the horizon at finite values of affine parameter. Show that, on the other hand, the horizon is not a singularity, by constructing the null coordinate frame, in which the metric on the horizon is explicitly regular.
\paragraph{Solution.} The geodesic equation is obtained from normalization condition
\begin{equation}
 \varepsilon^2=u^\mu u_\mu 
 	=f(r)\Big(\frac{dt}{d\lambda}\Big)^{2}
 	-f^{-1}(r)
 		\Big(\frac{dr}{d\lambda}\Big)^{2},
\end{equation}
where $\varepsilon^2=0$ or $\varepsilon^2=1$  for null and timelike geodesics respectively, and conservation equation due to the Killing vector
\begin{equation}
E=u^\mu \xi_\mu =u_0 
	=f(r)\frac{dt}{d\lambda}, \label{BH-energy}
\end{equation}
and reads
\begin{equation}
\frac{d\lambda}{dr}
	=\Big[E^2 -f(r)\varepsilon^2 \Big]^{-1/2}. \label{BH-lambda}
\end{equation}
The integral $\lambda(r)$ converges at $r=r_g$, so the value of the affine parameter at the horizon is finite.

In regard to regularity let us start from the exterior region. In terms of $u,v$ the metric has the overall conformal factor, which turns to zero at the horizon:
\begin{equation}
f(r)\sim (r-r_g)\sim e^{\ln (r-r_g)}
	\sim e^{x}\sim e^{(v-u)/2},
\end{equation}
so in order to eliminate that, we just need to use new null coordinates $(u',v')$, such that the conformal factor in
\begin{equation}
ds^{2}
	\sim \Big(e^{-u/2}\frac{du}{du'}\Big)
	\cdot \Big(e^{+v/2}\frac{dv}{dv'}\Big)
	\cdot	du' dv'
\end{equation}
is finite and does not turn to zero. This is achieved e.g. by coordinate transformation to
\begin{equation}
 u'=e^{u/2},\qquad v'=e^{-v/2}.
\end{equation}
After this one can carry out the second part of construction of the conformal diagram.

\item \emph{Piecing the puzzle.} Geodesic incompleteness means the full conformal diagram must be assembled from the parts corresponding to external and internal solutions by gluing them together along same values of $r$ (remember that each point of the diagram corresponds to a sphere). Piece the puzzle. 

Note that a) there are two variants of both external and internal solutions' diagrams, differing with orientation and b) the boundaries of the full diagram must go along either infinities or singularities.
\paragraph{Solution.} Starting from any one of the four pieces, there is only one way to assemble the diagram, and one has to use all the parts:
\begin{figure}[!ht]
\center
\includegraphics*[width=0.45\textwidth]{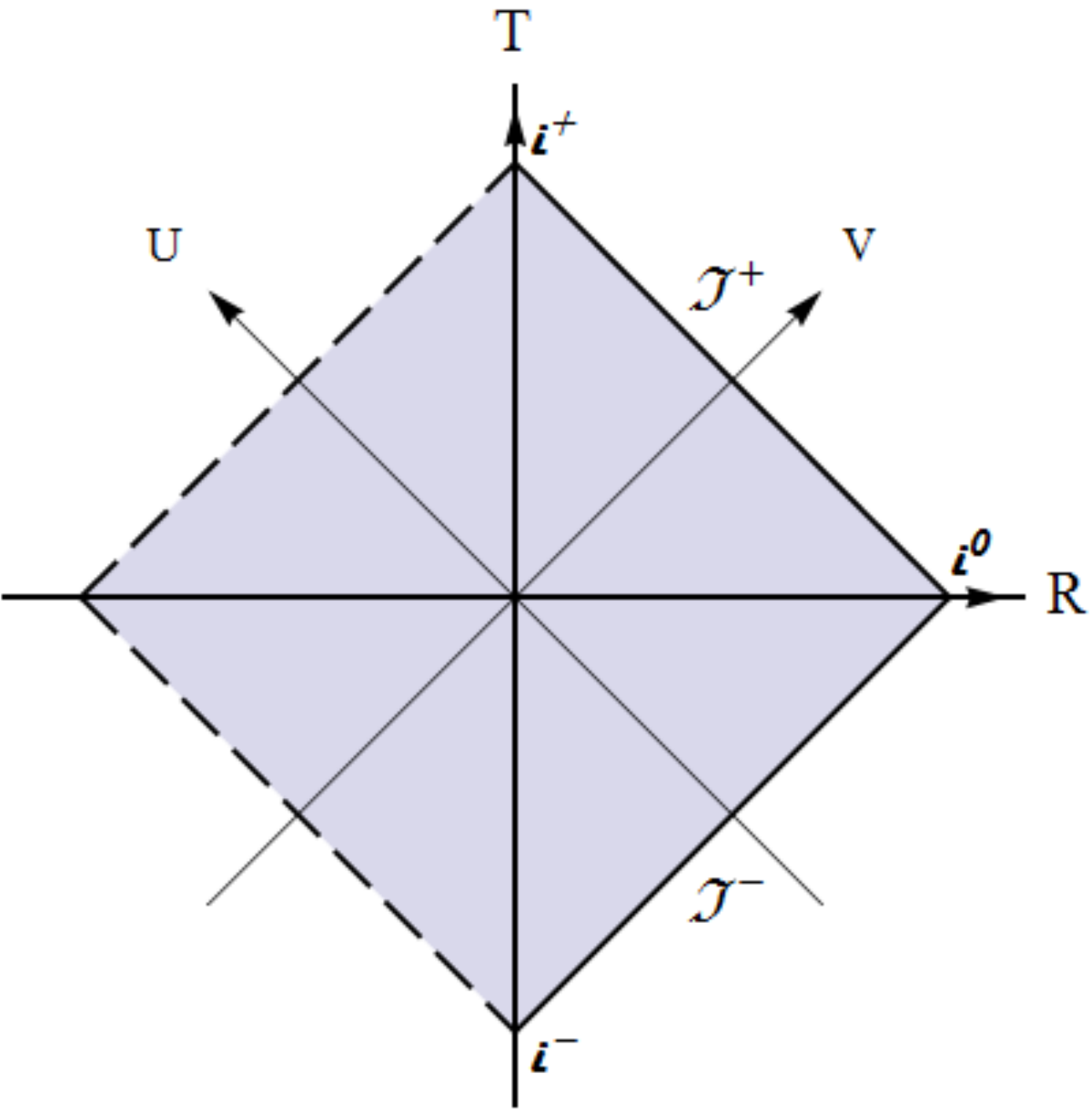}\hfill
\includegraphics*[width=0.45\textwidth]{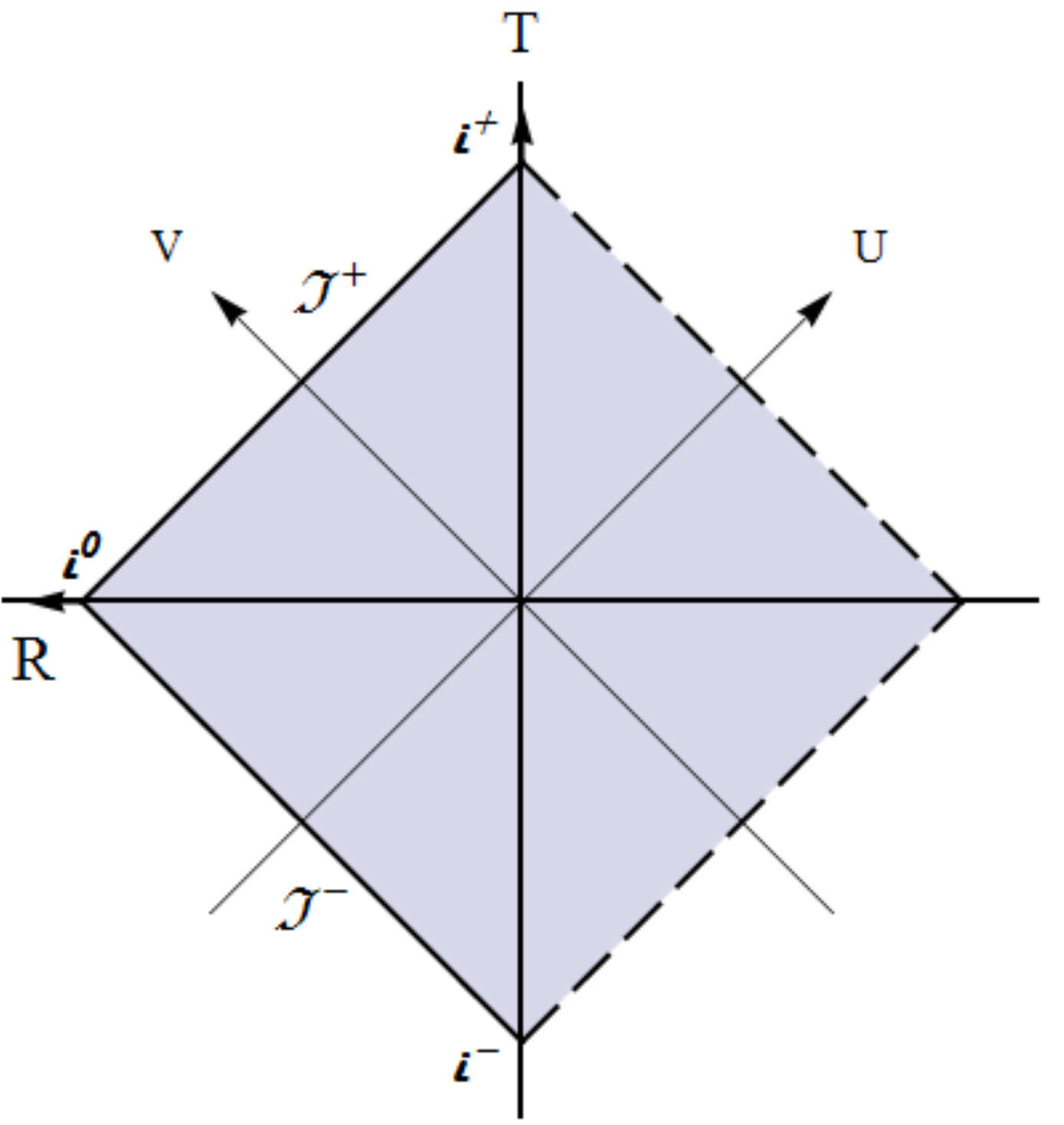}\hfill
\includegraphics*[width=0.45\textwidth]{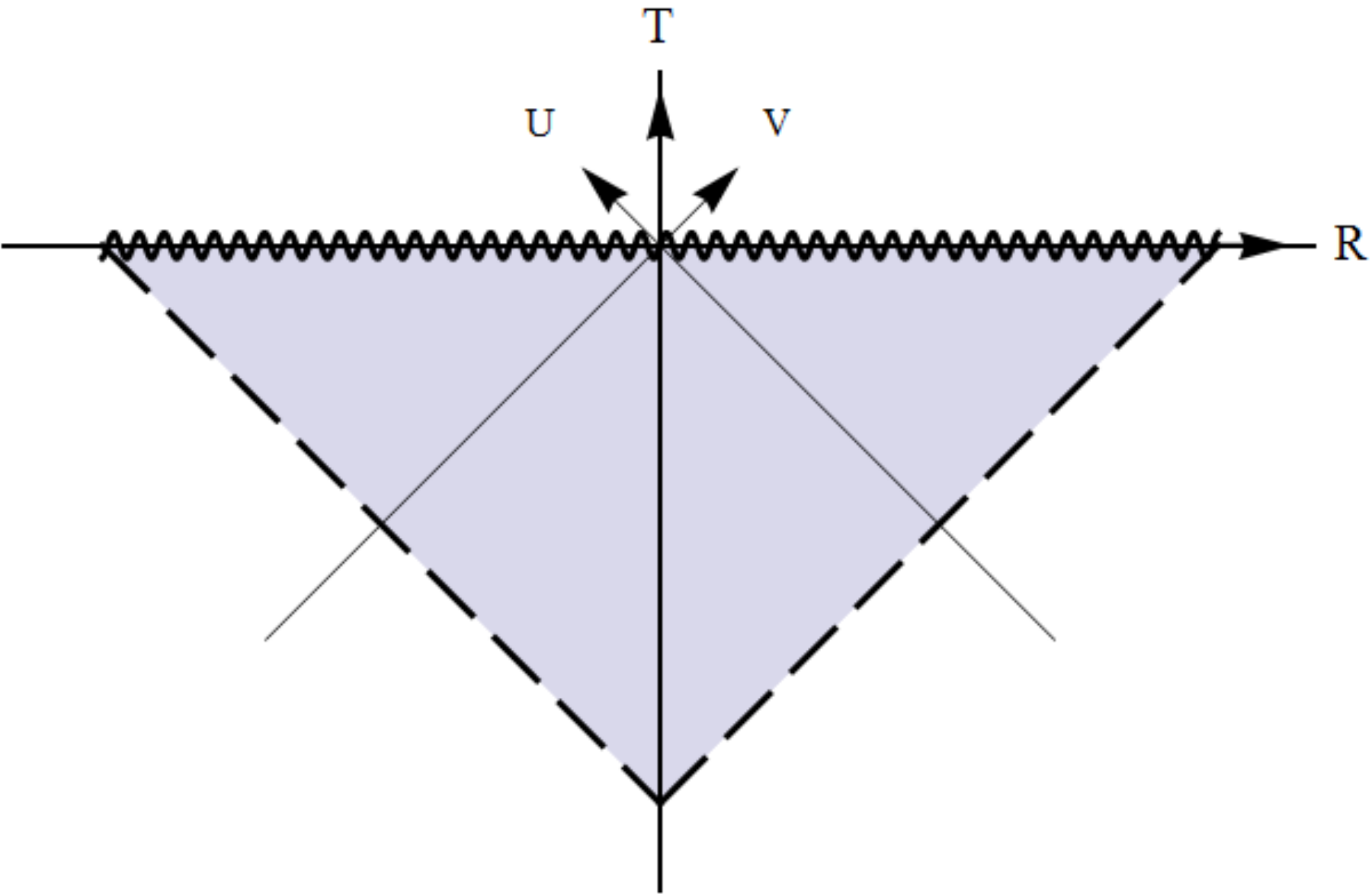}\hfill
\includegraphics*[width=0.45\textwidth]{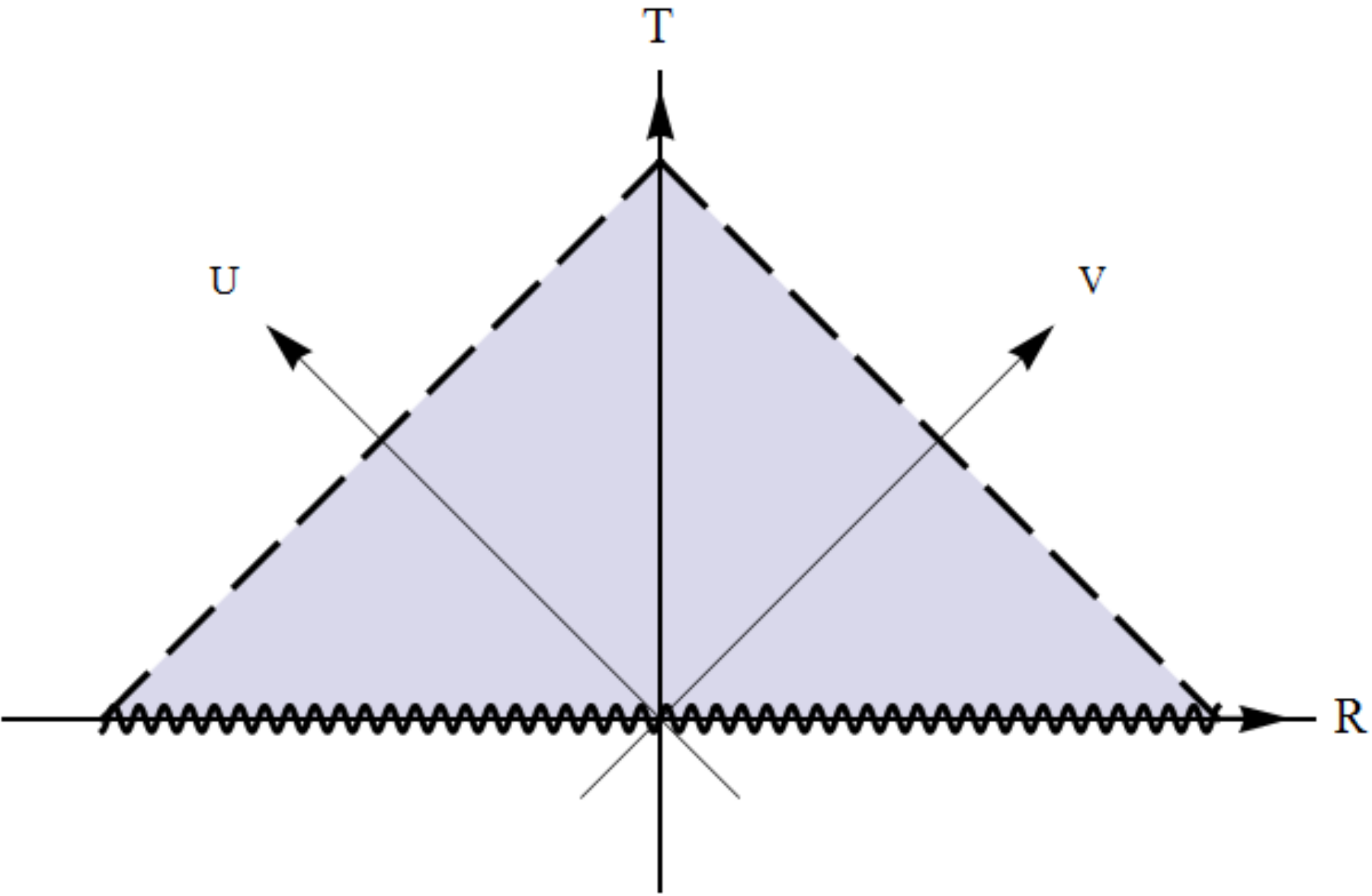}
\parbox{0.7\textwidth}{\caption{The puzzle pieces: two for the exterior region (above) and two for the interior region (below).}}
\end{figure}

The horizon structure is very similar to that of full de Sitter spacetime in open slicing or ``static'' coordinates (the notions and properties of R- and T-regions apply equally well), while the structure of infinity is the same as that of Minkowski, and additionally there are singularities.

Finally, one could note that the form of the conformal block for the exterior solution was determined from the start. First, we have asymptotic flatness, therefore the Minkowski's structure of infinity. This is already half of the block's boundary. 

\begin{figure}[!ht]
\center
\includegraphics*[width=0.8\textwidth]{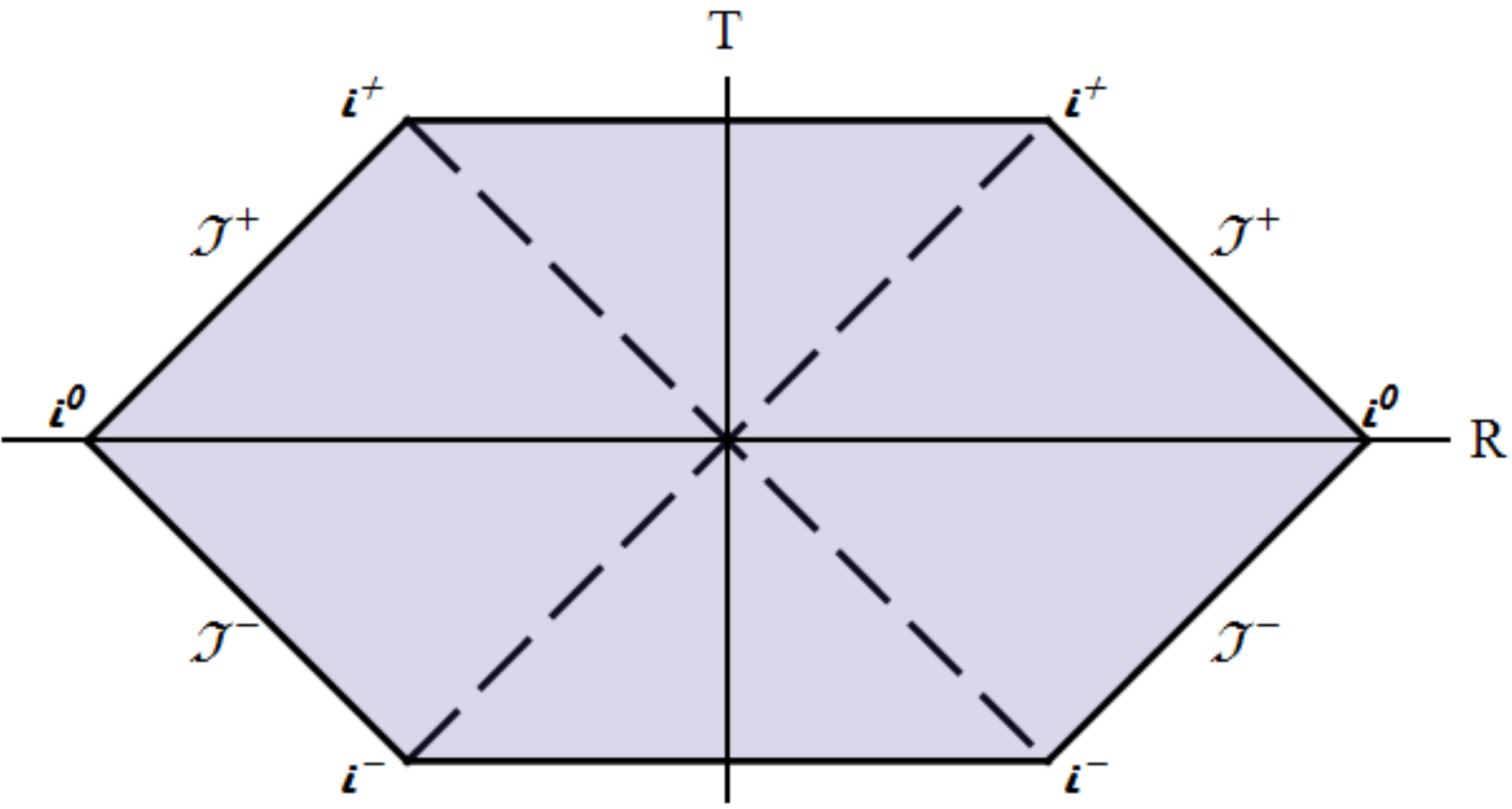}
\parbox{0.5\textwidth}{\caption{The full conformal diagram for the (maximally extended) Schwarzschild solution.}}
\end{figure}
\end{enumerate}

\subsection{Other spherically symmetric black holes}
The Schwarzschild metric (\ref{SchwBH}) is the vacuum spherical symmetric solution of Einstein's equation. If we add some matter, we will obtain a different solution, e.g. Reissner-Nordstr\"{o}m for an electrically charged black hole. In all cases, spherical symmetry means that in appropriately chosen coordinate frame the metric takes the same form (\ref{SchwBH}),
\begin{equation}
 ds^2 =f(\rho)dt^2 -\frac{d\rho^2}{f(\rho)}-r^2 (\rho)d\Omega^2 ,
\end{equation} 
but with some different function $f(\rho)$: one can always ensure that the metric functions have this form by choosing the appropriate radial coordinate\footnote{It is sometimes called the ``quasiglobal coordinate''} $\rho$. The angular part $\sim d\Omega^2$ does not affect causal structure and conformal diagrams. The zeros $\rho=\rho^\star$ of $f(\rho)$ define the surfaces, which split the full spacetime into R- and T-regions and are the horizon candidates.

\begin{enumerate}[resume]
\item \emph{Killing horizons.} Consider radial geodesic motion of a massive or massless particle. Make use of the integral of motion $E=-u^\mu \xi_\mu$ due to the Killing vector and find $\rho (t)$  and $\rho(\lambda)$, where $\lambda$ is the affine parameter (proper time $\tau$ for massive particles)
\begin{enumerate}
\item When is the proper time of reaching the horizon $\tau^\star =\tau (\rho^\star)$ finite?  
\item Verify that surface $\rho=\rho^\star$ is a Killing horizon
\end{enumerate}
\paragraph{Solution.} The formulas derived for Schwarzschild (\ref{BH-energy}) and (\ref{BH-lambda}) still work, just now $f(\rho)$ is not fixed. 
\begin{enumerate}
\item As for almost all particles $d\tau /d\rho$ stays bounded, the proper time of reaching the horizon candidate at $\rho=\rho^\star$ is finite if and only if $\rho^\star$ itself is finite. Otherwise we would have the ``horizon'' at infinite proper distance. It can be called a remote horizon, but extension across such a surface is not possible and not needed, as spacetime in that direction is already geodesically complete.
\item The Killing vector that forms the horizon is $\xi_t =\partial_t$, its norm $|\xi_t |^2 =g_{tt}=f$. Thus $\rho=\rho^\star$ is indeed a Killing horizon by definition.
\end{enumerate}

\item \emph{Extension across horizons.} Let us shift the $\rho$ coordinate so that $\rho^\star =0$, and assume that
\begin{equation}
f(\rho)=\rho^{q}F(\rho),\qquad q\in\mathbb{N},
\end{equation}
where $F$ is some analytic function, with $F(0)\neq 0$. Suppose we want to introduce a new radial ``tortoise'' coordinate $x$, such that the two-dimensional part of the metric in terms of $(t,x)$ has the conformally flat form:
\begin{equation}
ds^2_2 =f(\rho(x))\big[dt^2 -dx^2 \big].
\end{equation}
\begin{enumerate}
\item What is the asymptotic form of relation $\rho (x)$? 
\item Rewrite the metric in terms of null coordinates
\begin{equation}
V=t+x,\qquad W=t-x.
\end{equation}
Where is the horizon in terms of these coordinates? Is there only one?
\item Suppose we pass to new null coordinates $V=V(v)$ and $W=W(w)$. What conditions must be imposed on functions $V(v)$ and $W(w)$ in order for the mixed map $(v,W)$ to cover the past horizon and the map $(w,V)$ to cover the future horizon without singularities?
\end{enumerate}
\paragraph{Solution.} 
\begin{enumerate}
\item Integrating $dx=f^{-1}d\rho$, one obtains
\begin{equation}
\rho\sim \left\{\begin{array}{l}
e^{xF(0)},\quad \text{for}\quad q=1,\\
|x|^{-1/(q-1)}\quad\text{for}\quad q>1 .
\end{array}\right.
\end{equation}
We see that $x\to -\infty$ at the horizon. It converges for $q<1$, but in this case one can show that the Kretschmann scalar $R_{\mu\nu\rho\sigma}R^{\mu\nu\rho\sigma}$ diverges, so this is not a horizon but a singularity.
\item $ds^2 =f \; dV\;dW$; the horizon is at $x\to -\infty$ while $t$ is fixed, so in terms of $V$ and $W$ it is split into two parts: the future horizon at $V\to -\infty$ while $W$ is fixed, and the past horizon at $W\to +\infty$ while $V$ is fixed.
\item Near the past horizon in terms of $(v,W)$ the metric takes form
\begin{equation}
ds^2 =\Big(f \frac{dW}{dw}\Big)\; dV\; dw,
\end{equation}
while $W\approx -x$, so for it to be regular there we need
\begin{equation}
\frac{dW}{dw}\sim f^{-1}\sim \rho^{q} \sim \left\{
\begin{array}{l}
|x|^{(1-q)/q}\sim W ^{(1-q)/q}\quad\text{for}\quad q>1 ,\\
e^{xF(0)}\sim e^{-WF(0)}\quad\text{for}\quad q= 1,
\end{array}\right.
\end{equation}
and after integrating
\begin{equation}
W\sim \left\{\begin{array}{l}
w^{1-q}\quad\text{for}\quad q>1 ,\\
\ln w\quad\text{for}\quad q= 1.
\end{array}\right.
\end{equation}
The same way the asymptotic relation $V(v)$ is found near the future horizon. 
\end{enumerate}
\end{enumerate}
Thus a horizon candidate $\rho=\rho^\star$ is defined by condition $f(\rho^\star)=0$. If $\rho^\star =\pm\infty$, then the proper time of reaching it diverges, so it is not a horizon, but an infinitely remote boundary of spacetime.  It can be a remote horizon if $x^\star =\pm \infty$. In case $x^\star =O(1)$ the candidate is not a horizon, but a singularity, so again there is no continuation. An asymptotically flat spacetime by definition has the same structure of infinity as that of Minkowski spacetime; in general this is not always the case---recall the de Sitter and other cosmological spacetimes.

\begin{enumerate}[resume]
\item \emph{Infinities, horizons, singularities.} Draw the parts of conformal diagrams near the boundary that correspond to the limiting process
\begin{equation}
\rho\to\rho_0 ,\qquad |\rho_0|<\infty,
\end{equation}
under the following conditions:
\begin{enumerate}
\item spacelike singularity: $f(\rho_0)>0$, $|x_0 |<\infty$;
\item timelike singularity: $f(\rho_0)<0$, $|x_0 |<\infty$;
\item asymptotically flat infinity: $\rho_0 =\pm \infty$, $f(\rho)\to f_0>0$;
\item a horizon in the R region: $f(\rho)\to +0$, $|x_0 |\to \infty$;
\item a horizon in the T region: $f(\rho)\to -0$, $|x_0 |\to \infty$;
\item remote horizon in a T-region: $\rho_0 =\pm \infty$, $f(\rho)\to -<0$;
\end{enumerate}
Remember that any spacelike line can be made ``horizontal'' and any timelike one can be made ``vertical'' by appropriate choice of coordinates. 
\paragraph{Solution.}
\begin{enumerate}
\item Horizontal line, approached either from above or from below; it acts as the boundary of spacetime -- there is no continuation across (see Fig. \ref{Conf-BH-SingSpace});
\begin{figure}[!ht]
\center
\includegraphics*[width=0.4\textwidth]{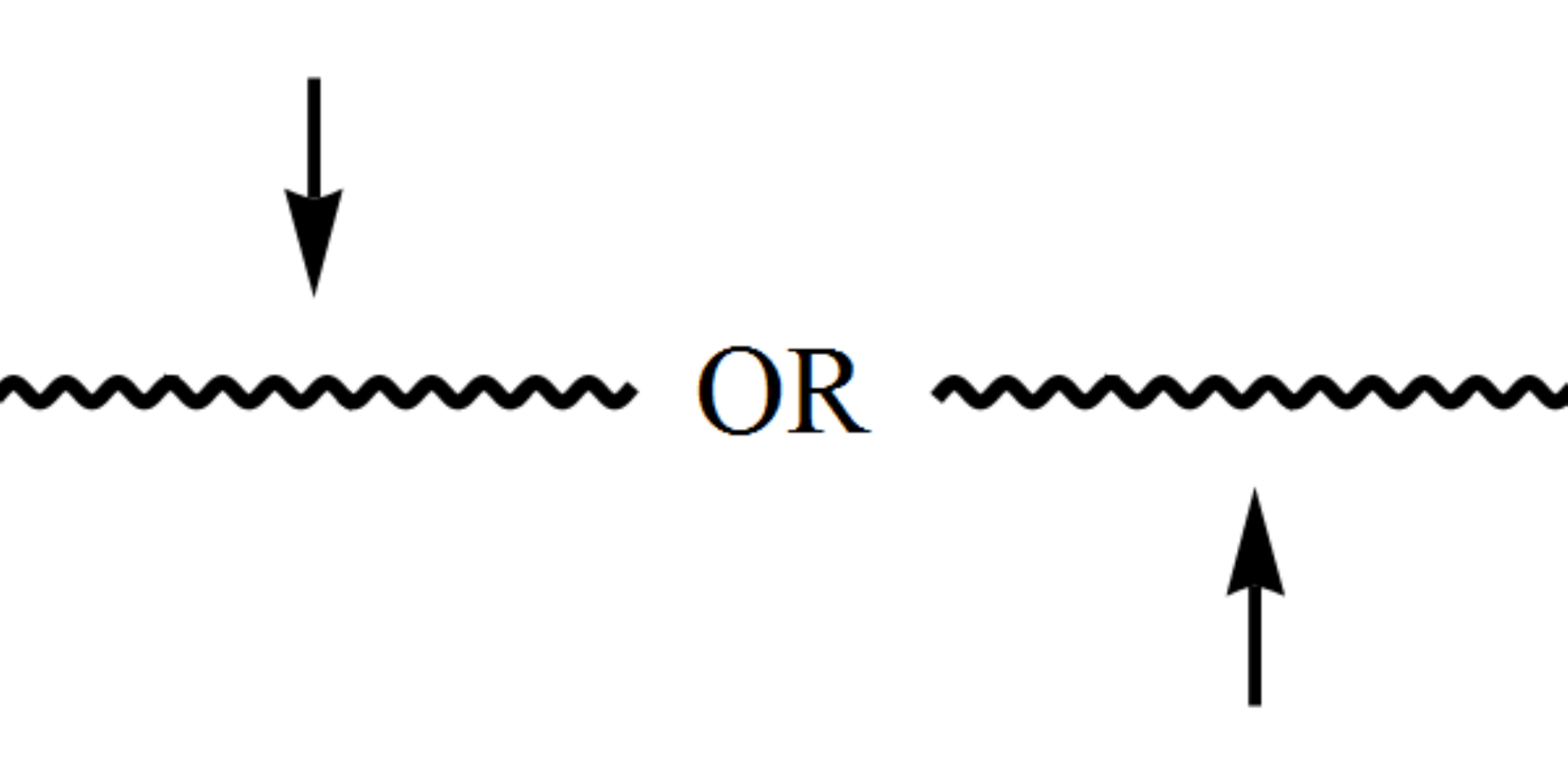}
\caption{\label{Conf-BH-SingSpace} Space-like singularity}
\end{figure}
\item Vertical line, approached either from the left or from the right; as it is singular, there is again no continuation (see Fig. \ref{Conf-BH-SingTime});
\begin{figure}[!ht]
\center
\includegraphics*[width=0.4\textwidth]{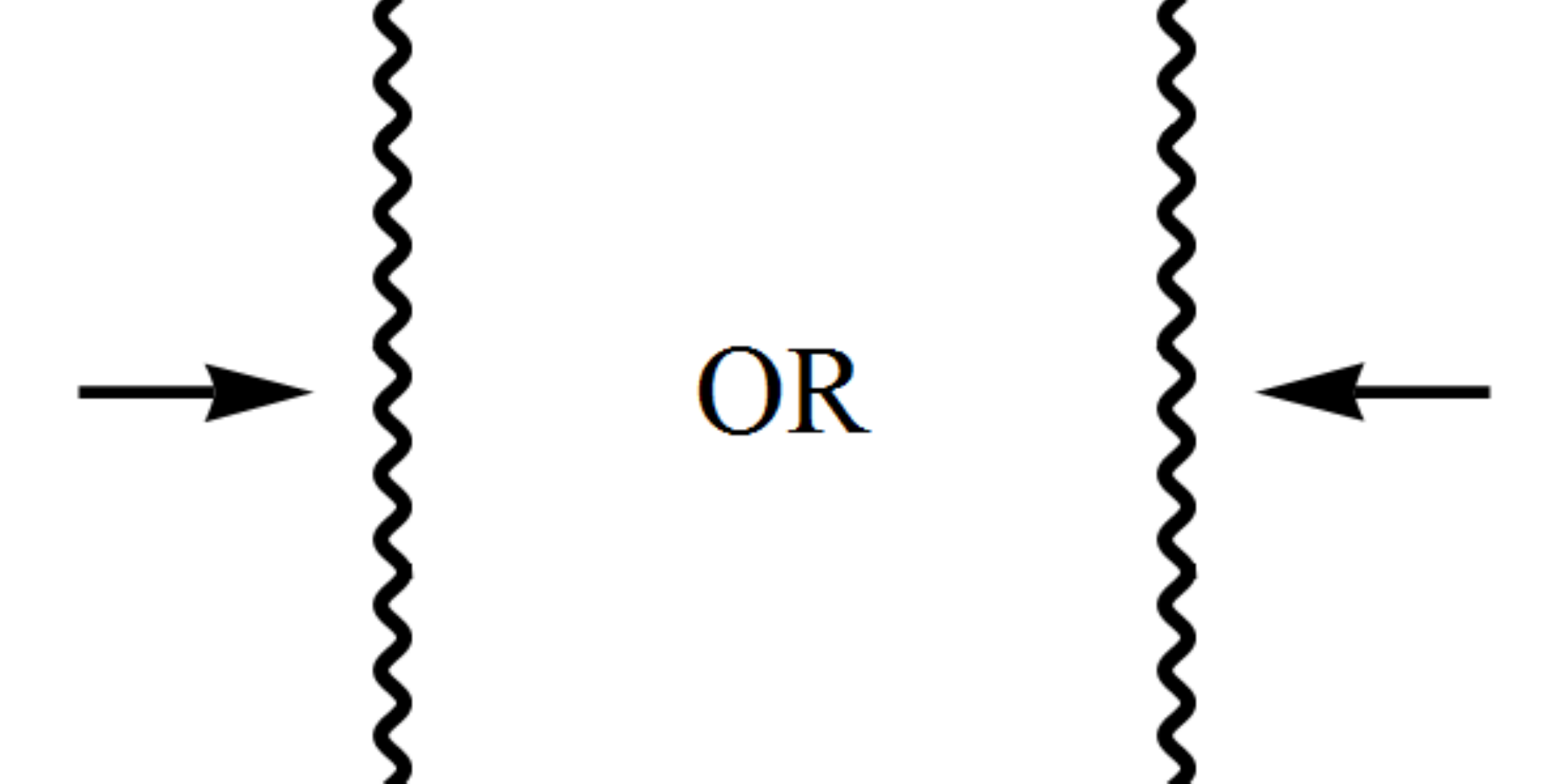}
\caption{\label{Conf-BH-SingTime} Time-like singularity}
\end{figure}
\item This is the infinitely remote part of Minkowski spacetime, from future infinity, to future null infinity, to spacelike infinity, to past null infinity, and finally to past infinity; it can constitute either the left or the right boundary of a conformal block and has the shape of ``$>$'' or ``$<$'' (see Fig. \ref{Conf-BH-InfMink});
\begin{figure}[!ht]
\center
\includegraphics*[width=0.4\textwidth]{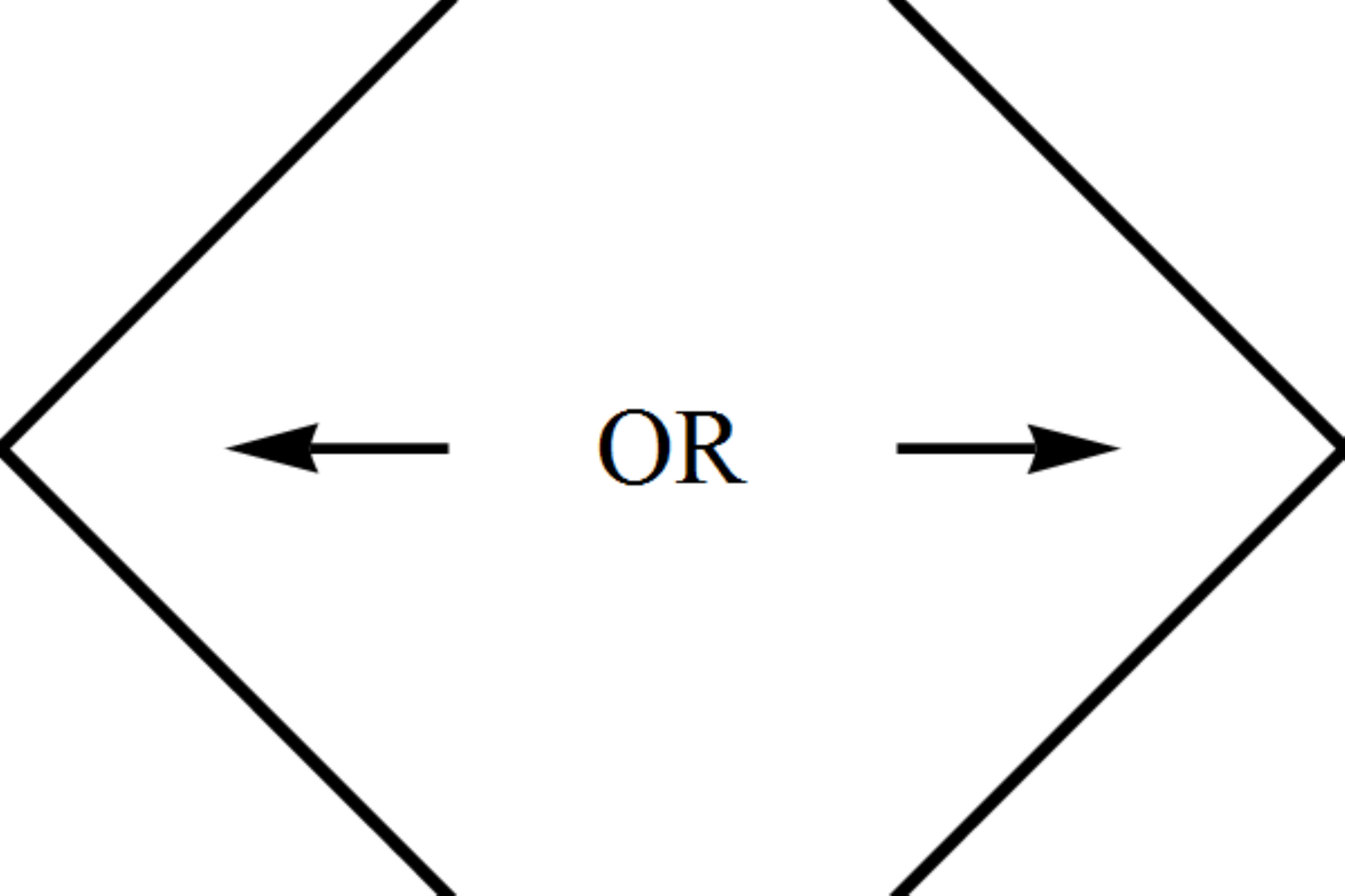}
\caption{\label{Conf-BH-InfMink} Minkowski infinity}
\end{figure}
\item The same as the Schwarzschild horizon as approached from the exterior region: the two sides of a triangle pointed left ``$<$'' or right ``$<$'' and approached from the inside; thus the shape is the same as above, but now the horizon allows continuation across (see Fig. \ref{Conf-BH-HorExt});
\begin{figure}[!ht]
\center
\includegraphics*[width=0.4\textwidth]{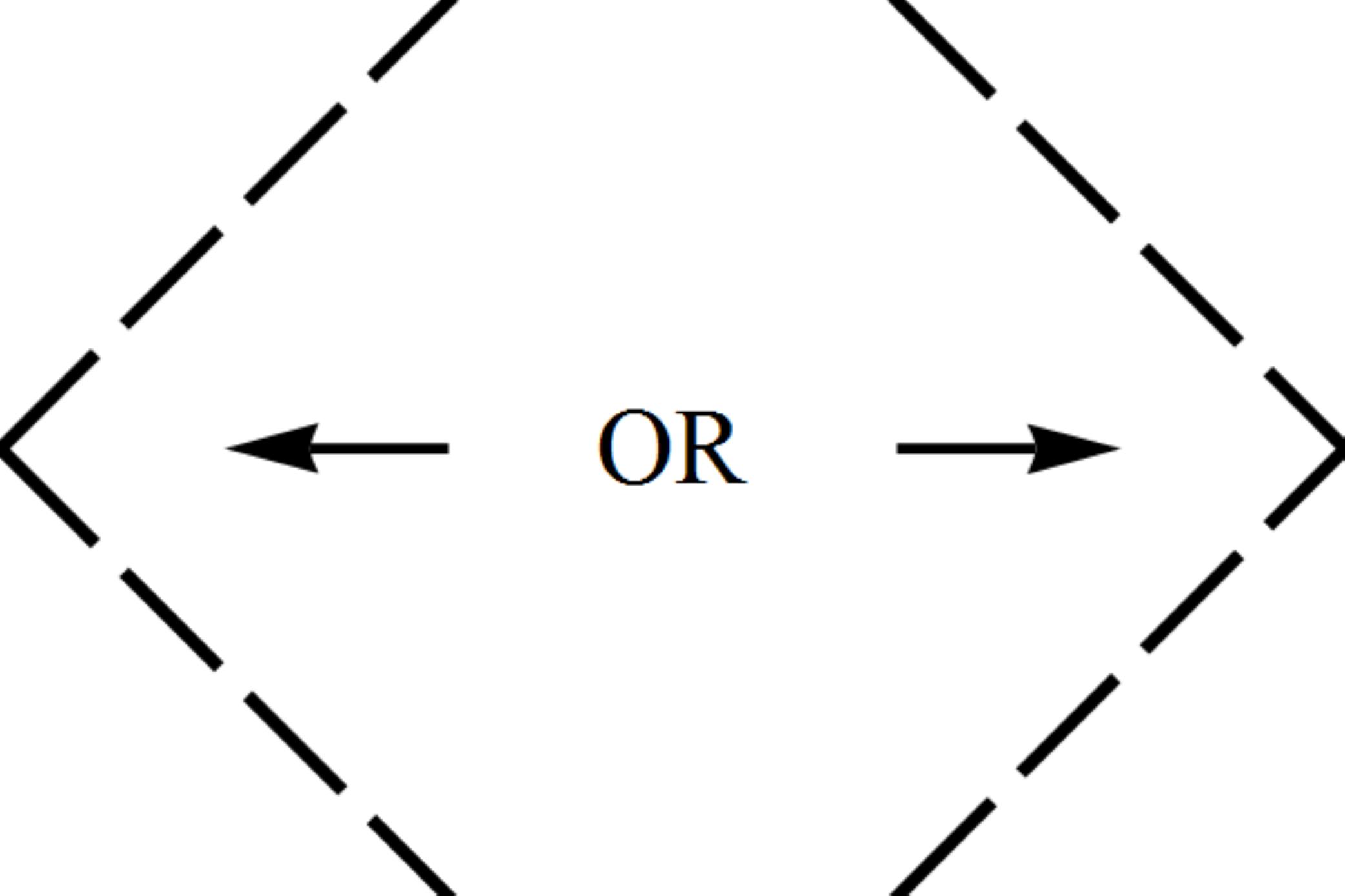}
\caption{\label{Conf-BH-HorExt} Horizon as boundary of an R-region}
\end{figure}
\item The same as Schwarzschild horizon as approached from the interior region: two sides of a triangle pointed up ``$\wedge$'' or down ``$\vee$'' and approached from the inside (see Fig. \ref{Conf-BH-HorInt}).
\begin{figure}[!ht]
\center
\includegraphics*[width=0.8\textwidth]{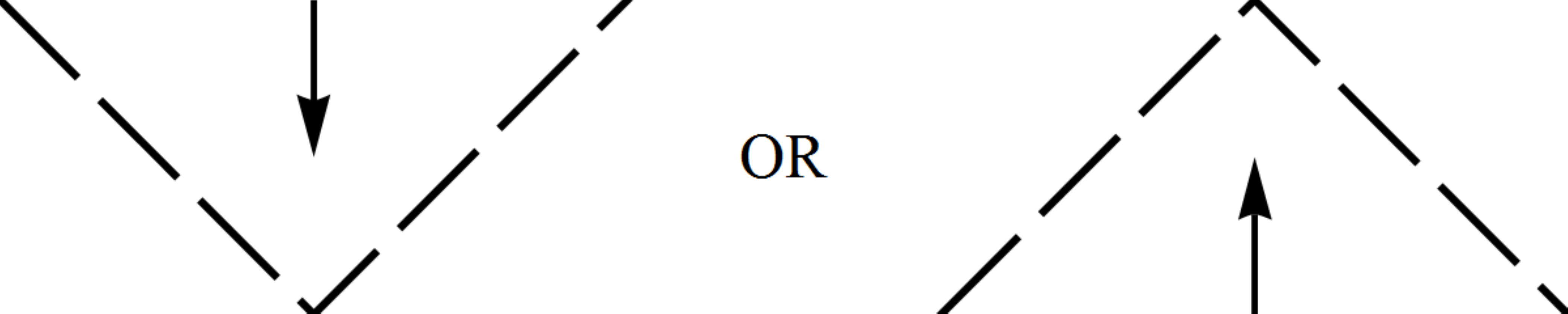}
\caption{\label{Conf-BH-HorInt} Horizon as boundary of a T-region}
\end{figure}
\item Two diagonal lines -- a corner of a triangle, -- pointed up or down, and approached from the inside. Thus the shape is the same as in the previous case, ``$\wedge$'' or ``$\vee$''; the difference is that now there is no continuation across (see Fig. \ref{Conf-BH-InfTime}).
\begin{figure}[!ht]
\center
\includegraphics*[width=0.8\textwidth]{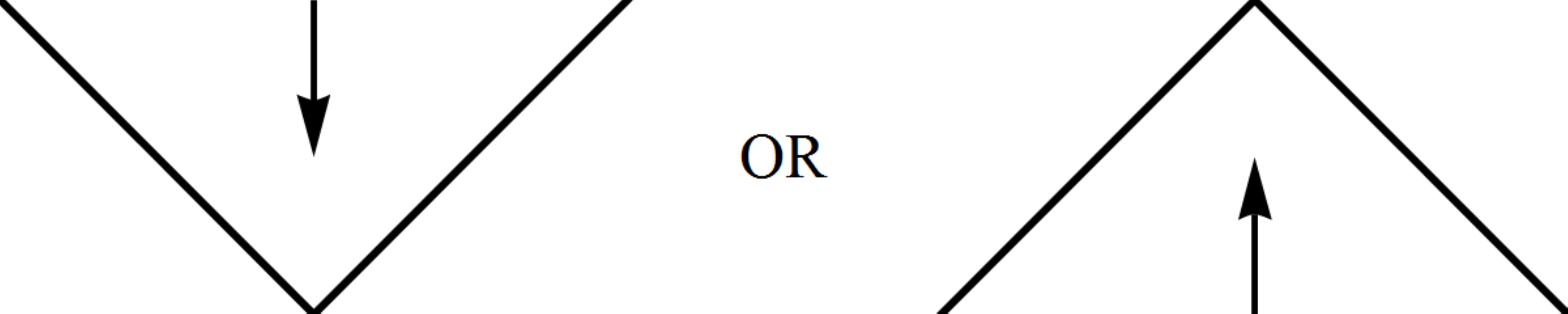}
\caption{\label{Conf-BH-InfTime} Remote horizon as a boundary of a T-region}
\end{figure}
\end{enumerate}
\end{enumerate}
For arbitrary $f(\rho)$ we can 
\begin{itemize}
\item split the full spacetime into regions between the zeros $\rho_i$ of $f(\rho)$;
\item draw for each region the conformal diagram, which is a square $\Diamond$ or half of it $\triangle$, $\nabla$, $\triangleleft$, $\triangleright$;
\item glue the pieces together along the horizons $\rho=\rho_i$, while leaving singularities and infinities as the boundary. 
\end{itemize}
The resulting diagram can turn out to be either finite, as for Schwarzshild, when singularities and infinities form a closed curve enclosing the whole spacetime, or not.

\begin{enumerate}[resume]
\item \emph{Examples.} Draw the conformal diagrams for the following spacetimes:
\begin{enumerate}
\item Reissner-Nordstr\"{o}m charged black hole:
\begin{equation}
f(r)=1-\frac{r_g}{r}+\frac{q^2}{r^2}, \qquad 0<q<r_{g},\qquad r>0;
\end{equation}
\item Extremal Reissner-Nordstr\"{o}m charged black hole 
\begin{equation}
f(r)=\Big(1-\frac{q}{r}\Big)^2, \qquad q>0,\qquad r>0;
\end{equation}
\item Reissner-Nordstr\"{o}m-de Sitter charged black hole with cosmological constant (it is not asymptotically flat, as $f(\infty)\neq 1$)
\begin{equation}
f(r)=1-\frac{r_g}{r}+\frac{q^2}{r^2}-\frac{\Lambda r^2}{3}\qquad q,\Lambda>0;
\end{equation}
analyze the special cases of degenerate roots.
\end{enumerate}
\paragraph{Solution.}
\begin{enumerate}
\item In the generic case there are two positive roots $+\infty>r_+ >r_- >0$, which give horizons, while $r=0$ is a singularity. This gives us three regions between the roots: (a) $r\in(r_+ ,+\infty)$ between the horizon and asymptotically flat infinity, thus having the same structure as the Schwarzschild exterior region and the shape of a ``diamond'' $\Diamond$; (b) $r\in (r_- ,r_+)$ is a T-region between the horizons (the shape is the same  $\Diamond$) and (c) $r\in (0,r_-)$ is again an R-region between the timelike singularity and the horizon. The timelike singularity turns the latter conformal diagram into a triangle, $\triangleright$ or $\triangleleft$. Gluing all the blocks together gives us the left diagram of Fig. \ref{Conf-BH-Examples}.
\item Due to degeneracy the T-region is absent, instead the two kinds of R-regions are separated by double horizons; the III-blocks on the right vanish, while on the left the singularity merges into a solid vertical line.
\begin{figure}[!ht]
\center
\hfill
\includegraphics*[height=0.7\textwidth, viewport=0 0 600 960]{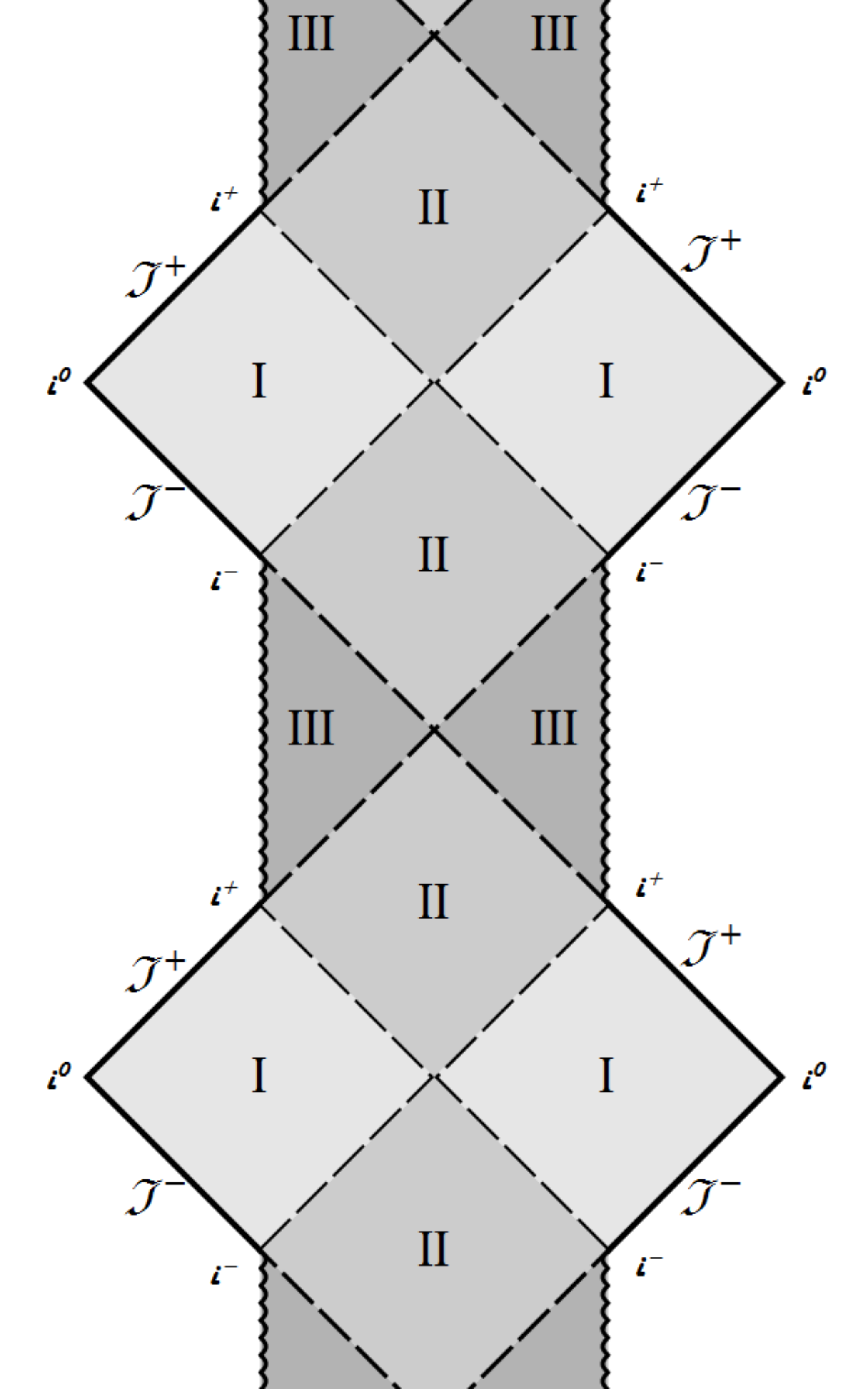}\hfill
\includegraphics*[height=0.7\textwidth, viewport=170 0 600 840]{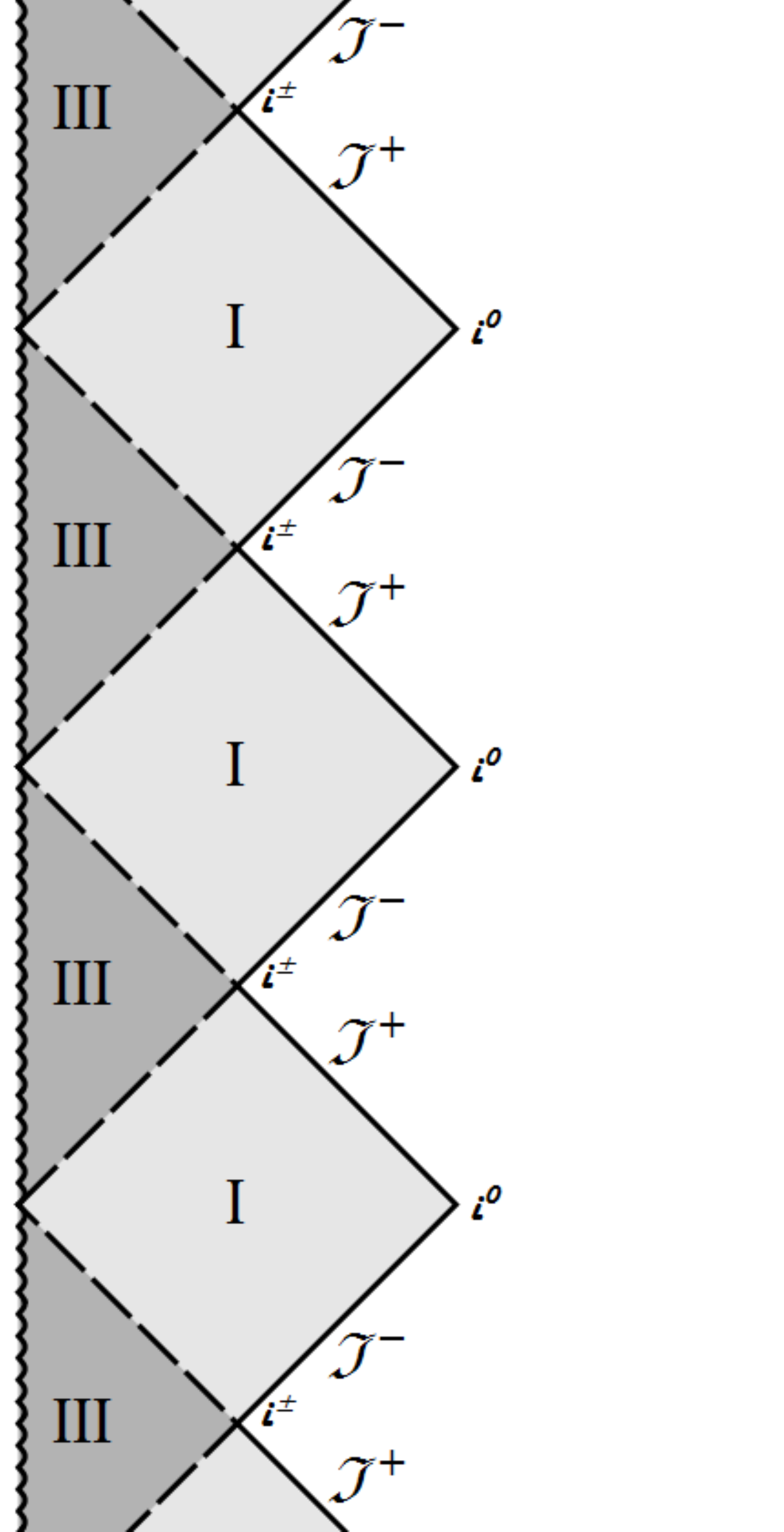}\hfill
\parbox{0.9\textwidth}{\caption{\label{Conf-BH-Examples} Conformal diagrams for the generic (with no degenerate horizons) Reissner-Nordstr\"{o}m black hole solution on the left and the extremal (with double horizon) Reissner-Nordstr\"{o}m on the right. Blocks I and III are R-regions, blocks II are T-regions. Horizons are denoted by dashed lines, infinities by solid thick lines, singularities by wriggling curves. Both diagrams are infinitely continued up and down.}}
\end{figure}
\item There can be up to three positive roots of $f$ in case $\Lambda>0$: $r_{-}<r_+<r_\Lambda$. Then there are four different conformal blocks: 
\begin{enumerate}
\item block IV for the R-region between the timelike singularity and the inner horizon $r\in (0,r_-)$: $\triangleleft,\triangleright$;
\item block III for the T-region between the two horizons $r\in(r_- ,r_+)$; it is the same as for Reissner-Nordstr\"{o}m: $\Diamond$;
\item block II for the R-region between the horizons $r\in (r_+ ,r_\Lambda)$; it differs from the exterior region of Reissner -Nordstr\"{o}m or Schwarzschild by replacement of asymptotic infinity with another pair of horizons. The shape is the same, $\Diamond$, but the boundary now allows continuation across it in all directions;
\item block I for the T-region between the external (cosmological) horizon and the de-Sitter-like infinity. The structure of infinity is determined by $f(r)$ for large $r$, where the $\sim 1/r$ and $1/r^2$ terms can be neglected, so effectively we have the de Sitter spacetime. Thus the infinity is spacelike and represented by one horizontal line. The block is the triangle: $\triangle,\nabla$ (the same as the upper and lower sectors of the exact full de Sitter spacetime).
\end{enumerate}
The full diagram is shown on Fig. \ref{Conf-BH-ExampleGen}.
\begin{figure}[!hbt]
\center
\includegraphics*[width=0.6\textwidth]{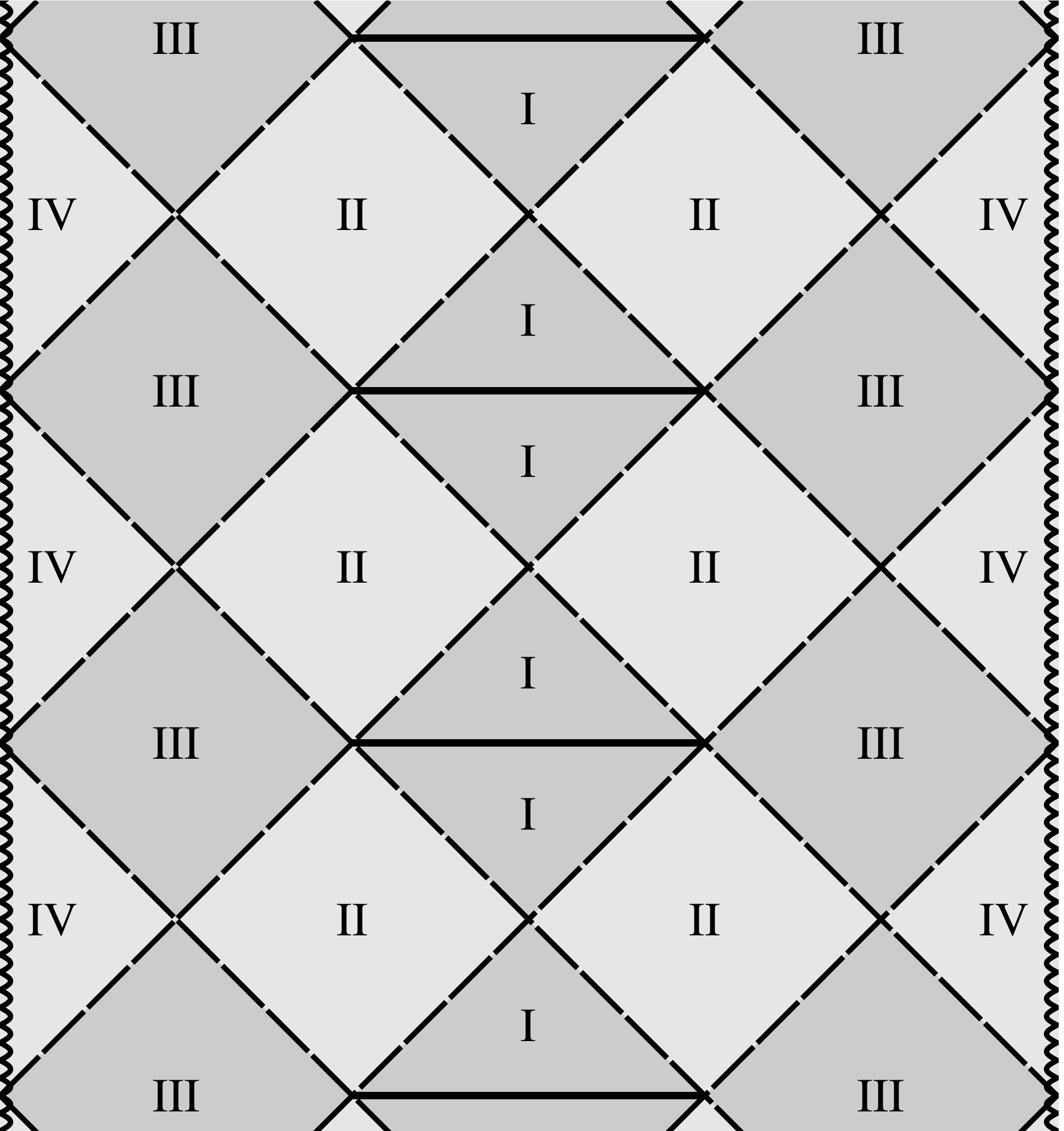}
\caption{\label{Conf-BH-ExampleGen} Conformal diagram for the Reissner-Nordstr\"{o}m-de Sitter black hole solution. T-regions are shaded with darker grey. The diagram is infinitely continued up and down.}
\end{figure}
\end{enumerate}
\end{enumerate}

\FloatBarrier

\section{Hubble sphere}
The Hubble radius is the proper distance $R_H (t)=c H^{-1}(t)$. The sphere of this radius is called the Hubble sphere. From definition, the Hubble recession ``velocity'' of a comoving observer on the Hubble sphere is $v(R_H)=H R_H =c$ and equal to the speed of light $c$. This is true, of course, at the same moment of time $t$, for which $H (t)$ is taken.

\begin{enumerate}[resume]
\item All galaxies inside the Hubble sphere recede subluminally (slower than light) and all galaxies outside recede superluminally (faster than light). This is why the Hubble sphere is sometimes called the ``photon horizon''.  Does this mean that galaxies and their events outside the photon horizon are permanently hidden from the observer's view? If that were so, the photon horizon would also be an event horizon. Is this correct? 

\paragraph{Solution.} At first glance, observation of galaxies beyond the Hubble sphere appears to be an unsolvable problem. As Eddington in 1933 wrote: ``Light is like a runner on an expanding track with the winning-post receding faster than he can run.'' But never give up!
In most models of the Universe  the Hubble parameter $H$ is not constant. In a decelerating universe the Hubble radius $R_H$ increases with time. Moreover, the Hubble sphere expands faster than the Universe, so that the edge of the Hubble sphere -- the photon horizon -- overtakes the receding galaxies.  Light rays outside the Hubble sphere moving toward us may therefore eventually be overtaken by the photon horizon.  They will then be inside the Hubble sphere and will at last start approaching us. Eddington's runner sees the winning-post receding, but he must keep running and not give up; the expanding track is slowing down and eventually the winning-post will be reached.

\item Show, by the example of static universe, that the Hubble sphere does not coincide with the boundary of the observable Universe.

\paragraph{Solution.} The observable Universe is limited by the particle horizon, and if the Hubble sphere and the observable Universe were the same, the latter in a static Universe would be infinitely large ($H_0 =0$, so $R_H =\infty$).  But static Universes of finite age have particle horizons at finite distance. So, the Hubble sphere cannot be the boundary of the observable Universe.

\item Estimate the ratio of the volume enclosed by the Hubble sphere to the full volume of a closed Universe.

\item Show that in a spatially flat Universe ($k=0$), in which radiation is dominating, the particle horizon coincides with the Hubble radius.
\paragraph{No solution}

\item Find the dependence of comoving Hubble radius $R_H /a$ on scale factor in a flat Universe filled with one component with the state equation $\rho=w p$.

\item Express the comoving particle horizon through the comoving Hubble radius for the case of domination of a matter component with state parameter $w$.

\item Show that
\begin{equation}
\frac{dR_H }{dt}=c(1+q),\label{dRHdt}
\end{equation}
where 
\begin{equation}
q=-\frac{\ddot{a}/a}{H^{2}}
\end{equation}
 is the deceleration parameter.
\paragraph{Solution.} As $R_H =H^{-1}=a/\dot{a}$, the answer is obtained after one differentiation.

\item Show that
\begin{equation}
\frac{dL_p}{dt}=1+\frac{L_p}{R_H}.
\end{equation}
\paragraph{Solution.} Using the definition $L_p =a \int^t dt a^{-1}$, we obtain this after differentiation. 

\item Show that in the Einstein-de Sitter Universe the relative velocity of the Hubble sphere and galaxies on it is equal to $c/2$.

\item Find $a(t)$ in a universe with constant positive deceleration parameter $q$.
\paragraph{Solution.} Using
\begin{equation}
\dot{H}=\frac{\ddot{a}}{a}-\frac{\dot{a}^2}{a^2},
\end{equation}
and the condition $q=const$, we get
\begin{equation}
\frac{d}{dt}H^{-1}=1+q ,
\end{equation}
thus $H^{-1}=(1+q)t$ and $a\sim t^{n}$ with
\begin{equation}
n=\frac{1}{1+q}.
\end{equation}
Also $q>0$ means $n<1$.

\item Show that in universes of constant positive deceleration $q$, the the ratio of distances to the particle and photon horizons is $1/q$.

\item Show that the Hubble sphere becomes degenerate with the particle horizon at $q=1$ and with the event horizon at $q=-1$. 

\item Show that if $q$ is not constant, comoving bodies can be inside and outside of the Hubble sphere at different times. But not so for the observable universe; once inside, always inside. 
\paragraph{Solution.} From (\ref{dRHdt}) we see that recession velocity of the Hubble sphere can be lesser or greater than $c$, depending on the sign of $q$. As recession velocity of matter on it is exactly $c$, be definition, this means that matter can cross the Hubble sphere in both directions at different times. 

``Horizons are like membranes; the photon horizon acts as a two-way membrane (comoving bodies can cross in both directions depending on the value of $q$), and the particle horizon acts like a one-way membrane (comoving bodies always move in and never out).'' E.~Harrison, Science of the Universe.
\end{enumerate}

\section{Proper horizons}
The following five problems are based on work by F. Melia \cite{Melia}.

Standard cosmology is based on the FLRW metric for a spatially homogeneous and isotropic three-dimensional space, expanding or contracting with time. In the coordinates used for this metric, $t$ is the cosmic time, measured by a comoving observer (and is the same everywhere), $a(t)$ is the expansion factor, and $r$ is an appropriately scaled radial coordinate in the comoving frame.

F. Melia  demonstrated the usefulness of expressing the FRLW metric in terms of an observer-dependent coordinate $R=a(t)r$, which explicitly reveals the dependence of the observed intervals of distance, $dR$, and time on the curvature induced by the mass-energy content between the observer and $R$; in the metric, this effect is represented by the proximity of the physical radius $R$ to the cosmic horizon $R_{h}$, defined by the relation
\[R_{h}=2G\,M(R_h).\]
In this expression, $M(R_h)$ is the mass enclosed within $R_h$ (which terns out to be the Hubble sphere). This is the radius at which a sphere encloses sufficient mass-energy to create divergent time dilation for an observer at the surface relative to the origin of the coordinates.
\begin{enumerate}[resume]
\item\label{dyn-Melia1} Show that in a flat Universe $R_{h}=H^{-1}(t)$.
\paragraph{Solution.} By definition
\[R_{h}=\Big(\frac{3}{8\pi G \rho}\Big)^{1/2}
	=\frac{1}{H(t)}.\] 

\item\label{dyn-Melia2} Represent the FLRW metric in terms of the observer-dependent coordinate $R=a(t)r$.
\paragraph{Solution.} It is convenient to recast FLRW metric using a new function $f(t)$
\[a(t)=e^{f(t)}.\]
In that case
\[d{s^2} = {c^2}d{t^2} - {e^{2f(t)}}\left( {d{r^2}
	+ {r^2}d{\Omega ^2}} \right).\]
Making the coordinate transformation to the radial coordinate $r=R\,e^{-t}$, we obtain
\[ds^2 = \Phi \Big[dt
	+\big(R\dot f\big){\Phi^{ - 1}}dR\Big]^2
 	-\Phi^{- 1} dR^2 - R^2 d\Omega^2,\]
where for convenience we have defined the function
\[\Phi=1-\big(R\dot f\big)^{2}.\]

It is easy to see, that the radius of the cosmic horizon for the observer at the origin is
\[R_{h}=1/\dot{f}=\frac{a}{\dot{a}}=\frac{1}{H}\]
and
\[\Phi=1-\frac{R}{R_h}.\]
Finally, we obtain
\begin{align*}
 ds^{2}&=\Phi\Big[dt+\frac{R}{R_h}\Phi^{-1}dR\Big]^{2}
	-\Phi^{-1}dR^2 -R^{2}d\Omega^2 =\\
	&=\Big(1-\frac{R}{R_h}\Big)
	\Big[dt+\frac{R/R_h}{1-R/R_h}dR\Big]^{2}
	-\frac{dR^2}{1-R/R_h}-R^2 d\Omega^2.
\end{align*} 

\item\label{dyn-Melia3} Show, that if we were to make a measurement at a fixed distance $R$ away from us, the time interval $dt$ corresponding to any measurable (non-zero) value of $ds$ must go to infinity as $r\to R_h$.
\paragraph{Solution.} Let us examine the behavior of the interval $ds$ connecting any arbitrary pair of spacetime events at $R$. For an interval produced at $R$ by the advancement of time only $dR=d\Omega=0$, metric obtained in previous problem gives
\[ds^{2}=\Phi dt^{2}.\]
Function  $\Phi\to 0$ as $R\to R_h$, thus for any measurable (non-zero) value of $ds$ the interval $dt$ must go to infinity as $R\to R_h$. In the context of black-hole physics (see Chapter 4), we recognize this effect as the divergent gravitational redshift measured by a static observer outside of the event horizon. 

\item\label{dyn-Melia4} Show that $R_h$ is an increasing function of cosmic time $t$ for any cosmology with $w>-1$.
\paragraph{Solution.} It is straightforward to demonstrate from Friedman equations that
\[\dot{R}_{h}=\frac{3}{2}(1+w).\]
Consequently, $\dot{R}_{h}>0$ for $w>-1$.

``$R_h$ is fixed only for de Sitter, in which $\rho$ is a cosmological constant and $w=-1$. In addition, there is clearly a demarcation at $w=-1/3$. When $w<-1/3$, $R_h$ increases more slowly than lightspeed ($c=1$ here), and therefore our universe would be delimited by this horizon because light would have traveled a distance $t_0$ greater than $R_{h}(t_0)$ since the big bang. On the other hand, $R_h$ is always greater than $t$ when $w>-1/3$, and our observational limit would then simply be set by the light travel distance $t_0$''. 

\item\label{dyn-Melia5} Using FLRW metric in terms of the observer-dependent coordinate $R=a(t)r$, find $\Phi(R,t)$ for the specific cosmologies:
\begin{enumerate}
\item[a)] the De Sitter Universe ;
\item[b)] a cosmology with  $R_h =t$, ($w=-1/3$);
\item[c)] radiation dominated Universe  ($w=1/3$);
\item[d)] matter dominated Universe ($w=0$).
\end{enumerate}
\paragraph{Solution.} 
Let us consider all cases separately
\begin{enumerate}
\item[a)] De Sitter
\[H=H_{0}=const,\quad a(t)=e^{H_{0}t},\quad
	f=\ln a(t)=H_{0}t.\]
In this case $\dot{R}_{h}=0$ and therefore $R_h$ is fixed
\[R_{h}=\frac{1}{H_0}.\]

\item[b)] an equation of state $w=-1/3$ is the only one for which the current age, $t_0$, of the Universe can equal the light-crossing time, $t_{h}=R_h$. In this case
\begin{align}
d{s^2}& = \Phi {\Big[ {cdt + \big(\frac{R}{t}\big)
	{\Phi ^{ - 1}}dR} \Big]^2} - {\Phi ^{ - 1}}d{R^2}
	- {R^2}d{\Omega ^2},\\
\Phi& = 1 - {\left( {\frac{R}{{ct}}} \right)^2}
\end{align}
The cosmic time $dt$ diverges for a measurable line element as $R\to R_h =t$.

\item[c)] In the case of radiation domination
\begin{align*}
a(t)& = {\left( {2{H_0}t} \right)^{1/2}},
\quad f(t) = \frac{1}{2}\ln \left( {2{H_0}t} \right),
\quad \dot f = \frac{1}{{2t}}\\
d{s^2} &= \Phi {\Big[ {dt + \left( {\frac{R}{{2t}}} \right){\Phi ^{ - 1}}dR} \Big]^2} - {\Phi ^{ - 1}}d{R^2} - {R^2}d{\Omega ^2},\\
\Phi & = 1 - \left( {\frac{R}{{2t}}} \right)^2.
\end{align*}
Thus, measurements made at a fixed $R$ and $t$ still produce a gravitationally-induced dilation of $dt$ as $R$ increases, but this effect never becomes divergent within that portion of the Universe (i.e., within $t_0$) that remains observable since the Big Bang.

\item[d)] Matter domination:
\begin{align*}
a(t) &= {\left( {3/2{H_0}t} \right)^{2/3}},\quad f(t) = \frac{2}{3}\ln \left( {3/2{H_0}t} \right),\quad \dot f = \frac{2}{{3t}}\\
d{s^2} &= \Phi {\Big[ {dt + \left( {\frac{R}{{3t/2}}} \right){\Phi ^{ - 1}}dR} \Big]^2} - {\Phi ^{ - 1}}d{R^2} - {R^2}d{\Omega ^2},\\
\Phi  &= 1 - {\left( {\frac{R}{{3t/2}}} \right)^2}
\end{align*}
The situation is similar to that for a radiation dominated universe, in that $R_h$ always recedes from us faster than lightspeed. Although dilation is evident with increasing $R$, curvature alone does not produce a divergent redshift.
\end{enumerate} 
\end{enumerate}

\section{Inflation}
\begin{enumerate}[resume]
\item Is spatial curvature important in the early Universe? Compare the curvature radius with the particle horizon.
\paragraph{Solution.} While three-space curvature radius is of the order of $a$, the particle horizon in the radiation-dominated early Universe is $\sim ct$. At Planck's time then
\begin{equation}
a_{Pl}\sim a_0 \frac{T_0}{T_{Pl}}\sim 10^{-4}cm,
\end{equation}
while the particle horizon is of the order of Planck length $\sim 10^{-33}cm$. Thirty orders of magnitude make spatial curvature utterly negligible. 

\item Comoving Hubble radius 
\begin{equation}
	r_H =\frac{R_H}{a}=\frac{1}{aH}=\frac{1}{\dot{a}}
\end{equation}
plays crucial role in inflation. Express the comoving particle horizon $l_e$ in terms of $r_H$. 
\paragraph{Solution.} Starting from definition,
\begin{equation}
l_p (t)=\int \limits_{0}^{t} \frac{dt}{a(t)}
	=\int\limits_0^a \frac{da}{Ha^2}=\int \frac{d\ln a}{Ha}
		=\int d\ln a \; r_H (a).
\end{equation}

\item Show that for the conventional Big Bang expansion (with $w\geq 0$) the comoving particle horizon and Hubble radius grow monotonically with time.
\paragraph{Solution.} As $\rho\sim a^{-3(1+w)}$, from the first Friedman equation
\begin{equation}
H^2 \sim \rho a^{-3(1+w)},\quad aH \sim a^{-(1+3w)/2},
	\quad r_H \sim a^{(1+3w)/2},
\end{equation}
and using the result of the previous problem,
\begin{equation}
l_p =\int\limits_{0}^{a}\frac{da}{a}\;r_H \sim a^{(1+3w)/2}.
\end{equation}
Note that the comoving particle horizon and comoving Hubble radius have the same dependence on scale factor (which is due to the power law in $a(t)$). Qualitatively this behavior depends on whether $(1+3w)$ is positive or negative. For radiation-dominated (rd) and matter-dominated (md) universes we find $l_p^{(rd)}\sim a$ and $l_p^{(md)}\sim a^{1/2}$ respectively.
\end{enumerate}

\textbf{The flatness problem}
\begin{enumerate}[resume]
\item The ``flatness problem'' can be stated in the following way: spacetime in General Relativity is dynamical, curving in response to matter in the Universe. Why then is the Universe so closely approximated by Euclidean space? Formulate the ``flatness problem'' in terms of the comoving Hubble radius.
\paragraph{Solution.} First, we rewrite the first Friedman equation
\[H^2 =\frac{8\pi G}{3}\rho(a)-\frac{k}{a^2}\]
in the form
\begin{equation}
1-\Omega(a)=-\frac{k}{a^2 H^2}=-k\,r_H^2,
	\qquad \Omega(a)=\frac{\rho(a)}{\rho_{cr}(a)}.
\end{equation}
In standard Big Bang cosmology the comoving Hubble radius $r_H$ grows with time and  the quantity $|\Omega-1|$ must also increase. The critical value $\Omega=1$ is an unstable  fixed point. Therefore, in standard Big Bang cosmology without inflation, the near-flatness observed today, $\Omega\sim 1$, requires extreme fine-tuning of $\Omega$ close to $1$ in the early Universe.

\item Inflation is defined as any epoch, in which scale factor grows with acceleration\footnote{Technically, this includes also the current epoch of cosmological history -- late-time accelerated cosmological expansion.}, i.e. $\ddot{a}>0$. Show that this condition is equivalent to the comoving Hubble radius decreasing with time.
\paragraph{Solution.} The comoving Hubble radius is
\begin{equation}
\frac{H^{-1}}{a}.
\end{equation}
The condition that it decreases with time is
\begin{equation}
\frac{d}{dt}\frac{1}{aH}=\frac{d}{dt}\frac{1}{\dot{a}}=-\frac{\ddot{a}}{\dot{a}^2}<0 ,
\end{equation}
which means $\ddot{a}>0$.

\item Show how inflation solves the flatness problem.
\paragraph{Solution.} The first Friedmann equation for a non-flat Universe can be rewritten as
\begin{equation}
|1-\Omega(a)|=\frac{1}{a^2 H^2}=r_H^2
\end{equation}
If the comoving Hubble radius decreases this drives the Universe toward  flatness. This solves the flatness problem.  The solution $\Omega=1$ (flat Universe) is an attractor during inflation! 
\end{enumerate}

\textbf{The horizon problem}
\begin{enumerate}[resume]
\item What should be the scale of homogeneity in the early Universe, as function of time, for the observed cosmological background (CMB) to be almost isotropic? Compare this with the functional dependence of comoving particle horizon on time. How can this be compatible with the causal evolution of the Universe? 
\paragraph{Solution.} As shown in previous problems, for the conventional Big Bang expansion the comoving particle horizon grows monotonically with time. This implies that comoving scales entering the horizon today have been far outside the horizon at CMB decoupling (which was at much much earlier times).  But the near-isotropy of CMB tells us that the universe was extremely homogeneous at the time of last-scattering, on scales encompassing many regions that should be causally independent. In order to explain this, we either need to violate causality, or assume another instance of extreme fine-tuning at the level of initial conditions. This is the horizon problem.

\item If CMB was strictly isotropic, in what number of causally independent regions temperature had had to be kept constant at Planck time?
\paragraph{Solution.} Currently the Universe is homogeneous and isotropic at the scales of the order of $ct_0$. The initial (at time $t_i$) size of an inhomogeneity is then $l_i \sim ct_0 a_i /a_0$. Comparing this with the scale of causality $l_{caus}\sim ct_i$, we get
\begin{equation}
\frac{l_i}{l_{caus}}\sim \frac{t_0}{t_i}\frac{a_i}{a_0}.
\end{equation}
At Planck scale $t_i \sim t_{Pl}$, taking into account that $aT\approx const$, we get
\begin{equation}
\frac{l_i}{l_{caus}}\Big|_{Pl}\sim \frac{t_0 T_0}{t_{Pl} T_{Pl}}\sim 	10^{28}.
\end{equation}
Therefore, in order to reproduce the observed today anisotropy of the CMB at the level of $\Delta T/T \sim 10^{-5}$, at Planck time temperature should have been constant to the same precision in $\sim 10^{84}$ causally disconnected regions.

\item Consider the case of dominating radiation. Show, that at any moment in the past, within the matter comprising today the observable Universe, one can find regions that are out of causal contact.
\paragraph{Solution.} The maximal size of a causally connected region is given by the particle horizon, which for dominating radiation behaves as $L_{p}\sim ct$. At the same time, the physical distance between comoving galaxies increases proportional to $a(t)\sim t^{1/2}$. Going into the past, we will find that the particle horizon decreases much faster than $a(t)$. Therefore among the galaxies that currently observable, and constitute the observable Universe today, at any moment of time in the past we can find some that are out of causal contact. This contradicts the established isotropy of the cosmological background radiation (on the level of $10^{-5}$), and constitutes the essence of the horizon problem.

\item Illustrate graphically the solution of the horizon problem by the inflation scenario.
\paragraph{Solution.} 
\begin{figure}[htb]
\center
\includegraphics*[width=0.45\textwidth]{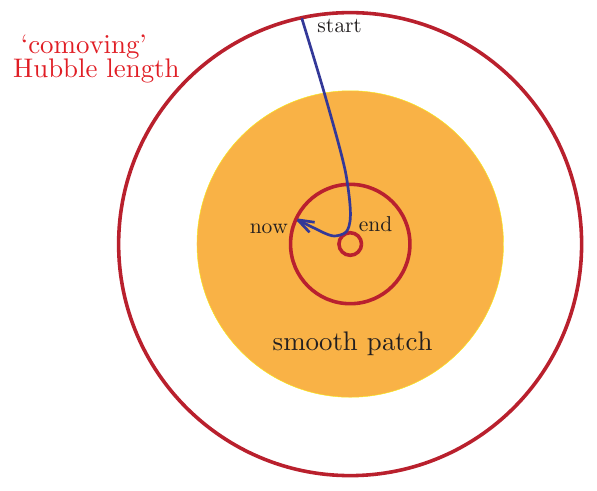}\hfill
\includegraphics*[width=0.5\textwidth]{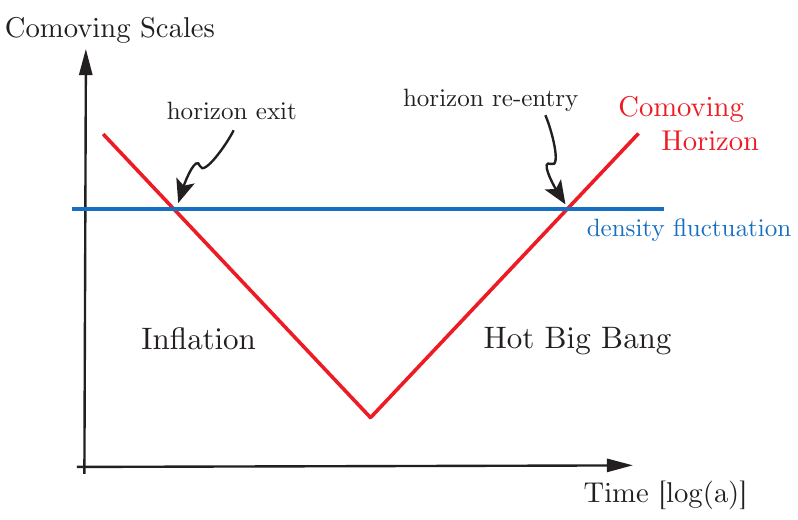}
\caption{\label{Conf-Baumann} Solution of the horizon problem by the inflation scenario \cite{Baumann}}
\end{figure}
On the left is evolution of the comoving Hubble radius in the inflationary universe. The comoving Hubble sphere shrinks during inflation and expands after inflation. What is important for the horizon problem is not the Hubble radius itself, but the particle horizon. In simple models of inflation, such as de Sitter inflation, they coincide. When the comoving particle horizon behaves as shown on the figure, inflation works as a mechanism to `zoom-in' on a smooth sub-horizon patch. 

Right: Solution of the horizon problem. All scales that are relevant to cosmological observations today were larger than the horizon scale until $a\sim 10^{-5}$. However, at sufficiently early times, these scales were smaller than the horizon and therefore causally connected. Similarly, the scales of cosmological interest came back within the horizon at relatively recent times. 
\end{enumerate}

\textbf{Growth of perturbations}
\begin{enumerate}[resume]

\item Show that any mechanism of generation of the primary inhomogeneities in the Big Bang model violates the causality principle.
\paragraph{Solution.} The wavelength of any perturbation $\lambda_P$, as any linear scale, grows as $\lambda_{P}\sim a(t)$. The Hubble radius is $R_H =H^{-1}=a/\dot{a}$. If $a\sim t^q$, then $R_H \sim t \sim a^{1/q}$, therefore
\begin{equation}
\frac{\lambda_P}{R_H}\sim a^{(q-1)/q}.
\end{equation}
Both for radiation ($q=1/2$) and dust ($q=2/3$), $q<1$, so $\lambda_P /R_H$ decrease with time. This leads to conclusion that initial perturbations must be correlated on the scales much larger than the Hubble radius. Therefore any mechanism of generation of the primary inhomogeneities in the Big Bang model will contradict the causality principle. If the mechanism is causal, then the corresponding scale must be less than the Hubble radius, i.e. $\lambda_P < R_H$. However, in case of domination of radiation or matter, for sufficiently small $a$ this cannot be true.

\item How should the early Universe evolve in order to make the characteristic size $\lambda_P$ of primary perturbations decrease faster than the Hubble radius, if one moves backward in time?
\paragraph{Solution.} The condition can be reformulated as
\begin{equation}
-\frac{d}{dt} \frac{\lambda_P}{R_H}<0 .
\end{equation}
As $\lambda_P \sim a$, and $R_H \sim a/\dot{a}$, this is equivalent to $\ddot{a}>0$, i.e. the expansion must be accelerated. In other words, if the mechanisms responsible for structure formation are causal, the Universe must have expanded with acceleration in the past. 
\end{enumerate}

\section{Holography}
In the context of holographic description of the Universe (see the minimal introduction on the subject in the corresponding Chapter)
 the Hubble sphere is often treated as  the holographic screen, and consequently called a horizon, although technically it is not.

\begin{enumerate}[resume]
\item Formulate the problem of the cosmological constant (see chapter on Dark Energy) in terms of the Hubble radius.
\paragraph{Solution.} If the dark energy in indeed the cosmological constant, then it introduces a fundamental length scale to the theory
\begin{equation}
 L_\Lambda \equiv H_\Lambda^{-1},
 \end{equation} 
 related to the constant dark energy density $\rho_\Lambda$ through the Friedman equation
 \begin{equation}
 H_\Lambda^2 =\frac{8\pi G}{3}\rho_\Lambda .
 \end{equation}
 This quantity can be interpreted as the Hubble radius in the universe filled only with this dark energy and no other matter. There is another fundamental length, the Planck length $l_P$. The dimensionless combination
 \begin{equation}
\Big( \frac{R_H}{l_P}\Big)^2 \sim 10^{123}.
 \end{equation}
Existence in the theory of two fundamental length scales --- different to this extent --- comprises the cosmological problem.

\item Choosing the Hubble sphere as the holographic screen, find its area in the de Sitter model (recall that in this model the Universe's dynamics is determined by the cosmological constant $\Lambda >0$).
\paragraph{Solution.} The area is
\begin{equation}
A=4\pi R_H^2 =\frac{12\pi}{\Lambda}.
\end{equation}

\item Find the Hubble sphere's area in the Friedman's Universe with energy density $\rho$.
\paragraph{Solution.} For a non-flat Universe the Hubble radius is ($c=1$)
\begin{equation}
R_H \equiv H^{-1}=\Big(\frac{8\pi G}{3} H^2 -\frac{k}{a^2} \Big)^{-1/2},
\end{equation}
so the area is
\begin{equation}
A=\frac{1}{\frac{2G}{3}\rho -\frac{4\pi k}{a^2}}.
\end{equation}

\item Show that in the flat Friedman's Universe filled with a substance with state equation $p=w\rho$ the Hubble sphere's area grows with the Universe's expansion under the condition $1+w >0$.
\paragraph{Solution.} Using the result of the previous problem, we have
\begin{equation}
A'=-\frac{3}{2G}\frac{\rho'}{\rho^3},
\end{equation}
while from the conservation equation
\begin{equation}
\frac{\rho'}{\rho}=-3(1+w)\frac{a'}{a}.
\end{equation}
Differentiating with respect to $a$, we get
\begin{equation}
\frac{dA}{da}=\frac{9}{2G}\frac{1+w}{a\rho},
\end{equation}
from which it follows that $A'$ has the same sign as $1+w$.

\item Estimate the temperature of the Hubble sphere $T_H$ considering it as the holographic screen.
\paragraph{Solution.} The temperature of a horizon is calculated with the help of Hawking radiation formula, which gives the temperature of a massless field near the horizon of a black hole of mass $M$:
\begin{equation}
 T_{BH}=\frac{\hbar c^3}{8\pi G\; k_B M}.
\end{equation} 
Now using it for the Hubble sphere instead, we treat $M$ as the mass of the enclosed volume, and using the Friedman equations to get $\rho (H)$, obtain
\begin{equation}
 T_{BH}=\frac{\hbar c^3}{8\pi k_B}
 	\frac{1}{\Big(\frac{8\pi G}{3}\rho\Big)\Big(\frac{c}{H}\Big)^3}
 		=\frac{\hbar H}{4\pi k_B}.
 \end{equation} 
The numerical value is
 \begin{equation}
 T_{BH}\sim 10^{-39}\; K.
 \end{equation}
 
\item Taking the Hubble sphere for the holographic screen and using the SCM parameters, find the entropy, force acting on the screen and the corresponding pressure.
\paragraph{Solution.} The entropy is
\begin{equation}
S_{H}=\frac{1}{4}\frac{A}{l_{P}^2},
\end{equation}
where $l_P =\sqrt{\hbar G /c^3}$ is the Planck length, so substituting $A=4\pi R_H^2$, we get
\begin{equation}
S_{H}=\frac{k_B c^3}{G\hbar}\cdot \pi R_H^2.
\end{equation}
The force acting on the holographic screen is
\begin{equation}
F_{H}=-\frac{dE}{d R_H} =-T_{H}\frac{dS_H}{dR_H}
	=-T_H\frac{2\pi k_B c^3}{G\hbar}\cdot R_H
		=-\frac{c^4}{2G}\sim 0.6\times 10^{44}\;N .
\end{equation}
It is a fundamental constant -- half of the Planck force. The pressure is $P=F_H /A$.

\item The most popular approach to explain the observed accelerated expansion of the Universe assumes introduction of dark energy in the form of cosmological constant into the Friedman equations. As seen in the corresponding Chapter, 
this approach is successfully realized in SCM. Unfortunately, it leaves aside the question of the nature of the dark energy. An alternative approach can be developed in the frame of holographic dynamics. In this case it is possible to explain the observations without the dark energy. It is replaced by the entropy force, which acts on the cosmological horizon (in this case it is the Hubble sphere) and leads to the accelerated expansion of the Universe. Show that the Hubble sphere's acceleration obtained this way agrees with the result obtained in SCM.
\paragraph{Solution.} From the Hubble law the acceleration of an object placed at distance $R$ from the observer is
\begin{equation}
\dot{V}=R (\dot{H}+H^2);
\end{equation}
while from the second Friedman equation
\begin{align}
\dot{H} + H^2 &=  - \frac{4\pi G}{3}(\rho  + p) \\
	&=  - \frac{4\pi G}{3}( - 2\rho_\Lambda + \rho _m)  \\
	&= \frac{8\pi G}{3}\big(\rho_\Lambda - \tfrac{1}{2}\rho_m \big)\\
	&= H_0^2 \big(\Omega_\Lambda - \tfrac{1}{2}\Omega_m \big).
\end{align}
Thus
\begin{equation}
\dot V = H_0^2 R \big(\Omega_\Lambda - \tfrac{1}{2}\Omega_m \big).
\end{equation}
For $R=4000 Mpc$ (the size of the observable Universe), $\dot{V}\approx 4\times 10^{-10} m/s^2$.

Acceleration of the holographic screen (Hubble sphere in this case) is given in holographic dynamics  by 
\begin{equation}
a_H =\frac{2\pi k_B c}{\hbar} T_H,
\end{equation}
so using the expression for temperature
\begin{equation}
T_H =\frac{\hbar H}{4\pi k_B},
\end{equation}
we get
\begin{equation}
a_H =\dot{V}_H =cH\sim 10^{-9}\;m/s^2 .
\end{equation}
The results agree by the order of magnitude with the ones obtained in the frame of the SCM.

\item Show that a pure de Sitter Universe obeys the holographic principle in the form\footnote{This and the next problem are based on \cite{Padmanabhan}, see there for more details.}
\begin{equation}
N_{sur}=N_{bulk}.
\end{equation}
Here $N_{sur}$ is the number of degrees of freedom on the Hubble sphere, and $N_{bulk}$ is the effective number of degrees of freedom, which are in equipartition at the horizon temperature $T_H =\hbar H /2\pi k_B$.
\paragraph{Solution.} The degrees of freedom of the Hubble sphere are given by
\begin{equation}
N_{sur}=\frac{1}{4}\frac{R_H^2}{l_P^2}=4\pi \Big(\frac{c/H}{l_P}\Big)^2 .
\end{equation}
On the assumption of equipartition, the number of bulk degrees of freedom is
\begin{equation}
N_{bulk}=\frac{|E|}{k_B T /2}.
\end{equation}
Taking $E$ as the ``energy'' $(\rho+3p)V_H c^2$ enclosed by the Hubble sphere 
\begin{equation}
N_{bulk}=-\frac{2 (\rho+3p)\cdot \frac{4\pi}{3}\frac{c^3}{H^3}\cdot c^2}{k_B T_H},
\end{equation}
and using Friedman equations to find $\rho+3p =-2\rho$ through $H$, one gets
\begin{equation}
N_{bulk}=4\pi \Big(\frac{c/H}{l_P}\Big)^2  ,
\end{equation}
which proves the initial statement.

\item If it is granted that the expansion of the Universe is equivalent to the emergence of space (in the form of availability of greater and greater volumes of space), then the law governing this process must relate the emergence of space to the difference 
\begin{equation}
 N_{sur}-N_{bulk}
\end{equation} 
(see previous problem). The simplest form of such a law is
\begin{equation}
\Delta V\sim ( N_{sur}-N_{bulk}) \Delta t .
\end{equation}
We could imagine this relation as a Taylor series expansion truncated at the first order. Show that this assumption is equivalent to the second Friedman equation.
\paragraph{Solution.} Reintroducing the Planck scale and passing to infinitesimal increments $\Delta V\to dV$, we get
\begin{equation}
\frac{dV}{dt}=l_P^{2} (N_{sur}-N_{bulk}).
\end{equation}
Then substituting $V=\frac{4\pi}{3}(c/H)^3$, $N_{sur} =4\pi (c/l_P H)^2$ and \begin{equation}
N_{bulk}=-\frac{2E}{k_B T}=-\frac{2(\rho+3p)Vc^2}{k_B T_H},
\end{equation}
we find
\begin{equation}
\frac{\ddot{a}}{a}=-\frac{4\pi l_P^2}{3}(\rho+3p).
\end{equation}
\end{enumerate}


\begin{thebibliography}{99}
\bibitem{Harrison} E. Harrison, \textit{Cosmology: the science of the Universe}, CUP (1981),  ISBN~978-0-521-66148-5.

\bibitem{Rindler} W. Rindler, Visual horizons in world-models, \href{http://adsabs.harvard.edu/abs/1956MNRAS.116..662R}{\textit{MNRAS} \textbf{116} (6), 662--677 (1956)}.

\bibitem{Ellis} G.F.R. Ellis and T. Rothman, Lost horizons, \textit{Am. J. Phys.} \textbf{61}, 883 (1993).

\bibitem{Muchanov} V.F. Muchanov, \textit{Physical foundations of cosmology}, CUP (2005), ISBN~0-521-56398-4.

\bibitem{BrRubin} K.A. Bronnikov and S.G. Rubin, \textit{Black holes, cosmology and extra dimensions}, WSPC (2012), ISBN~978-9814374200.

\bibitem{Melia} F. Melia, The cosmic horizon.  \textit{MNRAS} \textbf{382} (4), 1917--1921 (2007) [\href{http://arxiv.org/abs/0711.4181}{arXiv:0711.4181}];  F.~Melia and M.~Abdelqader, The Cosmological Spacetime,  \textit{Int. J. Mod. Phys. D} \textbf{18}, 1889 (2009) [\href{http://arxiv.org/abs/0907.5394}{arXiv:0907.5394}].

\bibitem{Baumann} D. Baumann, TASI Lectures on Inflation, [\href{http://arxiv.org/abs/0907.5424}{arXiv:0907.5424}].

\bibitem{Padmanabhan} T. Padmanabhan, Emergent perspective of Gravity and Dark Energy, \textit{Research in Astron. Astrophys.} \textbf{12}, No. 8, 891–916 (2012) [\href{http://arxiv.org/abs/1207.0505}{arXiv:1207.0505}].
\end{thebibliography}
\end{document}